\newif\ifarxiv\arxivtrue \documentclass[a4paper,UKenglish,autoref,thm-restate]{lipics-v2021}

\makeatletter
\@ifundefined{ifarxiv}{\newif\ifarxiv
\arxivfalse
}{}
\makeatother

\nolinenumbers
    \hideLIPIcs

\def\ArxivAppendixRef#1#2{\autoref{#2}}
\def\ArxivLemmaRef#1#2{\autoref{#2}}

\title{A Multi-Prover Interactive Protocol for Certification of Resolution Refutations}

\author{Philipp Czerner}{Department of Informatics, TU München, Germany}{czerner@in.tum.de}{https://orcid.org/0000-0002-1786-9592}{}{}
\unless\ifx\authoranonymous\relax \author{Adrian Krauss}{Department of Informatics, TU München, Germany}{adrian.krauss@tum.de}{https://orcid.org/0009-0003-2693-0740}{}{}
\author{Josia Pietsch}{Department of Informatics, TU München, Germany}{josia.pietsch@tum.de}{https://orcid.org/0009-0002-2634-3469}{}{}
\author{Javier Esparza}{Department of Informatics, TU München, Germany}{esparza@cit.tum.de}{https://orcid.org/0000-0001-9862-4919}{}{}
\fi

\authorrunning{P. Czerner, A. Krauss, J. Pietsch and J. Esparza} 

\Copyright{Philipp Czerner, Adrian Krauss, Josia Pietsch and Javier Esparza} 

\ccsdesc[500]{Theory of computation~Logic and verification} 

\keywords{Resolution, interactive proof systems}

\relatedversion{} \relatedversiondetails{Extended Version}{http://arxiv.org/abs/2611.14783} 

\supplement{}
\supplementdetails[subcategory={Source Code}, cite={}, swhid={}]{Software}{https://gitlab.lrz.de/i7/rescertify}

\funding{This work was supported by the DFG project 568501586 "Competitive Interactive Certification for Automated Reasoning".}

\EventShortTitle{ArXiV}
\nolinenumbers

\RequirePackage{mathtools}
\RequirePackage{makecell}

\RequirePackage[english,linesnumbered,ruled,vlined]{algorithm2e}
\DontPrintSemicolon
\SetKwProg{Procedure}{Procedure}{}{} \SetKw{Accept}{accept}               \SetKw{Reject}{reject}               \SetKw{Assert}{assert}               \SetKw{QueryOracle}{query oracle}    \SetKw{QueryProver}{query prover}    

\NewEnvironmentCopy{displaymathifenoughspace}{displaymath}

\makeatletter \newcommand{\FixEnumerateItemRefParentheses}[2][]{\renewcommand{\p@enumi}[1]{(#1##1)}#2}\makeatother 

\newcounter{customenumstar}
\newenvironment{enumerate*}[1][]{\unskip \setcounter{customenumstar}{0}\def\item{\stepcounter{customenumstar}(\roman{customenumstar})
    }}{}

\newcounter{reusethm}
\newcounter{savedtheorem}\newenvironment{reusetheoremnumber}[1]{\stepcounter{reusethm} \setcounter{savedtheorem}{\value{theorem}}\setcounter{theorem}{#1}\addtocounter{theorem}{-1}\let\savedtheHtheorem\theHtheorem \def\theHtheorem{reused.\thereusethm.theorem.\thetheorem}}{\let\theHtheorem\savedtheHtheorem \setcounter{theorem}{\value{savedtheorem}}}
\RequirePackage{xcolor}
\RequirePackage{stmaryrd}

\def\O{\mathcal{O}}

\newcommand{\NP}{\mathsf{NP}}
\newcommand{\coNP}{\mathsf{coNP}}

\newcommand{\NEXP}{\mathsf{NEXP}}

\newcommand{\Abs}[1]{\mathopen|#1\mathclose|}

\definecolor{nicebg}{HTML}{f6f0e4}
\definecolor{nicered}{HTML}{7f0a13}
\definecolor{nicebgred}{HTML}{f2e7e8}
\definecolor{niceblue}{HTML}{104354}
\definecolor{nicebgblue}{HTML}{e8edee}
\definecolor{nicegreen}{HTML}{217516}
\definecolor{nicebggreen}{HTML}{e9f1e8}
\definecolor{nicepurple}{HTML}{884bab}
\definecolor{nicebgpurple}{HTML}{f3edf7}
\definecolor{niceorange}{HTML}{d27c11}
\definecolor{nicebgorange}{HTML}{fbf2e8}
\definecolor{nicepink}{HTML}{e95f9f}
\definecolor{nicebgpink}{HTML}{fdeff6}
\definecolor{niceredlight}{HTML}{c9888d}
\definecolor{nicebluelight}{HTML}{78a4b8}
\definecolor{nicegreenlight}{HTML}{76de68}
\definecolor{nicepurplelight}{HTML}{bc87db}
\definecolor{niceredbright}{HTML}{bd0310}
\definecolor{nicebgredbright}{HTML}{f9e6e8}
\definecolor{nicebluebright}{HTML}{197b9b}
\definecolor{nicebgbluebright}{HTML}{e8f2f5}

\newenvironment{noitemize}{\unskip \let\item\undefined \newcommand{\item}[1][]{}}{}

\newenvironment{movetoappendix}{\unskip \comment
}{\endcomment }

\usepackage{IEEEtrantools} 
\RequirePackage{xparse}
\RequirePackage[most]{tcolorbox}

\DeclareDocumentEnvironment{DecisionProblem}{m}{\begin{tcolorbox}[adjusted title=#1,colback=white,coltitle=black,colbacktitle=white,boxrule=1pt,enhanced,attach boxed title to top left={xshift=0.5cm,yshift=-1.5mm},boxed title style={boxrule=1pt},lower separated=false]}{\end{tcolorbox}
}
\DeclareDocumentCommand{\ProblemInput}{m}{\textbf{Input:} #1\\
}
\DeclareDocumentCommand{\ProblemOutput}{m}{\textbf{Output:} #1
}
\RequirePackage{adjustbox}

\let\O\undefined
\newcommand{\O}{\mathcal{O}}
\newcommand{\Poly}{\operatorname{poly}}
 \newcommand{\sem}[1]{\llbracket #1 \rrbracket}
\newcommand{\Eval}[1]{\sem{{#1}}} \newcommand{\EvalBig}[1]{\big\llbracket #1 \big\rrbracket}
\newcommand{\pev}[1]{\pi_{[{#1}]}}
\newcommand{\Pev}[1]{{\mathrm{\Pi}}_{{#1}}}

\newcommand{\overwriteVars}[2]{\pev{#1 \coloneqq #2}}

\newcommand{\Alg}[1]{\textnormal{\textsc{#1}}}
\newcommand{\EACertify}{\Alg{EACertify}}
\newcommand{\ResCertify}{\Alg{ResCertify}}

\newcommand{\MIP}{\textnormal{\textsf{MIP}}}
\newcommand{\EACSAT}{\textnormal{\textsf{EAC-SAT}}}

\newcommand{\IndexPoly}[2]{\langle #2 \rangle_{#1}}
\newcommand{\MultilinEncode}[2]{\langle #2 \rangle_{#1}}

\makeatletter
\DeclareRobustCommand{\shortto}{\mathrel{\mathpalette\short@to\relax}}

\newcommand{\short@to}[2]{\clipbox{0 0 {.55\width} 0}{$\m@th#1\vphantom{+}{\leftarrow}$}}
\makeatother

\newcommand{\IndexA}[2]{#1{\shortto}\,#2}

\newcommand{\Lit}{\operatorname{Lit}} \newcommand{\Arith}[1]{\langle #1 \rangle} \newcommand{\ArithLit}[1]{\mathfrak{L}({#1})} \newcommand{\Wres}{W_{\scriptscriptstyle \text{\textup{RES}}}} \newcommand{\Wrup}{W_{\scriptscriptstyle \text{\textup{RUP}}}} 

\newcommand{\DegreeReduction}[1]{\delta_{#1}}

\newcommand{\maxdeg}{\operatorname{maxdeg}}
\newcommand{\BooleanDomain}{\mathbb{B}} \newcommand{\F}{\mathbb{F}} 

\newcommand{\Multilinear}[1]{\F[#1]_{\leq1}} \newcommand{\PartEval}[1]{\pi_{[#1]}}

\newcommand{\restrict}[1]{{\upharpoonright}_{#1}}

\newcommand{\qeq}{\overset{\smash{\scriptscriptstyle ?}}{=}}

\newcommand{\concat}{\mathbin{+\mkern-10mu+}} 

\tikzstyle{inode}=[minimum width=1cm, minimum height=0.5cm,text centered]

\newcommand{\Bit}[2]{#2_{#1}}
\newcommand{\ArithBit}[2]{\langle#2_{#1}\rangle}

\newcommand{\Res}{\operatorname{Res}} \newcommand{\RUPStep}{\operatorname{Step}}

\newcommand{\GadgetBinary}{\operatorname{\textup{\texttt{binary}}}}
\newcommand{\GadgetInitial}{\operatorname{\textup{\texttt{initialClauses}}}}
\newcommand{\GadgetEmptyClause}{\operatorname{\textup{\texttt{emptyClause}}}}
\newcommand{\GadgetFunction}{\operatorname{\textup{\texttt{function}}}}
\newcommand{\GadgetRegressive}{\operatorname{\textup{\texttt{regressive}}}}
\newcommand{\GadgetResolution}{\operatorname{\textup{\texttt{resolution}}}}
\newcommand{\GadgetRUP}{\operatorname{\textup{\texttt{rup}}}}
\newcommand{\GadgetSort}{\operatorname{\textup{\texttt{sort}}}}

\SetKwFunction{ProcedureUnifyClaims}{UnifyClaims}
\newcommand{\UnifyClaims}{\hyperref[alg:unify-claims]{\ProcedureUnifyClaims}}

\newcommand{\NewNotation}[1]{\phantomsection\label{#1}}

\tikzstyle{yindexnode} = []
\tikzstyle{yindexleaf} = [fill=lipicsYellow, rounded corners=0.5mm]
\tikzstyle{yindexedge} = [->, draw=lipicsGray, color=lipicsGray, shorten >=0.5mm, shorten <=0.5mm]

\tikzstyle{eacnode} = [rectangle, rounded corners, minimum height=.5cm, text centered, draw=black]
\tikzstyle{eacroot} = [eacnode]
\tikzstyle{eacleaf} = [eacnode]
\tikzstyle{eacedge} = [thick,->,>=stealth] \tikzstyle{eacnode} = [rectangle, rounded corners, minimum height=.5cm, text centered, draw=black]
\tikzstyle{eacevallabel} = [text=lipicsGray] \tikzstyle{eacnodelabel} = [text=lipicsGray] \tikzstyle{eacclaim} = [text=black, rounded corners=0.5mm, fill=lipicsYellow, font={\linespread{1.5}\tiny\selectfont}]

\tikzstyle{eacertifystep} = [thick,->,
dotted, shorten <=5mm, shorten >=5mm, color=lipicsGray, rounded corners ]
\tikzstyle{verifier} = [font={\linespread{0}\small\selectfont},
    align=left]
\tikzstyle{prover} = [font={\linespread{0}\small\selectfont},
    align=left]
\usetikzlibrary{positioning, arrows, shapes, arrows.meta, patterns.meta, fit}

\let\ifwip\iffalse

\newtheorem{fact}[theorem]{Fact}

    \let\todo\undefined \NewDocumentCommand{\todo}{ O{} O{} m O{}}{}
    \NewDocumentCommand{\todospace}{ O{} m}{}

\SetSymbolFont{stmry}{bold}{U}{stmry}{m}{n}

\begin{document}

\maketitle

 \begin{abstract}
A resolution refutation of a SAT-formula $\varphi$ is an independently checkable certificate proving that $\varphi$ is unsatisfiable. However, the refutations of $\varphi$ can be exponentially larger than $\varphi$. So, a verifier needs worst-case exponential time in the size of $\varphi$ to check their correctness. We present $\ResCertify$, an interactive probabilistic protocol in which the verifier interacts with multiple independent, \emph{untrusted} provers ($\MIP$) to check the refutation with very low probability of error. The verifier runs in linear time in the size of $\varphi$.
The provers require time $\O(n \cdot |\Phi|)$, where $\Phi$ is the resolution refutation and $n$ is the number of variables of  $\varphi$.
If they can be trusted to act as a multilinear oracle on certain queries,
a single prover suffices, and the runtime can be improved to $\O(|\Phi|)$.
Finally, while the focus of our work is theoretical, we also report on a proof-of-concept implementation of $\ResCertify$.
\end{abstract}

 \section{Introduction}
Certifying SAT-solvers produce either a satisfying assignment or a proof of unsatisfiability in some proof system, e.g.\ resolution (see \cite{Heule21} for a survey). The proof is usually called an UNSAT \emph{certificate}.

We consider a setting where the SAT-solver runs on a powerful untrusted machine capable of storing large certificates, but the certificate is checked by an independent trusted verifier only able to invest polynomial time in the size of the formula (not in the size of the certificate!).
In this case, checking the certificate, or even fetching it, is infeasible unless $\NP = \coNP$.\footnote{Note that a proof system with polynomial-size certificates verifiable in polynomial time exists if and only if $\text{UNSAT} \in \NP$.}

By introducing randomness,
interactive proof protocols provide a theoretical solution \cite{GoldwasserMicaliRackoff,Babai,LundFKN92,AroraBarak}.
Intuitively, instead of fetching and checking the certificate, the verifier asks questions about it of an agent, called a \emph{prover}, which we can imagine running on the same machine as the SAT-solver. The questions are selected following an \emph{interactive protocol}. The protocol guarantees that the verifier employs only randomized polynomial time, accepts every correct certificate (completeness), and rejects every faulty certificate with high probability (soundness).\footnote{The verifier rejects a certificate if{}f the answers given by the prover are incompatible with a correct certificate; this can happen because the certificate is faulty, or because it is correct but the provers give wrong answers to the verifier's questions.}
Compared to conventional certificate checkers,
interactive protocols incur a greater total computational cost but reduce the workload of the trusted verifier.

The main obstacle to making interactive proof protocols interesting to SAT-solving practice is the \emph{runtime overhead} of the prover. For example, the original UNSAT protocol of \cite{LundFKN92} essentially requires the prover to  compute the complete truth-table of the formula.

Czerner et al.~have recently proposed to study interactive proof protocols in which the prover has small overhead relative to SAT-solving algorithms more efficient in practice \cite{CzernerEK24}. As a first step, they present an interactive proof system for Davis-Putnam resolution proofs~\cite{Harrison09} enjoying the following property: Given an unsatisfiable formula $\varphi$ and a Davis-Putnam resolution proof $\Phi$ of $\varphi$, prover and verifier run in  $\O(n^2 |\Phi|)$ and $\O(n^2 + |\varphi|)$ time, respectively, with error probability $\O(\Poly(\Abs{\varphi})/|\F|)$, where $n$ is the number of variables in $\varphi$ and $\F$ a finite field. So, loosely speaking, a SAT-solver producing Davis-Putnam resolution proofs can be extended with fast interactive certification for the verifier while keeping the overhead for the prover very low.

The Davis-Putnam procedure is a significant improvement over brute force, but generates much larger resolution proofs than modern techniques. In this paper we address the following question: Is there an interactive proof system whose prover has linear overhead for \emph{arbitrary} resolution proofs?  We show that the answer is affirmative if we move from \emph{single-prover} to \emph{multiple-prover} interactive proof systems ($\MIP$).  In this setting, the verifier asks questions to multiple provers (in our case just two) who, crucially, are not allowed to exchange information.

\subparagraph{Related Work.}
To the best of our knowledge, no interactive protocol specific to automated reasoning has yet been proposed or implemented, besides the ones proposed by Couillard et al.~\cite{ip-pspace-practical-cav} and Czerner et al.~\cite{CzernerEK24}. There exist multiple works on verifying generic computations without re-executing them~\cite{WalfishB15, ZhangGKPP18,OzdemirBW22,YadavG23}, but with numerical applications in mind.
There is also extensive research on probabilistically checkable proofs (PCPs), which has produced PCP protocols for algebraic constraint satisfaction problems with quasi-optimal provers (see e.g. \cite{BenSassonCGT13}), with applications to cryptographic protocols. None of these papers considers the problem of certifying the output of a SAT-solver, and we are not aware of any implementations.
Other advances have been made in the field of zero-knowledge proofs~\cite{ZhangLWZSXZ21,LauferOB24, OzdemirWBB25}, which, in the context of SAT-solving, would amount to proving the existence of a satisfying assignment without revealing information about it.

\subparagraph{Our Contribution.}
We present a sound and complete $\MIP$-protocol, $\ResCertify$, between two provers and one verifier.\footnote{More precisely, one prover and one oracle suffice. Loosely speaking, an oracle is a ``memoryless'' prover whose answers do not depend on previous answers. For formal definitions see \autoref{sec:mip}.}
$\ResCertify$ consists of two parts:
A \emph{multilinearity test},
which amounts to a syntactic check that the resolution proof has the correct format,
and a \emph{zero test},
which checks that the resolution proof is correct.
We describe the protocol in terms of a verifier interacting with one prover and one \emph{oracle} (see \autoref{sec:mip} for details) and prove:

\begin{theorem}[informal]\label{thm:main}
	Let $\varphi$ be an unsatisfiable  CNF-formula with $n$ variables and let $\Phi$ be a resolution proof of $\varphi$.
    The verifier of $\ResCertify$ requires time  $\O(|\varphi| + \log(|\Phi|))$.
    To answer the questions asked by the verifier,
    the prover and the oracle need time  $\O(n \cdot |\Phi|)$ for the multilinearity test
    and $\O(|\Phi|)$ for the zero test.
\end{theorem}

\autoref{thm:main} is proved in three steps:
\begin{itemize}
    \item  \autoref{sec:gadgets} introduces \emph{polynomial refutations}, triples of polynomials encoding  resolution proofs.   Further, it defines \emph{extended arithmetic circuits} (EACs), arithmetic circuits that can contain \emph{function variables}, and their satisfiability problem $\EACSAT$. Finally, it shows how to transform a given CNF-formula $\varphi$ into an EAC $c_\varphi$ that is satisfiable if{}f $\varphi$ has  a polynomial refutation.
    \item  \autoref{sec:eacertify} presents $\EACertify$, an $\MIP$-protocol for $\EACSAT$.
    \item $\ResCertify$ constructs the EAC $c_\varphi$ and then invokes $\EACertify$.
       \autoref{sec:efficient-prover} describes a prover for $\ResCertify$.
\end{itemize}
An overview of $\ResCertify$ is shown in \autoref{fig:overview}.

While our work is theoretical, we also report on a proof-of-concept implementation of $\ResCertify$. In a set of formulas (taken mostly from satlib \cite{satlib}) with up to 250 variables and 1688 clauses, the runtime of the verifier is reduced by several orders of magnitude compared to \texttt{tracecheck}, a non-interactive checker for resolution proofs, and the speed-up factor grows rapidly with the size of the formula; the price to pay is a constant factor of $\approx 150$ in the provers' overhead when compared to \texttt{picosat}.

\begin{figure}[t]
	\centering
	\scalebox{0.85}{
        \begin{tikzpicture}
            \def\refColor#1{{\color{lipicsGray}#1}}
            \tikzstyle{protocol-component} = [rectangle, draw=black, rounded corners=1mm, text width=\boxWidth, align=center]
            \tikzstyle{protocol-component-label} = [rectangle, draw=black, fill=white, rounded corners=1mm, align=center,anchor=east]
            \tikzstyle{protocol-component-label-shifted} = [protocol-component-label, xshift=-0.75cm]
            \tikzstyle{protocol-component-step} = [rounded corners=1mm, text width=\boxWidth-0.5cm, align=center]
            \tikzstyle{trusted} = [fill=lipicsYellow]
            \tikzstyle{semitrusted} = [pattern={Lines[angle=45, distance=2pt, line width=1pt]}, pattern color=lipicsYellow]
            \tikzstyle{dataflow} = [->, shorten >=0.5mm, shorten <=0.5mm]
            \tikzset{interaction/.style={
                    postaction={draw, dataflow, transform canvas={yshift=2.5pt}},
                    postaction={draw, dataflow, <-, transform canvas={yshift=-2.5pt}},
            }}
            \def\proverTop{-1.1}
            \def\boxWidth{5cm}
            \node [align=center, anchor=north] (phi) at (0,0) {CNF $\varphi$\strut};
            \node [align=center, anchor=north west] (Phi) at (\boxWidth, 0)
                {RES-Refutation $\Phi$\strut};
            \node [align=center, anchor=north] at ([yshift=0.3cm]Phi.south)
                {\refColor{(\autoref{def:cnfres})}};

\node[protocol-component, minimum height=3.75cm, anchor=north]
                (prover) at (\boxWidth,\proverTop) {};
            \node[protocol-component-label-shifted]
                (labelProver) at (prover.north)
                {Prover};

\node[protocol-component, trusted, minimum height=3.75cm, anchor=north]
                (verifier) at (-\boxWidth,\proverTop) {};
            \node[protocol-component-label-shifted, trusted] (labelVerifier) at (verifier.north)
                {Verifier};
\draw[dataflow] (phi) -| ([xshift=-0.8cm]prover);
            \draw[dataflow] (Phi) -| ([xshift=-0.5cm]prover);
            \draw[dataflow] (phi) -| (verifier);

            \node[protocol-component-step] (polyRefutation) at (\boxWidth,\proverTop-1)
                {Construct polynomial refutation $\omega$ of length $M$\\\refColor{(\autoref{def:poly-ref})}};
            \node[protocol-component-step] (constructEACs) at (polyRefutation -| verifier.center)
                {Construct EAC $\psi$\\\refColor{(Section \ref{sec:eac-gadgets})}};
            \draw[dataflow] (polyRefutation -| prover.west) -> node [above] {$M$} (constructEACs-| verifier.east);

            \node[protocol-component-step] (multilinTest) at (-1*\boxWidth,\proverTop-2)
                {Check multilinearity of $\omega$};
            \node[protocol-component-step, semitrusted] (oracle) at (multilinTest -| prover.center)
                {Evaluate $\omega$};
            \path[interaction] (multilinTest -| verifier.east)
                -- node [rectangle, fill=white, text width=3cm, align=center] {Multilinearity Test}
                (multilinTest -| prover.west);

            \node[protocol-component-step] (eacSAT) at (-1*\boxWidth,\proverTop-3)
                {Check $\llbracket \psi \rrbracket_\omega = 0$\\\refColor{(\autoref{alg:verifier})}};
            \node[protocol-component-step] (proverStrategies) at (eacSAT -| prover.center)
                {Prover Strategies\\\refColor{(\autoref{sec:efficient-prover})}};
            \path[interaction] (eacSAT -| verifier.east)
                -- node [rectangle, fill=white] {Zero Test}
                (proverStrategies -| prover.west);

            \node[protocol-component, inner sep=1mm, densely dotted, draw=lipicsGray,
                fit=(multilinTest)(proverStrategies)(prover.south)(prover.east)(verifier.west)] (eacertify) {};
            \node[protocol-component-label, anchor=center, densely dotted, draw=lipicsGray, text width=2cm]
                (eacertifyLabel) at (eacertify.south)
                {$\EACertify$\\\refColor{(\autoref{sec:eacertify})}};
	\end{tikzpicture}}
    \caption{An overview of $\ResCertify$ for an unsatisfiable CNF formula $\varphi$. Trusted parts are shown in yellow.
        As input, the prover gets $\varphi$ and a (potentially very large) resolution refutation $\Phi$ of $\varphi$.
        $\Phi$ is turned into a polynomial refutation $\omega$, and the length $M$ of $\omega$ is sent to the verifier.
        From $\varphi$ and $M$, the verifier constructs a certain EAC, $\psi$.
        Then, using the protocol $\EACertify$,
        the prover convinces the verifier that $\psi$ is satisfiable.
        $\EACertify$ consists of two parts: a multilinearity test and a zero test.
        During the multilinearity test, the prover is trusted to act as a memoryless oracle.}
	\label{fig:overview}
\end{figure}
 \section{Preliminaries}
 \subsection{MIP}\label{sec:mip}

In a multi-prover interactive protocol for a language $L$,
a \emph{verifier} interacts with several \emph{provers}.
The verifier can ask each prover questions to decide $x \in L$ for a given $x$.
The goal is that if $x \in L$ holds and the provers are \emph{honest}, meaning that they give correct answers to the verifier's questions, then the verifier accepts.
If $x \notin L$ holds, then the verifier rejects with high probability, independently of whether the answers of the provers are correct or not.

Formally, the verifier is a deterministic polynomial time Turing machine $V$
that is provided with randomness $r$,
a bit string chosen uniformly at random at the start of the protocol.
The provers are two computationally unbounded Turing machines, $P_1$ and $P_2$.
The protocol consists of an exchange of messages between the verifier and each prover, taking place in rounds.
$P_1$'s response can depend only on $x$ and the messages previously exchanged with the verifier,
but not on the communication between the verifier and $P_2$ and vice versa.
Let $\Pr_r(V, P_1, P_2, x)$ denote the probability that
the verifier accepts the input $x$ after interacting with $P_1$ and $P_2$
when given randomness $r$.
\begin{definition}
    A multi-prover interactive proof system for a language $L$ consists of a verifier $V$ and provers $P_1$, $P_2$ with the following property:
    There exist polynomials $q_1$, $q_2$, $q_3$ such that
    on input $x \in \{0, 1\}^*$,
    the verifier runs in $q_1(|x|)$ time,
    uses randomness $r \in \{0,1\}^{q_2(|x|)}$,
    interacts with the provers in $q_3(|x|)$ rounds,
    and
    \begin{enumerate}
    	\item (\textit{Completeness}) $x \in L \implies \Pr_{r}(V, P_1, P_2, x) = 1$,
            and
        \item (\textit{Soundness}) for some constant $c < 1$, called the \emph{soundness parameter},
            $x \notin L \implies \forall Q_1,Q_2: \Pr_{r}(V, Q_1, Q_2, x) \leq c$.
    \end{enumerate}
    $\MIP$ denotes the set of languages that admit a multi-prover interactive proof system.
\end{definition}

Fortnow et al.\ showed that $\MIP$ can also be characterized via the interaction of a verifier with \emph{one} computationally unbounded \emph{oracle},
a memoryless prover whose answers do not depend on previous interactions \cite[Thm.~1]{power-of-mip}.
To pinpoint where exactly the requirement of memorylessness is needed, we
will describe our protocol in terms of a verifier interacting with one prover (where memorylessness is not assumed) and one oracle.

\begin{remark}
    \label{rem:oracle-trust-tradeoff}
In our analysis, we trust the prover to behave like an oracle for certain queries.
    Note that in practical applications of $\MIP$ protocols, a certain amount of trust is always required,
    since the verifier can neither check that a prover is memoryless nor that two provers are independent.
    In particular, $\MIP$ protocols can only provide \emph{safety} but are not suitable when \emph{security} is needed.
\end{remark}
 \subsection{Multilinear Binary Polynomials}
\label{subsection:multilinear-polynomials}
Consider a finite prime field
$\F$
\NewNotation{notation:F}
and let $\BooleanDomain \coloneqq \{0,1\} \subseteq \F$.\NewNotation{notation:B}
For a set of variables $X = \{x_0,\ldots,x_{n-1}\}$, let $\F[X]$\NewNotation{notation:polynomial_ring} denote the polynomial ring in  $X$ over $\F$.
We use the notation $\Pev{\sigma} p$\NewNotation{notation:Pev}
for the (partial) evaluation of polynomial $p \in \F[X]$ at
a (partial) assignment $\sigma$,
i.e., for $X' \subseteq X$ and $\sigma' \in \F^{X'}$,
we have $\Pev{\sigma'} p \in \F[X \setminus X']$,
and, in particular, $\Pev{\sigma} p \in \F$ for a total assignment $\sigma \in \F^X$.
For $\sigma \in \F^X$, $x \in X$ and $a \in \F$,
let $\sigma[a/x]$\NewNotation{notation:substitution}
denote the assignment obtained by substituting $a$ for $x$
and $\sigma[-/x] \in \F^{X \setminus \{x\}}$\NewNotation{notation:unbind} the assignment obtained by removing $x$ from the domain.
For a natural number $i \in  \{0,\ldots,2^{n}-1\}$,
let $\Bit{k}{i}$\NewNotation{notation:Bit}
be the  $k$-th bit of $i$,
i.e., $i = \sum_{k = 0}^{n-1} i_k 2^k$,
and for a sequence of bits $b_0,\ldots,b_{n-1} \in \BooleanDomain$,
let $(b_{n-1} \ldots b_0)_2$\NewNotation{notation:binary_number} denote the corresponding binary number,
i.e., $i = (i_{n-1} \ldots i_0)_2$.
We define an assignment corresponding to $i$, $(\IndexA{X}{i}) \in \BooleanDomain^X$,\NewNotation{notation:IndexA}
by $x_k \mapsto i_k$.

\subparagraph{Arithmetization.}
A propositional formula with variables $X = \{x_0,\ldots,x_{n-1}\}$ gives rise to a function $f\colon \BooleanDomain^X \to \BooleanDomain$.
For our interactive protocol, we need to extend such a function to a polynomial $\hat{f} \in \F[X]$,
i.e., find $\hat{f}$ with $\hat{f}\restrict{\BooleanDomain^X} = f$,
where $\hat{f}\restrict{\BooleanDomain^X}$\NewNotation{notation:restriction}
denotes the restriction of $\hat{f}$ to $\BooleanDomain^X$.
A polynomial is \emph{binary} if{}f it only takes values $0$ and $1$ on $\BooleanDomain^X$.
Note that for a function $\BooleanDomain^X \to \F$, there are many polynomials extending it.
We can enforce uniqueness by requiring the polynomial to be \emph{multilinear}, meaning that in every monomial the degree of every variable is at most $1$ (see e.g. \cite{ODonnell}).
Let $\Multilinear{X} \subseteq \F[X]$\NewNotation{notation:Multilinear} denote the set of all multilinear polynomials.
For $i \in \{0,\ldots,2^n-1\}$, the \emph{key polynomial} $\IndexPoly{X}{i} \in \Multilinear{X}$\NewNotation{notation:IndexPoly}
is defined as
\[
    \IndexPoly{X}{i} \coloneqq \prod_{k = 0}^{n - 1}
        \begin{cases}
            x_k &\text{ if } i_k = 1,\\
            1 - x_k&\text{ if } i_k = 0,
        \end{cases}
\]
i.e., it is the unique multilinear polynomial that is $1$ at the assignment $\IndexA{X}{i}$
and  $0$ for all other Boolean assignments.
Hence, the unique extension of a function $f\colon \BooleanDomain^X \to \F$ to a multilinear polynomial is
$\hat{f} \coloneqq \sum_{i \in  \{0,\ldots,2^{n}-1\}}  \big( \IndexPoly{X}{i} \cdot \Pev{\IndexA{X}{i}}f \big)$.

\begin{example}
	For $X = \{x_0, x_1, x_2, x_3\}$, we have $\IndexPoly{X}{13} = \IndexPoly{X}{(1101)_2} = x_3 x_2 (1 - x_1) x_0$.
\end{example}

Note that the product of binary polynomials is again binary.
However, in general, the product of multilinear polynomials is not multilinear.
To restore multilinearity, we need an operation called \emph{degree reduction}.
For a variable $x \in X$ and a value $a \in \F$,
the map $\PartEval{x \coloneqq a}\colon \F[X] \to \F[X\setminus \{x\} ]$,\NewNotation{notation:pev}
called \emph{partial evaluation},
sends a polynomial $p$ to the polynomial obtained by substituting $a$ for all occurrences of $x$ in $p$, i.e., $\Pev{\sigma}\pev{x\coloneqq a} p = \Pev{\sigma[a/x]}p$.
This allows us to define degree reduction as follows:
$\DegreeReduction{x}(p) \coloneqq (1-x) \cdot \PartEval{x \coloneqq 0}(p) + x \cdot \PartEval{x \coloneqq 1}(p)$.\NewNotation{notation:degx}
Indeed, applying $\DegreeReduction{x}$ has the effect of reducing the exponent of all occurrences of $x$ to $1$
while preserving the values at $\BooleanDomain^X$,
i.e., $\DegreeReduction{x}(p)\restrict{\BooleanDomain^X} = p\restrict{\BooleanDomain^X}$.
Set $\DegreeReduction{X} \coloneqq \DegreeReduction{x_0} \ldots \DegreeReduction{x_{n-1}}$.\NewNotation{notation:degX}
Then $q \coloneqq \DegreeReduction{X}(p) \in \Multilinear{X}$ is the unique multilinear polynomial
with $q\restrict{\BooleanDomain^X} = p\restrict{\BooleanDomain^X}$.

Analogous to $\PartEval{x \coloneqq a}$, for a \emph{variable} $y$
we let $\PartEval{x \coloneqq y} p$ denote the polynomial obtained by substituting $y$ for  $x$ in  $p$,
i.e., $\Pev{\sigma}\pev{x\coloneqq y} p = \Pev{\sigma[\sigma(y)/x]}p$.\NewNotation{notation:pev-var}
Moreover, for $|Y| = |Z|$, we set  $\overwriteVars{Y}{Z} \coloneqq \pev{y_1 \coloneqq z_1} \ldots \pev{y_{|Y|} \coloneqq z_{|Z|}}$.

\subparagraph{Arithmetizing CNF propositional formulas.}
For a set of variables $X = \{x_0,\ldots,x_{n-1}\}$, let  $\Lit(X) \coloneqq X \cup \{\lnot x : x \in X\}$\NewNotation{notation:Lit}
denote the set of \emph{literals} over $X$.
A \emph{clause} is a set $C \subseteq \Lit(X)$ of literals which does not contain a variable and its negation.
It represents the disjunction of these literals.
We arithmetize literals as
$\Arith{\ell} \coloneqq x_i$ if $\ell = \lnot x_i$
and  $\Arith{\ell} \coloneqq  1-x_i$ if $\ell = x_i$.\NewNotation{notation:Arith}
Let $\ArithLit{X} \coloneqq \{\Arith{\ell} : \ell \in \Lit(X)\} = X \cup \{1 -x : x \in X\} \subseteq \Multilinear{X}$
denote the set of arithmetized literals.\NewNotation{notation:ArithLit}
Clauses are arithmetized as $\Arith{C}\coloneqq 1 - \prod_{\ell \in C} \Arith{\ell}$.\NewNotation{notation:ArithClause}
Note that $\Arith{C} \in \Multilinear{X}$ by our definition of clauses.
Moreover, $\Arith{C}$  is a binary polynomial
and $\sigma \in \BooleanDomain^X$ is a satisfying assignment of $C$ iff $\Pev{\sigma}\Arith{C} = 1$.
A sequence of clauses $\varphi = (C_0,\ldots, C_{m-1})$, which usually represents a CNF formula,
is arithmetized as $\Arith{\varphi}_Y \coloneqq  \sum_{i = 0}^{m-1} \IndexPoly{Y}{i} \Arith{C_i} \in \Multilinear{X, Y}$,
where $Y$ is a sufficiently large set of variables.\NewNotation{notation:ArithCNF}
\begin{example}
	Let $\varphi = (x_0 \lor x_1 \lor \neg x_2) \land (\neg x_0 \lor x_1) \land (\neg x_1 \lor x_2)$ be a formula in CNF and $Y = \{y_0, y_1\}$.
	Then,
    $
        \Arith{\varphi}_Y
		= \IndexPoly{Y}{0} (1 - (1 - x_0) (1 - x_1) x_2) + \IndexPoly{Y}{1} (1 - x_0 (1- x_1)) + \IndexPoly{Y}{2} (1 - x_1 (1 -x_2)).
    $
\end{example}

\subparagraph{Counting Satisfying Assignments.}
\label{sec:multilinear:counting}

Let $\sigma^{X}_{2^{-1}} \in \F^{X}$\NewNotation{notation:assignment:const}
denote the constant assignment
that maps every variable of $X$ to $2^{-1}$,
where the multiplicative inverse is taken in $\F$.
Then, for every $i \in  \{0,\ldots,2^{n}-1\}$, we have $\Pev{\sigma^X_{2^{-1}}}\IndexPoly{X}{i} = 2^{-|X|}$.
Thus, the number of satisfying assignments of a formula (modulo the characteristic of $\F$)
can be obtained from the unique multilinear expansion as follows:
\begin{fact}
    \label{fact:count-satisfying-assignments}
    Let $f\colon \BooleanDomain^X \to \BooleanDomain$ and let $\hat{f} \in \Multilinear{X}$ be
    the unique expansion of $f$ to a multilinear polynomial, i.e., $\hat{f}\restrict{\BooleanDomain^X} = f$.
    Then $|f^{-1}(\{1\})| = 2^{|X|} \cdot \hat{f}\left( \sigma^{X}_{2^{-1}} \right)$.
\end{fact}

\subparagraph{Arithmetizing Sets and Relations.}

We also use multilinear polynomials to encode sets and relations.
Let $X = \{x_0,\ldots,x_{|X|-1}\}$ and $Y = \{y_0,\ldots,y_{|Y|-1}\}$.
Recall that $i \in \{0,\ldots,2^{|X|}-1\}$ corresponds to the binary assignment
$\IndexA{X}{i}$.
Using this, we identify the set of binary assignments, $\BooleanDomain^X$,
and $\{0,\ldots,2^{|X|}-1\}$.

For a set $S \subseteq \BooleanDomain^X$
we define $\MultilinEncode{X}{S} \coloneqq \sum_{m \in S} \IndexPoly{X}{m} \in \Multilinear{X}$.\NewNotation{notation:MultilinEncodeSet}
We encode relations $R \subseteq \BooleanDomain^X \times \BooleanDomain^Y$ as $\MultilinEncode{X,Y}{R} \coloneqq \sum_{(a,b) \in R} \IndexPoly{X}{a} \IndexPoly{Y}{b} \in \Multilinear{X,Y}$.
Viewing functions as relations, those can be encoded in the same way.\NewNotation{notation:MultilinEncode}
\begin{example}
	\label{example:join}
	Consider a function $f \subseteq \BooleanDomain^X \times \BooleanDomain^Y$ and a set $S \subseteq \BooleanDomain^Y$.
	Since $\DegreeReduction{Y}\left( \IndexPoly{Y}{i} \IndexPoly{Y}{j}\right) = 0$ for all $i \neq j$, the encoding of the preimage $f^{-1}(S)$ is given by
	\[
	\MultilinEncode{X}{f^{-1}(S)} = 2^{|Y|} \Pev{\sigma_{2^{-1}}^Y}
	\DegreeReduction{Y}\left( \MultilinEncode{X,Y}{f} \MultilinEncode{Y}{S} \right).
	\]
\end{example}

\ArxivAppendixRef{D}{sec:notation} provides a compendium of the notation used in this article.
 \section{From Resolution Proofs to EACs}
\label{sec:gadgets}

We reduce the problem of checking the correctness of a resolution proof
to the satisfiability problem of  \emph{extended arithmetic circuits} (EACs). 
 \autoref{sec:polynomial-res-refutations} encodes correct resolution proofs as \emph{polynomial refutations}, triples of polynomials satisfying certain conditions (\autoref{def:poly-ref}). \autoref{subsec:EAC} defines EACs and their satisfiability problem. \autoref{sec:eac-gadgets} reduces checking the conditions of \autoref{def:poly-ref} to satisfiability of EACs.
 \subsection{Polynomial Resolution-Refutations}
\label{sec:polynomial-res-refutations}

We represent CNF-clauses as \emph{sets} of literals (i.e., it is tacitly assumed that a clause does not contain duplicate literals) and CNF formulas as sequences of clauses. We assume that tautological clauses are removed,
i.e., clauses never contain a literal and its negation at the same time.
For clauses $C_1$, $C_2$ and a literal $x$ with  $x \in C_1$, $\lnot x \in C_2$,
the \emph{resolvent} of $C_1$ and $C_2$ is defined as the clause $\Res_x (C_1, C_2) \coloneqq (C_1 \setminus \{x\}) \cup (C_2 \setminus \{\neg x\})$. Clearly, $C_1 \land C_2$ entails $\Res_x (C_1, C_2)$.
It is well known that a CNF-formula is unsatisfiable if{}f the empty clause can be derived through iterative computing of resolvents \cite{Harrison09}.
Formally, we represent such refutations as follows (we assume that the empty clause is not already part of the formula):
\begin{definition}
    \label{def:cnfres}
	Let $\varphi=(C_0,\ldots,C_{m-1})$\NewNotation{notation:m} be a CNF-formula.
    A \emph{RES-refutation} of $\varphi$ is a sequence of clauses $\Phi=(C_0,\ldots,C_{M-1})$\NewNotation{notation:M} with $M >m$, such that
    \FixEnumerateItemRefParentheses{\begin{enumerate}[(a)]
        \item\label{def:res-refutation:initial} the first $m$ clauses of $\varphi$ and $\Phi$ agree,
        \item\label{def:res-refutation:empty} $C_{M-1}$ is the empty clause, and
        \item\label{def:res-refutation:support}
            for every $i = m,\ldots, M-1$,
            there exist $j, k < i$ and a variable $x$
            such that
            $C_i = \Res_x (C_j, C_k)$.
            $C_j$ and  $C_k$ are the \emph{antecedents}
            of  $C_i$ and we call the pair of indices $(j, k)$ an \emph{antecedent list} for $C_i$.
	\end{enumerate}}
\end{definition}

We encode RES-refutations using multilinear polynomials.
To this end, we fix a set of function variables $\Wres \coloneqq \{p, u_1, u_2\}$.\NewNotation{notation:Wres}
The idea is to encode the sequence $\Phi$ in $\omega(p)$
and, for each clause, the indices of its two antecedents in $\omega(u_1)$ and $\omega(u_2)$, respectively.
Recall that for $0 \le i < 2^{|Y|}$,
$\IndexA{Y}{i}$ denotes the assignment mapping $y_k$ to the $k$-th bit of $i$.

\begin{definition}\label{def:poly-ref}
	Let $X,Y,Z$ denote disjoint sets of variables, with $\Abs{X}=n$ and $\Abs{Y}=\Abs{Z}$.
	Let $\varphi=(C_0,...,C_{m-1})$ be a CNF-formula with variables $X$.
    A map $\omega\colon \Wres \to \F[X, Y, Z]_{\le 1}$
    is a \emph{polynomial refutation}
    iff it satisfies the following properties:\makeatletter
    \def\property#1{\item\phantomsection\edef\@currentlabel{#1}}
    \makeatother
    \FixEnumerateItemRefParentheses[V]{\begin{enumerate}[(V1)]
        \item\label{valid:binary} $\omega(p) \in \F[X, Y]_{\le 1}, \omega(u_1), \omega(u_2) \in \F[Y, Z]_{\le 1}$ are binary and multilinear polynomials.
        \item\label{valid:prefix} $\Pev{\IndexA{Y}{j}}\,\omega(p)=\Arith{C_j}$ for all $0 \leq j \leq m-1$.
        \item\label{valid-derives-empty} $\Pev{\IndexA{Y}{M-1}}\,\omega(p)=\Arith{\emptyset} \;(= 0)$ for some $M > m$.
            $M$ is called the \emph{length} of the refutation.
        \item\label{valid-clause-mapping} There exist $U_1, U_2\colon \{m,...,M-1\}\rightarrow\{0,...,2^{\Abs{Z}}-1\}$ s.t.~$\omega(u_i)=\MultilinEncode{Y,Z}{U_i}$.
        \item\label{valid:regressive} $U_1(j) <j$ and $U_2(j)<j$ for all $m \leq j \leq M-1$.
        \item\label{valid:resolution} $\Bigl(\Pev{\IndexA{Y}{U_1(j)}}\,\omega(p) \cdot\Pev{\IndexA{Y}{U_2(j)}}\,\omega(p)\cdot\bigl(1-\Pev{\IndexA{Y}{j}}\,\omega(p)\bigr)\Bigr)\restrict{\BooleanDomain^X} = 0$ for all $m \leq j \leq M-1$.
	\end{enumerate}}
\end{definition}

Property \ref{valid:binary} ensures that the polynomials have the right form.
The next two, \ref{valid:prefix} and \ref{valid-derives-empty}, correspond to properties
\ref{def:res-refutation:initial} and \ref{def:res-refutation:empty} of \autoref{def:cnfres}. They establish that $p$ encodes a list starting with the clauses of $\varphi$ and ending with the empty clause. (There might be additional clauses encoded after the empty clause.)

We intend $\omega(u_1)$ and $\omega(u_2)$ to encode functions $U_1$ and $U_2$,
s.t.\ for every $m \leq j < M$ the pair $(U_1(j), U_2(j))$ is an antecedent list for $C_j$.
Property \ref{valid-clause-mapping} ensures that they do indeed encode functions
and \ref{valid:regressive} enforces that these functions are regressive,
i.e., a clause can only be derived from \emph{earlier} clauses, ruling out circular reasoning.
Finally, \ref{valid:resolution} states that, for all $m \leq j < M$, the conjunction of $C_{U_1(j)}$ and $C_{U_2(j)}$ entails $C_j$.
Note that this is less restrictive than requiring  $C_j$ to be a resolvent of  $C_{U_1(j)}$ and $C_{U_2(j)}$.
(However, our quasi-optimal prover will work only with polynomial refutations arising from resolution proofs.)
It is straightforward to prove that polynomial refutations are a sound and complete method for proving unsatisfiability:

\begin{restatable}{lemma}{polyResComplete}
	\label{lem:polynomial-refutation-complete}
	Let $\varphi$ denote a CNF-formula. There exists a polynomial refutation for $\varphi$ iff $\varphi$ is unsatisfiable.
\end{restatable}

\begin{remark}
	\autoref{def:poly-ref} is not limited to resolution proofs.
    Indeed, it is only required that a derived clause is \emph{implied}
    by the conjunction of its antecedents, not that it is a \emph{resolvent}.
	Hence, the same idea can be applied to other implicational proof systems.
    An example are \emph{Reverse Unit Propagation (RUP)} proofs.
    RUP is a more succinct proof format than resolution.
    Instead of storing every resolution step,
    several steps are combined
    and thus derived clauses can have up to $n+1$ antecedents.
    We use polynomials $u_0,\ldots, u_{n}$ to store them.
    \ref{valid-clause-mapping} and \ref{valid:regressive}
    have to be checked for more polynomials, and \ref{valid:resolution}
    changes to account for the higher number of antecedents.
    In \ArxivAppendixRef{C}{sec:rup}, we describe how our prover can be adapted to this setting.
\end{remark}
 \subsection{Extended Arithmetic Circuits}
\label{subsec:EAC}
Extended Arithmetic Circuits are a generalization of arithmetic circuits since we allow for more types of gates: partial evaluation and function variables.
The latter serve as placeholders for multilinear polynomials.
\begin{definition}
	Let $\F$ denote a finite prime field,
    $X$ a set of variables, and $W$ a set of function variables.
    An \emph{Extended Arithmetic Circuit} (EAC) is a rooted DAG
    with leaves
    labeled by
    \newcounter{enumiresume}
    \FixEnumerateItemRefParentheses{\begin{enumerate}[(i)]
        \item\label{def:eac:node-const} $b$ for some value $b \in \F$, or
        \item\label{def:eac:node-var} $x$, where $x \in X$ is a variable, or
        \item\label{def:eac:node-fvar} $f$, where $f \in W$ is a function variable
        \setcounter{enumiresume}{\value{enumi}}
    \end{enumerate}}
    and  internal nodes
    labeled by
    \FixEnumerateItemRefParentheses{\begin{enumerate}[(i)]
        \setcounter{enumi}{\value{enumiresume}} \item\label{def:eac:node-addmul}
             $+$ or $\cdot$, in which case the node has exactly two children, or by

        \item\label{def:eac:node-pev} $\PartEval{x \coloneqq b}$, where $x \in X$ and $b \in \F \cup X$, in which case the node has one child.
    \end{enumerate}}
    We let $|\phi|$ denote the number of nodes of an EAC $\phi$.
    An EAC is an \emph{arithmetic circuit} if only nodes of types
        \ref{def:eac:node-const}, \ref{def:eac:node-var}
        and \ref{def:eac:node-addmul} occur.
\end{definition}
An \emph{interpretation}
is a mapping $\omega\colon W \to \left( \F^{X} \to \F\right)$\NewNotation{notation:omega}
assigning functions to the function variables.
An interpretation $\omega$ is  \textit{multilinear} iff $\omega(f)\colon \F^X \to \F$ is a multilinear polynomial for all $f \in W$.
Note that a polynomial refutation is a multilinear interpretation for $W = \Wres$.
Given an interpretation $\omega$, we inductively assign to
every node $v$ of $\phi$ a map $\Eval{v}_{\omega}\colon \F^X \to \F$ as follows:\NewNotation{notation:Eval}
\[
    \Eval{v}_{\omega} \coloneqq \begin{cases}
    b
        &\text{ if $v$ is a leaf labeled with a constant $b \in \F$},\\
    x
        &\text{ if $v$ is a leaf labeled with a variable $x \in X$},\\
    \omega(f)
        &\text{ if $v$ is a leaf labeled with a function variable $f \in W$},\\
        \Eval{c_1}_{\omega} \circledast \Eval{c_2}_{\omega}
        &\text{ if $v$ is labeled with $\circledast \in \{+,\cdot \} $ and has children $c_1$ and $c_2$},\\
\PartEval{x \coloneqq b}\left( \Eval{c}_{\omega}  \right)
        &\text{ if $v$ is labeled $\PartEval{x \coloneqq b}$ with child $c$ and $b \in \F \cup X$}.
\end{cases}
\]
Finally, we set  $\Eval{\phi}_{\omega} \coloneqq \Eval{r}_{\omega}$,
where $r$ is the root of $\phi$.

Note that when $\omega$ is a \emph{multilinear} interpretation we can view the maps $\Eval{v}_{\omega}\colon \F^X \to \F$ as polynomials in $\F[X]$.
For later proofs, we are interested in the maximum degree of the polynomial represented by an EAC.
To this end, let $\deg_x(p)$\NewNotation{notation:deg}
denote the maximum degree of variable $x$ in the polynomial $p \in \F[X]$.
We define the degree of $\phi$ under multilinear interpretation $\omega$ as $\deg_x(\phi, \omega) \coloneqq \deg_x( \Eval{\phi}_{\omega})$ and
$\maxdeg(\phi)$ as the maximum of $\deg_x(\psi, \omega)$ over all nodes $\psi$ of $\phi$, multilinear interpretations $\omega$, and  variables $x$.

Degree reduction $\DegreeReduction{x}(p)$ was defined in \autoref{subsection:multilinear-polynomials}
as a shorthand for $(1- x) \cdot \PartEval{x \coloneqq 0}(p) + x \cdot \PartEval{x \coloneqq 1}(p)$.
We use nodes labeled $\DegreeReduction{x}$ with a single child as a shorthand for the same construction, which can be directly represented as an EAC.
These degree reduction nodes will be of importance later,
as they allow us to keep the degrees of the EACs and intermediate polynomials bounded.
\begin{figure}[t]
        \centering
    \def\DrawEAC#1#2{\node (#1root) at #2 [eacroot] { $\cdot$ };
        \node (#1top) [above=0.01cm of #1root] {};
        \node (#1add) [eacnode, below left=of #1root]{$+$};
        \node (#1pev) [eacnode, below right=of #1root]{$\pev{x \coloneqq 3}$};
        \node (#1f)   [eacnode, below=of #1pev]{$f$};
        \node (#1x2)  [eacnode, below=of #1add]{$x$\vphantom{$f$}};
        \draw [eacedge] (#1top) --  (#1root);
        \draw [eacedge] (#1root) -- (#1add);
        \draw [eacedge] (#1add) --  (#1f);
        \draw [eacedge] (#1add) --  (#1x2);
        \draw [eacedge] (#1root) -- (#1pev);
        \draw [eacedge] (#1pev) --  (#1f);
    }
    \def\DrawEACLabels#1{\node (#1rootlabel) [eacnodelabel,above left=-0.2cm and -0.1cm of #1root] {$v_0$};
        \node (#1addlabel) [eacnodelabel, above left=-0.2cm and -0.1cm of #1add] {$v_1$};
        \node (#1pevlabel) [eacevallabel, above right=-0.2cm and -0.1cm of #1pev] {$v_2$};
        \node (#1x2label) [eacevallabel,  above left=-0.2cm and -0.1cm of #1x2] {$v_3$};
        \node (#1flabel) [eacevallabel, above right=-0.2cm and -0.1cm of #1f] {$v_4$};
    }

    \begin{tikzpicture}[node distance=0.3cm]
        \DrawEAC{eac1}{(0,0)}
        \DrawEACLabels{eac1}
        \node (eac1rootclaim) [eacclaim,above right=-0.2cm and 0.1cm of eac1root] {${\scriptstyle\Pev{\sigma_0}\Eval{v_0}_\omega \qeq 0}$};

        \DrawEAC{eac2}{(8,0)}
        \node (eac2addclaim) [eacclaim,above left=0.05cm and -0.2cm of eac2add] {${\scriptstyle \Pev{\sigma_0}\Eval{v_1}_\omega \qeq 0}$};
        \node (eac2pevclaim) [eacclaim,above right=0.05cm and -1cm of eac2pev] {${\scriptstyle\Pev{\sigma_0}\Eval{v_2}_\omega \qeq 4}$};

        \draw [eacertifystep] (eac1pev) --
            node[verifier,above left=0mm and -10mm] {$\QueryProver(\Pev{\sigma_0}\Eval{v_1}_{\omega})$,\\$\QueryProver(\Pev{\sigma_0} \Eval{v_2}_{\omega})$}
            node[prover, below left=0mm and -10mm] {Response{:} $0, 4$\\$\Assert(0 \cdot 4=0)$}
            (eac2add);

        \DrawEAC{eac3}{(4,-3)}
        \node (eac3pevclaim) [eacclaim,above right=0.05cm and -1cm of eac3pev] {${\scriptstyle\Pev{\sigma_0}\Eval{v_2}_\omega \qeq 4}$};
        \node (eac3fclaim) [eacclaim,right=0.05cm of eac3f] {${\scriptstyle \Pev{\sigma_0}\Eval{v_4}_\omega \qeq 2}$};
        \node (eac3x2claim) [eacclaim,left=0.05cm of eac3x2] {${\scriptstyle \Pev{\sigma_0}\Eval{v_3}_\omega \qeq 5}$};

        \draw [eacertifystep] (eac2pev) -| (11,-2.2) -- (-2,-2.2) |-
            node[verifier, above right=0mm and 5mm] {$\QueryProver(\Pev{\sigma_0}\Eval{v_3}_{\omega})$,\\$\QueryProver(\Pev{\sigma_0} \Eval{v_4}_{\omega})$}
            node[prover, below right=0mm and 5mm] {Response{:} $5, 2$\\$\Assert(5 + 2 = 0)$}
            (eac3add);

        \DrawEAC{eac4}{(0,-6)}
        \node (eac4fclaim) [eacclaim,below right=0.05cm and -1cm of eac4f, align=left] {${\scriptstyle \Pev{\sigma_0}\Eval{v_4}_\omega \qeq 2}$\\
            ${\scriptstyle \Pev{\sigma_1}\Eval{v_4}_\omega \qeq 4}$};
        \node (eac4x2claim) [eacclaim,below left=0.05cm and -1cm of eac4x2] {${\scriptstyle \Pev{\sigma_0}\Eval{v_3}_\omega \qeq 5}$};

        \draw [eacertifystep] (eac3pev) -|
            node[verifier, above left=0mm and 10mm] {
                $\sigma_1 \gets \sigma_0[ 3 / x ]$
}
            (11,-5.2) -- (-2,-5.2) |-
            (eac4add);

        \DrawEAC{eac5}{(8,-6)}
        \node (eac5fclaim) [eacclaim,below right=0.05cm and -1cm of eac5f, align=left] {${\scriptstyle \Pev{\sigma_2}\Eval{v_4}_\omega \qeq 5}$};
        \node (eac5x2claim) [eacclaim,below left=0.05cm and -1cm of eac5x2] {${\scriptstyle \Pev{\sigma_0}\Eval{v_3}_\omega \qeq 5}$};

        \draw [eacertifystep] (eac4pev) --
            node[verifier, above, align=center] {\ProcedureUnifyClaims\\
                $\QueryProver(\Pev{\sigma_0[- / x]}\Eval{v_4}_{\omega})$}
            node[prover, below ] {Response{:} $6 \cdot x$\\
                $\Assert(6 \cdot \sigma_0(x) = 2)$\\
                $\Assert(6 \cdot \sigma_1(x) = 4)$\\
                $\sigma_2 \gets \sigma_0[2/x]$\\
                $6 \cdot \sigma_2(x) = 5$
            }
            (eac5add);
    \end{tikzpicture}
    \caption{
        An EAC $\psi$ with nodes $v_0,\ldots,v_4$
        over the field $\F_7$ is shown in the top left.
        The diagram illustrates a run of $\EACertify$ on $\psi$ for the interpretation $\omega(f) \coloneq 6x$.
        Rounds are visualized as dotted arrows.
        The verifier keeps track of claims (depicted in yellow), which are updated in each round.
        Note that the round for \ProcedureUnifyClaims is depicted simplified due to space constraints.
}
    \label{fig:example-run-eacertify}
\end{figure}
Lastly, we define the satisfiability problem for EACs:
\begin{DecisionProblem}{$\EACSAT$}
    \label{def:eac-sat}
    \ProblemInput{An EAC  $\phi$ with variables $X$ and function variables $W$.}
    \ProblemOutput{Does a multilinear interpretation
        $\omega\colon W \to \Multilinear{X}$ exist such that $\Eval{\phi}_{\omega}$ is the constant zero function,
        i.e., $\forall \sigma. \Eval{\phi}_{\omega}(\sigma) = 0$?
    }
\end{DecisionProblem}

\begin{example}
	\autoref{fig:example-run-eacertify} shows an EAC $\psi$ over the field $\F_7$
    with one variable $x$ and one function variable $f$. We have
    $\Eval{\psi}_{\omega} = (x + \omega(f)) \cdot \pev{x \coloneqq 3} \omega(f)$.
The EAC is satisfiable,
    e.g.~$\omega_0(f) \coloneqq 0$ and $\omega_1(f) \coloneqq 6\cdot x$
    are satisfying multilinear interpretations.
\end{example}

\begin{remark}
	$\EACSAT$ is $\NEXP$-complete. This can be shown via a reduction from the satisfiability problem for \emph{Dependency Quantified Boolean Formulas (DQBF)}~\cite{PetersonRA01}.
\end{remark}
 \subsection{EAC Gadgets}
\label{sec:eac-gadgets}

In \autoref{sec:eacertify}, we will introduce $\EACertify$, an interactive protocol for $\EACSAT$.
To turn this into a protocol for \emph{polynomial refutations}, we apply $\EACertify$
to certain gadgets:
Let $\varphi=(C_0,\ldots,C_{m-1})$ denote a CNF-formula.
Our goal is to construct an EAC $\psi$,
such that $\psi$ is satisfiable iff there exists a polynomial refutation of $\varphi$ of length $M$.
We will go through the properties \ref{valid:binary}\,--\,\ref{valid:resolution} and provide gadgets for each,
in the form of EACs that are satisfiable iff the corresponding property holds.
\autoref{tab:gadgets-overview} provides an overview of these gadgets.
We only present \ref{valid:regressive} in detail here,
as this features most of the relevant ideas.
Explanations and proofs of correctness for all other gadgets,
    the constructions $\overwriteVars{Y}{Z}$,
    $[Y < m]$ and  $[m \le Y < M]$,
as well as how to combine the gadgets into one final EAC,
can be found in \autoref{appendix:proof:gadgets}.

We construct a circuit $[Y\le Z]$
which represents a binary multilinear polynomial
such that
$\Pev{\IndexA{Y}{i}} \Pev{\IndexA{Z}{j}} [Y \le Z] = 1$
iff $i \le j$.
This can be done inductively:
Let $\alpha_0 \coloneqq 1$,
\[
    \alpha_{i+1} \coloneqq  (1 - y_i) z_i + ((1-y_i)(1-z_i) + y_i z_i) \cdot \alpha_{i}.
\]
and $[Y\le Z] \coloneqq \alpha_{|Y|}$ (remember that $|Y| = |Z|$).
Let $\GadgetRegressive(u_i) \coloneqq  \DegreeReduction{Y \cup Z}(u_i\cdot[Y\le Z])$.

\begin{lemma}
    \label{lem:gadget:regressive}
    Let $\omega$ be a multilinear interpretation such that
	$\omega(u_i)$ is binary and
    encodes a function $U_i$ as in
    \ref{valid-clause-mapping}.
	Then,
    $\Eval{\GadgetRegressive(u_i)}_{\omega} = 0$ iff
    the function $U_i$ is regressive,  i.e.,
    iff $\omega(u_i)$ fulfils \ref{valid:regressive}.
\end{lemma}
\begin{proof}
    Since $\Eval{\GadgetRegressive(u_i)}_\omega$ is multilinear and binary,
	we have $\Eval{\GadgetRegressive(u_i)}_\omega\ne0$ iff there are indices $j,k < 2^{\Abs{Y}}$ with $\Pev{\IndexA{Y}{j}}\Pev{\IndexA{Z}{k}}\Eval{\GadgetRegressive(u_i)}_\omega=1$.
    Then $\Pev{\IndexA{Y}{j}}\Pev{\IndexA{Z}{k}}(\omega(u_i)\cdot[Y\le Z])=1$,
	as $\DegreeReduction{Y\cup Z}$ does not change the evaluation on binary assignments.
	Both factors are binary and thus $1$, so we obtain $U_i(j)=k$ and $j\le k$, violating \ref{valid:regressive}.
	
	Conversely, assume that \ref{valid:regressive} does not hold.
	Then, there is a $j$ with $U_i(j)\ge j$
    and thus
    $\Pev{\IndexA{Y}{j}}\Pev{\IndexA{Z}{U_i(j)}}\Eval{\GadgetRegressive(u_i)}_\omega=1$.
\end{proof}
\begin{movetoappendix}
Lastly, we state some properties about these gadgets and their implications on our $\MIP$-protocol to certify them:
\begin{fact}
\begin{enumerate}
	\item There are $2n + 6 |Y| + 5 |Z|$-many rounds for degree reductions.
	\item The number of non degree reductions rounds is constant. \item \UnifyClaims{} is only called six times in total.
	\item The verifier needs time $\O(n + |\varphi| + \log_2 M)$ to certify all gadgets.
\end{enumerate}
\end{fact}
All facts directly follow from counting the corresponding nodes of each gadget.
These observations will be helpful in \autoref{sec:evaluation} when comparing our implementation with other certificate formats and checkers.
\end{movetoappendix}

	\begin{table}[t]
		\centering
		\caption{Overview of the Gadgets}
		\label{tab:gadgets-overview}
		\begin{tabular}{rl}
			Property & Gadget                                                                                                         \\
			\ref{valid:binary}
			         & $\GadgetBinary(p) = \DegreeReduction{X}\DegreeReduction{Y} (p (1-p))$,
			$\GadgetBinary(u_i) = \DegreeReduction{Z}\DegreeReduction{Y} (u_i (1-u_i))$                                               \\
			\ref{valid:prefix}
                     & $\GadgetInitial(p) = \Arith{\phi}_Y - \DegreeReduction{Y}\left( [Y < m] \cdot p \right)$ \\
			\ref{valid-derives-empty}
			         & $\GadgetEmptyClause(p) = \Pev{\IndexA{Y}{M-1}} p$                                                                \\
			\ref{valid-clause-mapping}
                     & $\GadgetFunction(u_i) = (2^{\Abs{Z}}\Pev{\sigma^{Z}_{2^{-1}}}u_i)-[m \le Y < M]$                       \\
			\ref{valid:regressive}
                     & $\GadgetRegressive(u_i) = \DegreeReduction{Z}\DegreeReduction{Y}\left( u_i \cdot [Y \le Z] \right)$ \\
			\ref{valid:resolution}
			         & \makecell[l]{$\GadgetResolution(p, u_1, u_2) = $                                                               \\ $\DegreeReduction{X} \DegreeReduction{Y} \left(
                         \left(2^{\Abs{Z}}\Pev{\sigma^{Z}_{2^{-1}}}\DegreeReduction{Z}(u_1\cdot\overwriteVars{Y}{Z}p)\right)
                     \cdot \left(2^{\Abs{Z}}\Pev{\sigma^{Z}_{2^{-1}}}\DegreeReduction{Z}(u_2\cdot\overwriteVars{Y}{Z}p)\right)
			\cdot  (1-p) \right)$}                                                                                                    \\
		\end{tabular}
	\end{table}
 \section{\texorpdfstring{$\EACertify$}{EACertify}: A Protocol for \texorpdfstring{$\EACSAT$}{EAC-SAT}}
\label{sec:eacertify}
In this section we present an $\MIP$-protocol for  $\EACSAT$.
Let $\phi$ be an EAC with set of variables $X$ and $\maxdeg(\phi) \leq d$.
The goal is to convince the verifier of the fact that
there exists a multilinear interpretation $\omega$ such that $\Eval{\phi}_{\omega} = 0$.
As discussed in \autoref{sec:mip}, we describe our protocol in terms of a
verifier interacting with one prover and one oracle.
The oracle is used only to evaluate function variables.
Let $\omega$ be the interpretation \emph{induced}
by an oracle $O$, i.e., $(\omega(f))(\sigma)$ is the response of $O$ on query $\Pev{\sigma}f$.
Note that this is indeed an interpretation, since $O$ is memoryless.

\subparagraph{Multilinearity Test.}
We first need to verify that $\omega$ is a multilinear interpretation.
This can be achieved using so-called \emph{low-degree tests}.
We say that a function $f\colon \F^X \to \F$
is $\varepsilon$-close to a multilinear function
if $\min_{g \in \Multilinear{X}} d(f,g) < \varepsilon$,
where $d(f,g)$ denotes the \emph{normalized Hamming distance},  $\frac{|\{x \in \F^n: f(x) \neq g(x)\}|}{|\F^n|}$.
For brevity we only state the following result here:
\begin{fact}[{\cite[Cor.~3.2]{multilin-test2}}]
    \label{fact:multilinearity-test}
    Let $\varepsilon > 0$.
    Suppose that $|\F| > \frac{16 |X|}{\varepsilon}$.
    An $\O(\frac{|X|}{\varepsilon})$-fold pairwise independent repetition of
    the so-called \emph{line test} makes $\O(\frac{|X|}{\varepsilon})$
    queries to the oracle.
    It accepts all multilinear polynomials $\Multilinear{X}$ and
    rejects
    functions $f\colon \F^{X} \to \F$
    which are not $\varepsilon$-close to a multilinear function
    with probability at least $\frac{1}{2}$.
\end{fact}

Note that it suffices to check that the interpretation $\omega$ induced by the oracle is $\varepsilon$-close to a multilinear function, for $\varepsilon$ small enough:
In the analysis we can use the unique multilinear interpretation $\tilde{\omega}$ closest to $\omega$.
The probability that the verifier asks a query where $\omega$ and $\tilde{\omega}$ differ is at most $q\cdot \varepsilon$,
where $q$ is the number of queries (in our case $q = 3$).
Thus, for the \emph{zero test} we simply assume that $\omega$ is a multilinear interpretation.

\subparagraph{Zero Test.}
The goal of the zero test is to check that $\Eval{\phi}_{\omega} = 0$.
This part of our protocol is very similar to the protocol presented in \cite{ip-pspace-practical-cav}.
We modify it to accommodate function variables.
Since this is the central part of $\EACertify$, we provide a full description.
The verifier is described in \autoref{alg:verifier}.
We will first give an informal overview and then make it precise in
\autoref{lem:unify-claims} and \autoref{thm:eacertify}:

During the protocol, the verifier sends queries of the form $\Pev{\sigma}\Eval{v}_{\omega}$
to the prover, where $v$ is a node of  $\phi$ and either
\begin{enumerate*}[(i)]
    \item $\sigma \in \F^X$ is an assignment,
        in which case the prover must respond with a value $a \in \F$, or
    \item $\sigma \in \F^{X \setminus \{x\}}$ assigns a value to all variables but one,
        in which case the prover responds with a univariate polynomial $p \in \F[x]$.
\end{enumerate*}
Since $\Eval{\phi}_{\omega}$ is a polynomial of degree at most $d$ in every variable,
by the Schwartz-Zippel lemma~\cite{schwartz-paper,zippel-paper} it suffices to check $\Pev{\sigma}\Eval{\phi}_{\omega} = 0$
for a single assignment $\sigma \in \F^X$ chosen uniformly at random.

For every node $v$ of $\phi$, the verifier keeps track of a set of claims that still need to be checked.
Initially, there is only one claim for the root node, $\Pev{\sigma}\Eval{\phi}_{\omega} \qeq 0$.
The nodes of $\phi$ are then traversed in topological order.
When processing a node with multiple associated claims,
the verifier first replaces all of these claims by a single claim by invoking \ProcedureUnifyClaims (see \autoref{alg:unify-claims}).
If at least one of these claims was incorrect,
then with high probability either
\begin{enumerate*}[(i)]
    \item the verifier rejects immediately or
    \item the new claim is incorrect as well.
\end{enumerate*}
When processing an internal node, the verifier interacts with the prover to compute claims about the children of that node,
such that if the original claim was incorrect, at least one of the claims of the children is incorrect as well with high probability.
Finally, claims about leaves labeled with a variable or constant can be checked locally by the verifier,
without interacting with the prover.
For leaves labeled with function variables the oracle is used to check the claim.

\begin{example}
	\autoref{fig:example-run-eacertify} shows a run of the protocol.
    Initially, the verifier selects the assignment $\sigma_0(x) \coloneqq 5$.
	Each dotted arrow depicts a round of $\EACertify$.
	The prover answers the queries honestly and uses $\omega(f) = 6x$ for the evaluations.
    In the last round, \ProcedureUnifyClaims is used to combine two claims about $v_4$ into a single claim.
    Finally, two claims concerning the leaves, $v_3$ and $v_4$,
    remain. 
    The claim about $v_3$ is checked locally by asserting that $\sigma_0(x) = 5$.
    For the claim about $v_4$, an oracle query is used.
\end{example}

Next, we turn to the correctness of the zero test for EACs. 
First, we only examine \ProcedureUnifyClaims in \autoref{lem:unify-claims} which is the only chance of a dishonest prover to turn a previous false claim into a correct one.
Then, we use this result to prove correctness and soundness for the complete zero test in \autoref{thm:eacertify}.

\long\def\AlgorithmLayout#1#2{\begin{figure}[t]
        #1
    \end{figure}
    \begin{figure}[t]
        #2
    \end{figure}
}

\AlgorithmLayout{\begin{algorithm}[H]
        \caption{$\EACertify$ Zero Test (Verifier)}
         \label{alg:verifier}
Choose $\sigma \in \F^X$ uniformly at random\;
         $C_r \gets \{ \Pev{\sigma}\Eval{ r }_{\omega} \qeq 0 \}$\;
         $C_v \gets \emptyset$ for all nodes $v \neq r$\;
         \ForEach{node $v$ of $\phi$ in topological order}{
             $\left(\Pev{\hat{\sigma}}\Eval{ v }_{\omega} \qeq \hat{a} \right)\gets \ProcedureUnifyClaims(C_v)$\;
             \uIf{$v$ is labeled $b \in \F$} {
                 $\Assert(b = \hat{a})$
             }
             \uElseIf{$v$ is labeled $x \in X$} {
                 $\Assert(\hat{\sigma}(x) = \hat{a})$
             }
             \uElseIf{$v$ is labeled $f \in W$}{
                 $\tilde{a} \gets \QueryOracle(\Pev{\hat{\sigma}}f)$\;
                 $\Assert(\tilde{a} = \hat{a})$\;
             }
             \uElseIf{$v$ is labeled  $\star \in \{+, \cdot\}$ with children $c_1,c_2$}{
                 $a_i \gets \QueryProver(\Pev{\hat{\sigma}}\Eval{ c_i }_{\omega})$, $i = 1,2$\;
                 $\Assert(a_1 \star a_2 = \hat{a})$\;
                 $C_{c_i} \gets C_{c_i} \cup \{\Pev{\hat{\sigma}}\Eval{ c_i }_{\omega} \qeq a_i\}$, $i = 1,2$
             }
             \ElseIf{ $v$ is labeled  $\PartEval{x \coloneqq z}$ with child $c$}{
                 \uIf{$z \in \F$}{$\hat{\sigma}(x) \gets z$}
                    \Else(\tcp*[h]{$z \in X$ is a variable})
                    {$\hat{\sigma}(x) \gets \hat{\sigma}(z)$
                    }
                                  $C_c \gets C_c \cup \{\Pev{\hat{\sigma}}\Eval{ c }_{\omega} \qeq \hat{a}\}$
             }
         }
         \Accept\;
     \end{algorithm}
 }{\begin{algorithm}[H]
         \caption{Unify Claims\vphantom{$\EACertify$ (Verifier)}}
         \label{alg:unify-claims}
         \Procedure{\ProcedureUnifyClaims}{
             \KwIn{Claims
             $\{\Pev{\sigma_i}\Eval{ v }_{\omega} \qeq a_i \}_{i=0}^n$
              concerning a node $v$.}
             \KwOut{Single claim $\Pev{\hat{\sigma}}\Eval{ v }_{\omega} \qeq \hat{a}$}
         $d \gets \maxdeg (v)$\;
         \While{$\exists x \in X.~|\{\sigma_0(x), \ldots, \sigma_n(x)\}| > 1$}{
             choose $z \in \F$ uniformly at random\;
             \For{$i = 0,\ldots, n$}{
             $p_i \gets \QueryProver(
                 \Pev{\sigma_i[- / x]}\Eval{ v }_{\omega}
             )$\;
             \Assert$(p_i \in \F[x]_{\le d})$\;
             \Assert$(p_i(\sigma_i(x)) = a_i)$\;
             $\sigma_i(x) \gets z$\;
             $a_i \gets p_i(z)$\;
             }
         }
         \Assert($a_0 = \ldots = a_n$)\;
         \Return{$(\Pev{\sigma_0}\Eval{ v }_{\omega} \qeq a_0)$}
     }
     \end{algorithm}
}
\begin{lemma}[{implicit in \cite[Prop.~2]{ip-pspace-practical-cav}}]
    \label{lem:unify-claims}
    Let $v$ be a node of an EAC $\phi$ with variables $X$
    such that $\maxdeg (v) \le d$
    and let $\omega$ be a multilinear interpretation.
    Suppose that the set
    $\{\Pev{\sigma_i}\Eval{ v }_{\omega} \qeq a_i \}_{i=0}^n$
    contains at least one incorrect claim.
    Then
    $\ProcedureUnifyClaims\left( \{\Pev{\sigma_i}\Eval{ v }_{\omega} \qeq a_i \}_{i=0}^n\right)$
    returns a correct claim with probability at most
    $\frac{|X| \cdot d}{|\F|}$.
\end{lemma}
\begin{proof}
    Consider one iteration of the while loop where the assignments are modified at variable $x \in X$.
    Suppose that $\sigma_i$ and $a_i$ are such that
    $\Pev{\sigma_i}\Eval{ v }_{\omega} \neq a_i$ is an incorrect claim.
    Let $p_i \in \F[x]_{\le d}$ be the prover's response
    to the query $\Pev{\sigma_i[- / x]}\Eval{ v }_{\omega}$.
    We may assume that $p_i(\sigma_i(x)) = a_i$, as otherwise the verifier rejects immediately.
    Thus, $p_i \neq \Pev{\sigma_i[- / x]}\Eval{ v }_{\omega}$,
    i.e., $p_i - \Pev{\sigma_i[- / x]}\Eval{ v }_{\omega}$
    has at most $d$ zeros.
    Since $z$ is chosen uniformly at random,
    the new claim
    $\Pev{\sigma_i[z / x]}\Eval{ v }_{\omega} \qeq p_i(z)$
    holds with probability at most ${d} / {|\F|}$.
    In particular, the probability that after the iteration all claims are correct
    is bounded by $d / |\F|$.
    There are at most $|X|$ iterations of the while loop,
    so we can conclude by applying the union bound over these iterations.
\end{proof}

\begin{theorem}[{modification of \cite[Prop.~2]{ip-pspace-practical-cav}}]
    \label{thm:eacertify}
    Let $\phi$ be an EAC with set of variables $X$ and $\maxdeg(\phi) \le d$.
    If $\phi \in \EACSAT$,
    then an honest prover convinces the verifier with probability $1$.
    Otherwise, for every multilinear interpretation and every prover, the verifier accepts with probability at most
    $\frac{d \cdot (|\phi|+1) \cdot |X|}{|\F|}$.
    The verifier queries the oracle at most $|W|$ times.
\end{theorem}
\begin{proof}
    If $\phi \in \EACSAT$, it is easy to see that the verifier will always accept,
    provided that the prover answers all queries truthfully.
    Suppose $\phi \not\in \EACSAT$.
    Let $\omega$ denote the multilinear interpretation induced by the oracle.
    Then $\Eval{\phi}_{\omega} \in \F[X]_{\le d}$ is not the zero polynomial
    and thus,
    by the Schwartz-Zippel lemma,
    $\Eval{\phi}_{\omega}(\sigma) = 0$ with probability at most $\frac{d \cdot |X|}{|\F|}$.

    Suppose that the verifier chooses $\sigma$ such that $\Pev{\sigma}\Eval{\phi}_{\omega} \neq 0$.
    If the verifier wrongly accepts, then there is at least one round
    where a node $v$ is processed such that $C_v$ contains an incorrect claim,
    but no new incorrect claim is added to its children.
    Note that this can only happen when the result of $\ProcedureUnifyClaims(C_v)$ is a correct claim.
    So, by \autoref{lem:unify-claims},
    the probability that the verifier wrongly accepts is bounded by
    $\frac{d \cdot |X|}{|\F|} + |\phi|\cdot \frac{d \cdot |X|}{|\F|}$.
\end{proof}
 \section{\texorpdfstring{$\ResCertify$}{ResCertify} and a Linear-Time Prover}
\label{sec:efficient-prover}

$\ResCertify$ consists of sending the proof length $M$ to the verifier, constructing the EAC gadgets from \autoref{tab:gadgets-overview}, and invoking  $\EACertify$.
It is clear that the verifier of $\EACertify$ can be implemented such that the runtime is linear in the size of the EAC,
i.e., in the case of $\ResCertify$, in the size of the gadgets constructed in \autoref{sec:gadgets}.
In this section we sketch a linear-time prover for $\ResCertify$.
Note that evaluating the EACs naively takes exponential time.
For example, when evaluating a degree reduction node, the child needs to be evaluated for two different assignments, so a chain of $n$ such nodes requires  $2^n$-fold evaluation of the node below.
Instead, our prover is tailored to $\ResCertify$ and can certify only the gadgets from \autoref{sec:gadgets} in linear time.
We present only the central ideas here.
The full construction,
which ultimately yields a prover with runtime linear in the size of the resolution proof,
can be found in \ArxivAppendixRef{B}{sec:appendix:prover-strategies}.
Finally, we give a formal version of \autoref{thm:main} and outline the proof.
 \subparagraph{Overview}

The main difficulty is to compute the intermediate values for verifying degree reductions.
When the verifier wants to check a claim of the form $\Pev{\sigma}\DegreeReduction{X} \phi \qeq a$ for a set of variables $X = \{x_0,\ldots,x_{n-1}\} $,
the prover needs to compute the univariate polynomials
$p_0(x_0) \coloneqq \Pev{\sigma_0} \DegreeReduction{\{x_1,\ldots,x_{n-1}\}} \phi$,
$p_1(x_1) \coloneqq  \Pev{\sigma_1} \DegreeReduction{\{x_2,\ldots,x_{n-1}\}} \phi$,
$\ldots$,
and
$p_{n-1}(x_{n-1}) \coloneqq \Pev{\sigma_{n-1}} \phi$
in this order.
The assignment $\sigma_i$ is indeterminate at $x_i$,
$\sigma_0 \coloneqq \sigma[-/x_0]$,
and consecutive assignments
$\sigma_i$ and  $\sigma_{i+1}$ differ only on $\{ x_i, x_{i+1}\}$;
$\sigma_{i+1}(x_i)$ being the new random element $z \in \F$ chosen by the verifier in the procedure \UnifyClaims.
Thus, the prover can cache intermediate results.
However, observe that when computing $p_i$,
it is crucial that the \emph{next} assignment $\sigma_{i+1}$
is not yet known to the prover,
hence the evaluation must be conducted through an online algorithm.
To simplify the presentation,
we only describe how to evaluate a sequence
$p_i(z_i) = \Pev{\tilde{\sigma}_i} \DegreeReduction{\{x_{i+1},\ldots,x_{n-1}\}} \phi$,
where $\tilde{\sigma_i} \coloneqq  \sigma_i[z_i / x_i]$ is a total assignment.
The univariate polynomials $p_i$ can be computed analogously (or recovered using interpolation).

We employ different strategies for the individual degree reduction blocks.
For example, for the gadget $\GadgetBinary(p) = \DegreeReduction{X}\DegreeReduction{Y}(p(1-p))$,
the prover needs to respond to queries of the forms
\begin{enumerate*}[(i)]
    \item $\DegreeReduction{Y'}(p(1-p))$, where $Y' \subseteq Y$, and
    \item $\DegreeReduction{X'}\DegreeReduction{Y}(p(1-p))$, where $X' \subseteq X$.
\end{enumerate*}
For (i), the most common strategy,
\autoref{lem:y-index:product:efficient}, is used,
whereas (ii) requires a special strategy, \autoref{lem:prover:evaluate-binary-p}.
\autoref{tab:gadgets-prover-overview} provides an overview of the strategies for each gadget and the names of the corresponding parts of our implementation.
\autoref{lem:y-index:product:efficient} and \autoref{lem:prover:evaluate-binary-p} showcase the central ideas.
Most of the other strategies are simple variations.
The only exception is the strategy for $\DegreeReduction{Z}$ in $\GadgetRegressive(u_i)$, which is rather technical.
All details can be found in \ArxivAppendixRef{B}{sec:appendix:prover-strategies}.

\ifx\authoranonymous\relax \def\codeLink#1#2{{\texttt{#2}}}\else
    \def\codeLink#1#2{\href{https://gitlab.lrz.de/i7/rescertify/-/blob/main/src/#1}{\texttt{#2}}}\fi

\begin{table}[htpb]
	\centering
	\caption{Overview of Prover Strategies}
	\label{tab:gadgets-prover-overview}
    \begin{tabular}{rllm{6.5cm}}
		Gadget
		& $\DegreeReduction{}$
        & Lemma
		& Implementation
		\\
		$\GadgetBinary(p)$
		& $Y$
        & \autoref{lem:y-index:product:efficient}
		& \codeLink{prover/strategies/ProductDegreeReductionStrategy.hpp}{ProductDegreeReductionStrategy}
		\\
		& $X$
        & \autoref{lem:prover:evaluate-binary-p}
        & {\hbadness=1000\codeLink{prover/strategies/CNFArithmetizationBinaryDegreeReductionStrategy.hpp}{CNFArithmetizationBinary}\-\texttt{DegreeReductionStrategy}}
		\\
		$\GadgetBinary(u_i)$
		& $Y$
        & \autoref{lem:y-index:product:efficient}
        & \codeLink{prover/strategies/ProductDegreeReductionStrategy.hpp}{ProductDegreeReductionStrategy}
		\\
		& $Z$
        & \ArxivLemmaRef{35}{lem:squared-index}
        & \codeLink{prover/strategies/SquaredIndexDegreeReductionStrategy.hpp}{SquaredIndexDegreeReductionStrategy}
		\\
		$\GadgetInitial(p)$
		& $Y$
        & \autoref{lem:y-index:product:efficient}
        & \codeLink{prover/strategies/ProductDegreeReductionStrategy.hpp}{ProductDegreeReductionStrategy}
		\\
		$\GadgetRegressive(u_i)$
		& $Y$
        &  \autoref{lem:y-index:product:efficient}
        & \codeLink{prover/strategies/ProductDegreeReductionStrategy.hpp}{ProductDegreeReductionStrategy}
		\\
		& $Z$
        & \ArxivLemmaRef{36}{lem:efficient_v5z}
        & \codeLink{prover/strategies/RegressiveDegreeReductionStrategy.hpp}{RegressiveDegreeReductionStrategy}
		\\
		$\GadgetResolution(p, u_1, u_2)$
		& $Z$
        & \autoref{lem:y-index:product:efficient}
        & \codeLink{prover/strategies/ProductDegreeReductionStrategy.hpp}{ProductDegreeReductionStrategy}
		\\
		& $Y$
        & \autoref{lem:y-index:product:efficient}
        & \codeLink{prover/strategies/ThreefoldProductDegreeReductionStrategy.hpp}{ThreefoldProductDegreeReductionStrategy}
		\\
		& $X$
        & \ArxivLemmaRef{33}{lem:y-index:degree-reduction-x:optimized}
        & \codeLink{prover/strategies/ResolutionDegreeReductionStrategy.hpp}{ResolutionDegreeReductionStrategy}
		\\
	\end{tabular}
\end{table}

 \subsection{\texorpdfstring{$Y$}{Y}-Indices}
\todo[inline]{
}

Central to our analysis will be the notion of a \emph{$Y$-index}:
Let $P$ denote a polynomial over variables $X\cup Y$ which is multilinear in $Y$.
We can write $P$ as $P=\sum_{i} \IndexPoly{Y}{i} p_i$,
where $p_i \in \F[X]$.
The idea is that each $p_i$ is simple and has a short representation (e.g., as an arithmetic circuit).
We call $P = \sum_i \IndexPoly{Y}{i} p_i$ a \emph{$Y$-index}
and store it as a list $(p_0,\ldots,p_{k-1})$.
It has \emph{size} $\Abs{P}\coloneqq \sum_i\Abs{p_i}$,
where $\Abs{p_i}$ is the size of some appropriately chosen representation of $p_i$.

\begin{figure}[t]
	\centering
	\scalebox{0.85}{
        \begin{tikzpicture}
            \def\rootHeight{0}
            \def\layerAHeight{-1}
            \def\layerBHeight{-2}
            \def\layerCHeight{-3.5}
            \def\leafHeight{-4.2}
            \node[yindexnode] (r0) at (0,\rootHeight)      {$1$};
            \node[yindexnode] (r1) at (-4,\layerAHeight)    {$\IndexPoly{\{y_2\}}{0}$};
            \node[yindexnode] (r2) at (4,\layerAHeight)     {$\IndexPoly{\{y_2\}}{1}$};
            \node[yindexnode] (r3) at (-6,\layerBHeight)    {$\IndexPoly{\{y_1,y_2\}}{0}$};
			\node[yindexnode] (r4) at (-2,\layerBHeight)    {$\IndexPoly{\{y_1,y_2\}}{1}$};
			\node[yindexnode] (r5) at (2,\layerBHeight)     {$\IndexPoly{\{y_1,y_2\}}{2}$};
			\node[yindexnode] (r6) at (6,\layerBHeight)     {$\IndexPoly{\{y_1,y_2\}}{3}$};
            \node[yindexnode] (r7) at (-7,\layerCHeight)    {$\IndexPoly{Y}{0}$};
            \node[yindexnode] (r8) at (-5,\layerCHeight)    {$\IndexPoly{Y}{1}$};
            \node[yindexnode] (r9) at (-3,\layerCHeight)    {$\IndexPoly{Y}{2}$};
            \node[yindexnode] (r10) at (-1,\layerCHeight)   {$\IndexPoly{Y}{3}$};
            \node[yindexnode] (r11) at (1,\layerCHeight)    {$\IndexPoly{Y}{4}$};
            \node[yindexnode] (r12) at (3,\layerCHeight)    {$\IndexPoly{Y}{5}$};
            \node[yindexnode] (r13) at (5,\layerCHeight)    {$\IndexPoly{Y}{6}$};
            \node[yindexnode] (r14) at (7,\layerCHeight)    {$\IndexPoly{Y}{7}$};
            \node[yindexleaf] (l1) at (-7,\leafHeight) {$p_0$};
			\node[yindexleaf] (l2) at (-5,\leafHeight) {$p_1$};
			\node[yindexleaf] (l3) at (-3,\leafHeight) {$p_2$};
			\node[yindexleaf] (l4) at (-1,\leafHeight) {$p_3$};
			\node[yindexleaf] (l5) at (1,\leafHeight) {$p_4$};
			\node[yindexleaf] (l6) at (3,\leafHeight) {$p_5$};
			\node[yindexleaf] (l7) at (5,\leafHeight) {$p_6$};
			\node[yindexleaf] (l8) at (7,\leafHeight) {$p_7$};
			\draw[yindexedge] (r0) edge node [above left] {$(1-y_2)$} (r1);
			\draw[yindexedge] (r0) edge node [above right] {$y_2$} (r2);
			\draw[yindexedge] (r1) edge node [above left] {$(1-y_1)$} (r3);
			\draw[yindexedge] (r1) edge node [above right] {$y_1$} (r4);
			\draw[yindexedge] (r2) edge node [above left] {$(1-y_1)$} (r5);
			\draw[yindexedge] (r2) edge node [above right] {$y_1$} (r6);
			\draw[yindexedge] (r3) edge node [left] {$(1-y_0)$} (r7);
			\draw[yindexedge] (r3) edge node [right] {$y_0$} (r8);
			\draw[yindexedge] (r4) edge node [left] {$(1-y_0)$} (r9);
			\draw[yindexedge] (r4) edge node [right] {$y_0$} (r10);
			\draw[yindexedge] (r5) edge node [left] {$(1-y_0)$} (r11);
			\draw[yindexedge] (r5) edge node [right] {$y_0$} (r12);
			\draw[yindexedge] (r6) edge node [left] {$(1-y_0)$} (r13);
			\draw[yindexedge] (r6) edge node [right] {$y_0$} (r14);
	\end{tikzpicture}}
    \caption{A $Y$-index $P = \sum_{i=0}^{7} \IndexPoly{Y}{i} p_i$ viewed as a tree ($Y = \{y_0, y_1, y_2\}$)}
	\label{fig:cache-init}
\end{figure}

Let $Y =\{y_0,\ldots,y_{|Y|-1}\}$.
We visualize $Y$-indices as binary trees of height $|Y|$,
see \autoref{fig:cache-init}.
At an inner node of level $|Y| - j$, we can decide between $y_{j}$ and $(1 - y_{j})$.
Thus, the path from the root to the $i$-th leaf represents the key polynomial $\IndexPoly{Y}{i}$.
Each leaf is annotated with its corresponding arithmetic circuit $p_i$.

\begin{lemma}\label{lem:y-index:eval}
    Let $P = \sum_{i} \IndexPoly{Y}{i} p_i$
    be a $Y$-index,
    and $\sigma\colon X \cup Y \to \F$ an assignment.
    The evaluation $\Pev{\sigma} P$
    can be computed in time $\O(|P| + 2^{|Y|})$.
\end{lemma}
\begin{proof}
    First, we evaluate all circuits $p_i$, which requires time $\O(|P|)$.
    Computing each $\Pev{\sigma} \IndexPoly{Y}{i}$ individually would take time $\O(|Y| \cdot 2^{|Y|})$.
    However, we can avoid duplicate work by evaluating them following the tree structure depicted in \autoref{fig:cache-init}.
\end{proof}

 \subsection{Degree Reductions in Y}

Recall that in our construction, the set of variables $Y$ was used for key polynomials that encode the indices of clauses.
Moreover, note that in our gadgets, degree reductions in $Y$ are always applied to products
of two (or, in the case of $\GadgetResolution$, three) factors that are multilinear in $Y$.
We will store most multilinear polynomials as $Y$-indices,
making the degree reductions in $Y$ easiest to compute:
\begin{lemma}
    \label{lem:y-index:product}
    Let $P = \sum_i \IndexPoly{Y}{i} p_i$ and $Q = \sum_i \IndexPoly{Y}{i} q_i$ be $Y$-indices.
    A $Y$-index $R=\sum_i \IndexPoly{Y}{i}r_i$ with $R=\DegreeReduction{Y}({P}\cdot{Q})$ can be computed in time $\O(\Abs{P} + \Abs{Q})$,
    and $\Abs{R}\in\O(\Abs{P}+\Abs{Q})$.
\end{lemma}
\begin{proof}
    Since
    $\DegreeReduction{Y}\left( \IndexPoly{Y}{i} \cdot \IndexPoly{Y}{j} \right) = 0$
    for $i \neq j$,
    and
    $\DegreeReduction{Y}\left( \IndexPoly{Y}{i}^2 \right) = \IndexPoly{Y}{i}$,
	we can set $r_i\coloneqq p_i\cdot q_i$.
\end{proof}
The case of three factors is analogous.
However, the prover needs to compute a \emph{sequence} of $|Y|$ degree reductions in both cases,
since the verifier also queries intermediate values.
We sketch how a linear runtime can be achieved by caching intermediate results:

Consider a small case where $Y = \{y_0,y_1,y_2\}$.
First, the verifier asks about $\Pev{\sigma_0}\DegreeReduction{Y} {P} \cdot {Q}$ for some assignment $\sigma_0$.
The prover evaluates the leaves of the $Y$-indices $P$ and $Q$ at $\sigma_0$
and then uses \autoref{lem:y-index:product} and \autoref{lem:y-index:eval} to compute the response.

\begin{figure}[t]
	\centering
	\scalebox{0.75}{
        \def\rootHeight{0}
        \def\layerAHeight{-1}
        \def\layerBHeight{-2}
        \def\leafHeight{-2.75}
		\begin{tikzpicture}
            \node[yindexnode] (r0) at (0,\rootHeight)      {$1$};
            \node[yindexnode] (r1) at (-4,\layerAHeight)    {$\IndexPoly{\{y_2\}}{0}$};
            \node[yindexnode] (r2) at (4,\layerAHeight)     {$\IndexPoly{\{y_2\}}{1}$};
            \node[yindexnode] (r3) at (-6,\layerBHeight)    {$\IndexPoly{\{y_1,y_2\}}{0}$};
			\node[yindexnode] (r4) at (-2,\layerBHeight)    {$\IndexPoly{\{y_1,y_2\}}{1}$};
			\node[yindexnode] (r5) at (2,\layerBHeight)     {$\IndexPoly{\{y_1,y_2\}}{2}$};
			\node[yindexnode] (r6) at (6,\layerBHeight)     {$\IndexPoly{\{y_1,y_2\}}{3}$};
            \node[yindexleaf] (l1) at (-6,\leafHeight) [xshift=-0.5cm] {$(1-y_0) \Pev{\sigma_0}p_0 + y_0 \Pev{\sigma_0}p_1$};
			\node[yindexleaf] (l2) at (-2,\leafHeight) [xshift=-0.2cm] {$(1-y_0) \Pev{\sigma_0} p_2 + y_0 \Pev{\sigma_0}p_3$};
			\node[yindexleaf] (l3) at (2,\leafHeight) [xshift=0.2cm] {$(1-y_0) \Pev{\sigma_0}p_4+ y_0 \Pev{\sigma_0}p_5$};
			\node[yindexleaf] (l4) at (6,\leafHeight) [xshift=0.5cm] {$(1-y_0) \Pev{\sigma_0}p_6+ y_0 \Pev{\sigma_0}p_7$};
            \draw[yindexedge] (r0) edge node [above left] {$(1-y_2)$} (r1);
			\draw[yindexedge] (r0) edge node [above right] {$y_2$} (r2);
			\draw[yindexedge] (r1) edge node [above left] {$(1-y_1)$} (r3);
			\draw[yindexedge] (r1) edge node [above right] {$y_1$} (r4);
			\draw[yindexedge] (r2) edge node [above left] {$(1-y_1)$} (r5);
			\draw[yindexedge] (r2) edge node [above right] {$y_1$} (r6);
        \end{tikzpicture}
    }
        \caption{Update of $P$ from \autoref{fig:cache-init} after the first query of the verifier.}
	\label{fig:cache-step}
\end{figure}
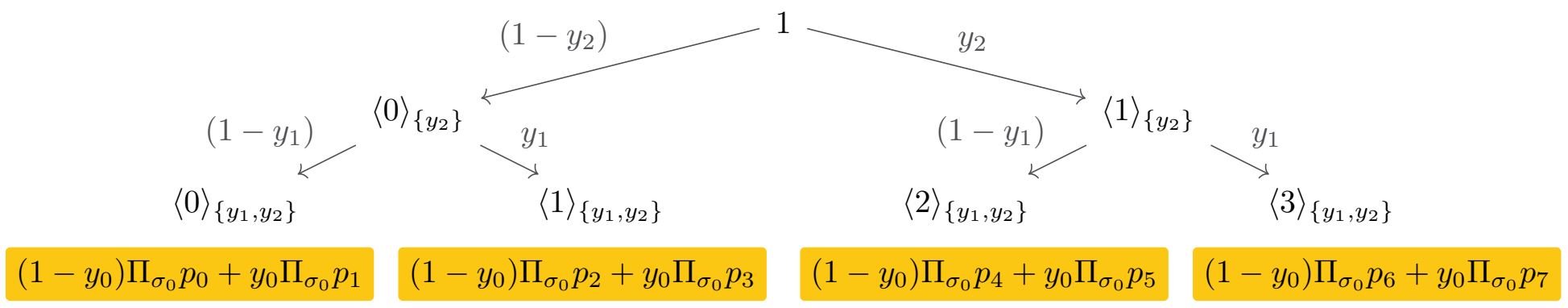

Next, the verifier asks for a univariate polynomial in $y_0$ to check the degree reduction $\DegreeReduction{y_1} \DegreeReduction{y_2} {P} \cdot {Q}$.
Note that $\DegreeReduction{y_1}\DegreeReduction{y_2}{P}\cdot{Q}$ is now only a multilinear polynomial in $\{y_1, y_2\}$.
To account for that, we update the leaves of both our $Y$-indices accordingly,
turning them into $\{y_1, y_2\}$-indices whose leaves are polynomials in $y_0$.
In \autoref{fig:cache-step} this is shown for ${P}$.
The univariate polynomial for $\DegreeReduction{y_1} \DegreeReduction{y_2} {P} \cdot {Q}$ can again be computed using \autoref{lem:y-index:product} and \autoref{lem:y-index:eval}.

After that, the verifier provides a new value for $y_0$ and asks for a univariate polynomial in $y_1$ to check $\DegreeReduction{y_2} {P}\cdot {Q}$.
The prover evaluates the leaves of both $\{y_1,y_2\}$-indices, then updates them to obtain $\{y_2\}$-indices
whose leaves are polynomials in $y_1$,
and so on.
As the size of $P$ and $Q$ is halved in each step,
the resulting total runtime is $\O(|P| + |Q| + 2^{|Y|})$.
Generalizing this argument, we obtain:
\begin{restatable}{lemma}{degYLinear}
	\label{lem:y-index:product:efficient}
	Let $Y = \{y_0,\ldots, y_{|Y|-1}\}$.
	Consider two $Y$-indices
	${P} = \sum_{i} \IndexPoly{Y}{i} {p_i}$
	and
	${Q} = \sum_{j} \IndexPoly{Y}{j} {q_j}$,
	where ${p_i}, {q_j} \in \F[X]$.
	Let $\sigma_0, \sigma_1,\ldots, \sigma_{|Y|}$ be assignments $Y \cup X \to \F$ s.t.\
    $\sigma_{k+1} = \sigma_k[a_k/y_{k}]$ for some $a_k \in \F$.
	Then the list of evaluations
	$\Pev{\sigma_0} \DegreeReduction{Y} ({P} \cdot {Q}), \Pev{\sigma_1} \DegreeReduction{\{y_1,\ldots, y_{|Y|-1}\}} ({P} \cdot {Q}),
	\ldots, \Pev{\sigma_{|Y|}} ({P} \cdot {Q})$ can be computed online
	in total time $\O(|P| + |Q| + 2^{|Y|})$.
\end{restatable}

Note that in our construction, $Y$-indices are always based on the resolution proof and thus their size is limited by the size of the resolution proof.
 \subsection{Degree Reductions in X}

The variables $X$ are the variables of the CNF formula.
Consider the degree reduction $\DegreeReduction{X}$ in the gadget
$\GadgetBinary(p) = \DegreeReduction{X}\DegreeReduction{Y}(p \cdot (1-p))$.
We have
$\DegreeReduction{Y}(p \cdot (1-p)) = \sum_i \IndexPoly{Y}{i} \Arith{ C_i } (1 - \Arith{ C_i } )$ by \autoref{lem:y-index:product},
where $C_i$ is the $i$-th clause of the refutation.
Recall that $\ArithLit{X}\subseteq \Multilinear{X}$
denotes the set of arithmetized literals in $X$.

\begin{lemma}
	\label{lem:prover:evaluate-binary-p}
    Let $P = \sum_i \IndexPoly{Y}{i} p_i$ denote a $Y$-index
    such that there is a set $L_i \subseteq \ArithLit{X}$ with
	$p_i=(1-\prod_{\ell\in L_i}\ell)\prod_{\ell\in L_i}\ell$ for each $i$.
    Consider a subset of variables $X^\ast \subseteq X$.
    A $Y$-index $Q^{X^\ast} = \DegreeReduction{X^*} P$ can be computed in time $\O(\Abs{P})$.
\end{lemma}
\begin{proof}
	Fix an index $i$ and let $L_i$ be such that
    $p_i=(1-\prod_{ \ell \in L_i} \ell)\prod_{\ell\in L_i}\ell$.
	Note that a variable $\hat{x}\in X^*$ which does not appear in $L_i$ does not appear in $p_i$
	and can thus be ignored.
	Suppose that the variable $\hat{x}\in X^*$ occurs in $L_i$.
	Consider
	$\DegreeReduction{\hat{x}} {p_i} = (1-\hat{x}) \pev{\hat{x}\coloneqq 0} {p_i} + \hat{x} \pev{\hat{x}\coloneqq 1} {p_i}$.
	Note that
	\begin{enumerate*}[(i)]
		\item if $\hat{x} \in L_i$ then $\pev{\hat{x} \coloneqq 0} {p_i} = 0$
		and
		\item if $(1- \hat{x}) \in L_i$ then $\pev{\hat{x} \coloneqq  1} {p_i} = 0$.
	\end{enumerate*}
	Thus,
    we obtain a $Y$-index $P' = \sum_i \IndexPoly{Y}{i} p_i'$ with ${P'} = \DegreeReduction{\hat{x}} {P}$
	by setting
\[p_i' \coloneqq
	\begin{cases}
		(1 - \hat{x}) \cdot \pev{\hat{x} \coloneqq 0} p_i &\text{ if } (1- \hat{x}) \in L_i,\\
		\hat{x} \cdot  \pev{\hat{x} \coloneqq 1} p_i &\text{ if } \hat{x} \in L_i,\\
		p_i &\text{ otherwise.}
	\end{cases}
	\]
So when $\hat{x}$ occurs in the literal $r \in L_i$,
	we have
    $p_i'
	= r \cdot (1 - \prod_{\ell \in L_i \setminus \{r\}} \ell) \prod_{\ell \in L_i \setminus \{r\}} \ell
= (1 - \prod_{\ell \in L_i \setminus \{r\}} \ell) \prod_{\ell \in L_i} \ell$.
	Iterating this argument,
    we obtain that the $Y$-index $Q^{X^\ast} =\sum_i \IndexPoly{Y}{i} q_i^{(X^\ast)}$ given by
    $q_i^{X^\ast} \coloneqq (1- \prod_{\ell \in L_i \setminus \ArithLit{X^\ast}} \ell) \prod_{\ell \in L_i} \ell$
	satisfies ${Q^{X^\ast}} = \DegreeReduction{X^\ast} {P}$.
\end{proof}
Applying \autoref{lem:prover:evaluate-binary-p} naively only yields a quadratic runtime for the prover.
However, the runtime can again be improved.
The basic idea is similar to that of \autoref{lem:y-index:product:efficient};
see \ArxivAppendixRef{B.2.1}{sec:prover:evaluate-binary-p:optimized}.
 \subsection{Main Theorem} Summarizing, we obtain:
\begin{reusetheoremnumber}{1}
\begin{theorem}\label{thm:main:formal}
	Let $\varphi$ be a CNF-formula with $n$ variables and $\Phi$ a resolution proof of $\varphi$.
    In the \emph{multilinearity test} of $\ResCertify$,
	\begin{noitemize}
        \item the prover is not used,
		\item $\O(n)$ queries are sent to the oracle, and
		\item the verifier runs in time $\O(n)$, with soundness parameter $1/2$.\footnote{Error reduction by repeating this step is of course possible.}
	\end{noitemize}
    In the \emph{zero test},
	\begin{noitemize}
        \item the prover can answer all queries in time $\O(|\Phi|)$,
        \item the oracle is queried $3$ times, and
		\item the verifier takes time $\O(|\varphi| + \log(|\Phi|))$, with soundness parameter $\O(\mathrm{|\varphi| \cdot n / |\F|})$.
	\end{noitemize}
    Each oracle query can be answered in $\O(|\Phi|)$ time.
\end{theorem}
\end{reusetheoremnumber}
\begin{proof}[Proof (sketch)]
    The honest prover turns the resolution refutation $\Phi$
    into a polynomial refutation
    $\omega$,
see Appendix \ref{sec:appendix:res-refutation-complete}.
    The multilinearity test, \autoref{fact:multilinearity-test}, is applied to the function variables
    $p$, $u_1$ and $u_2$.
Then the verifier constructs the gadgets from \autoref{tab:gadgets-overview},
    which have size $\O(|\varphi| + \log(|\Phi|))$.
    The claims about the verifier follow from
    the correctness of $\EACertify$
    (\autoref{thm:eacertify}),
    the fact that polynomial refutations are sound and complete (\autoref{lem:polynomial-refutation-complete}),
    and the correctness of the gadgets
    (\autoref{lem:gadget:regressive}
    and Appendix \ref{appendix:proof:gadgets}).
    The prover can answer the queries
	using the strategies from \autoref{tab:gadgets-prover-overview}.
\end{proof}

 \section{Implementation and Evaluation}\label{sec:evaluation}

\begin{figure}[t]
	\begin{subfigure}[t]{0.475\textwidth}
		\centering
		\includegraphics[width=\textwidth]{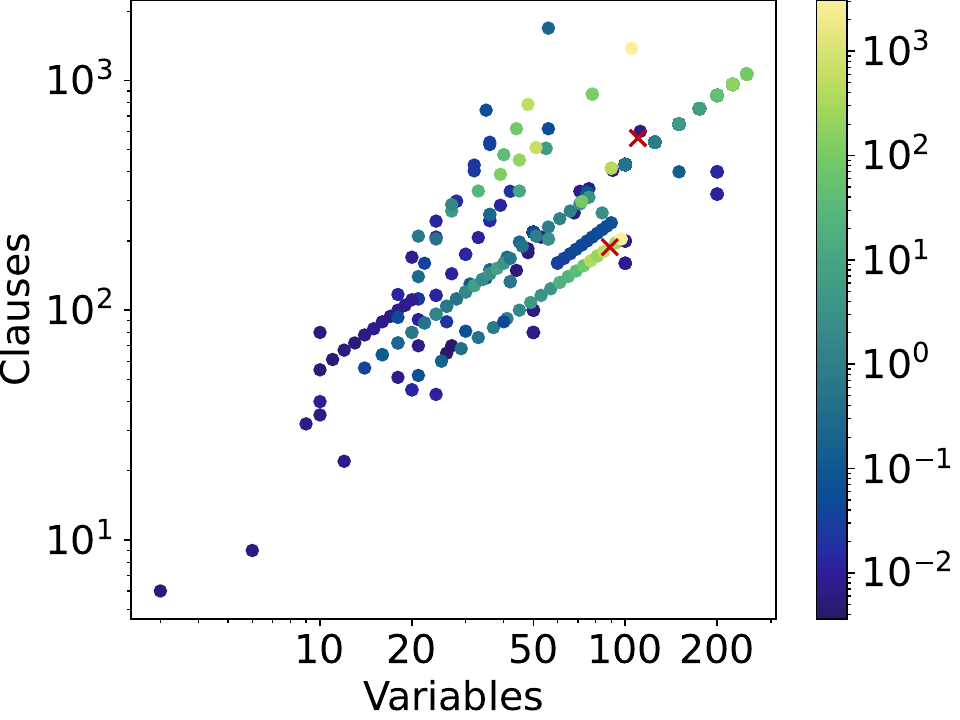}
		\subcaption{Distribution of our benchmark instances. Colors indicate the runtime of $\ResCertify$ in seconds.}
		\label{fig:benchmark-stats}
	\end{subfigure}
	\hfill
	\begin{subfigure}[t]{0.475\textwidth}
		\centering
		\includegraphics[width=\textwidth]{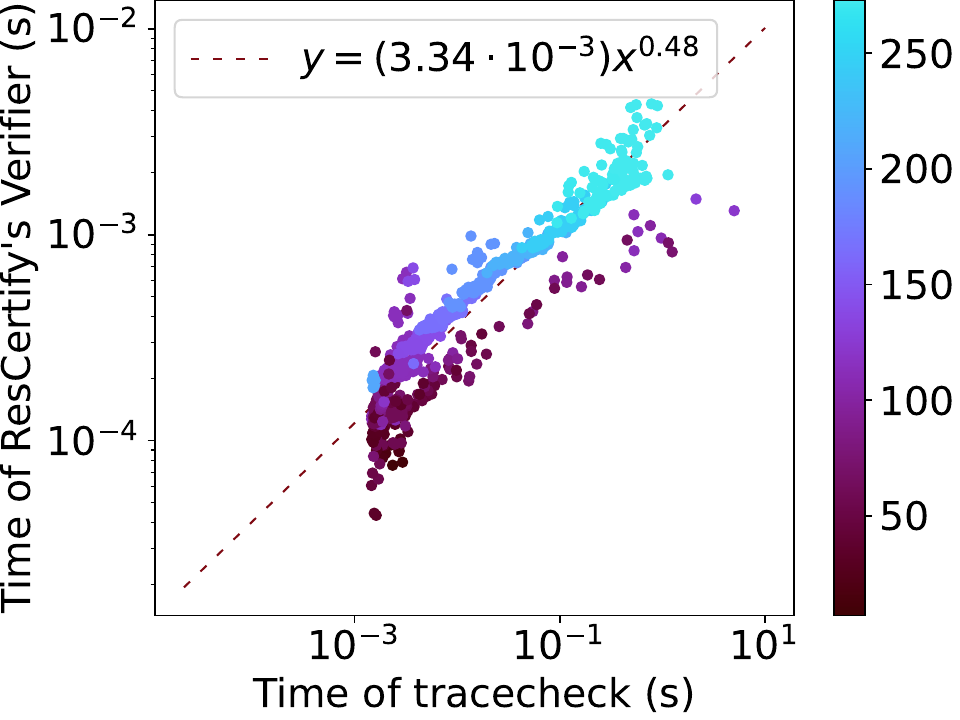}
		\subcaption{Runtime of our verifier and \texttt{tracecheck} where the color of each instance represents $n + \log(M)$}
		\label{fig:verifier-tracecheck}
	\end{subfigure}
    \par\medskip
    \begin{subfigure}[t]{0.475\textwidth}
		\centering
		\includegraphics[width=\textwidth]{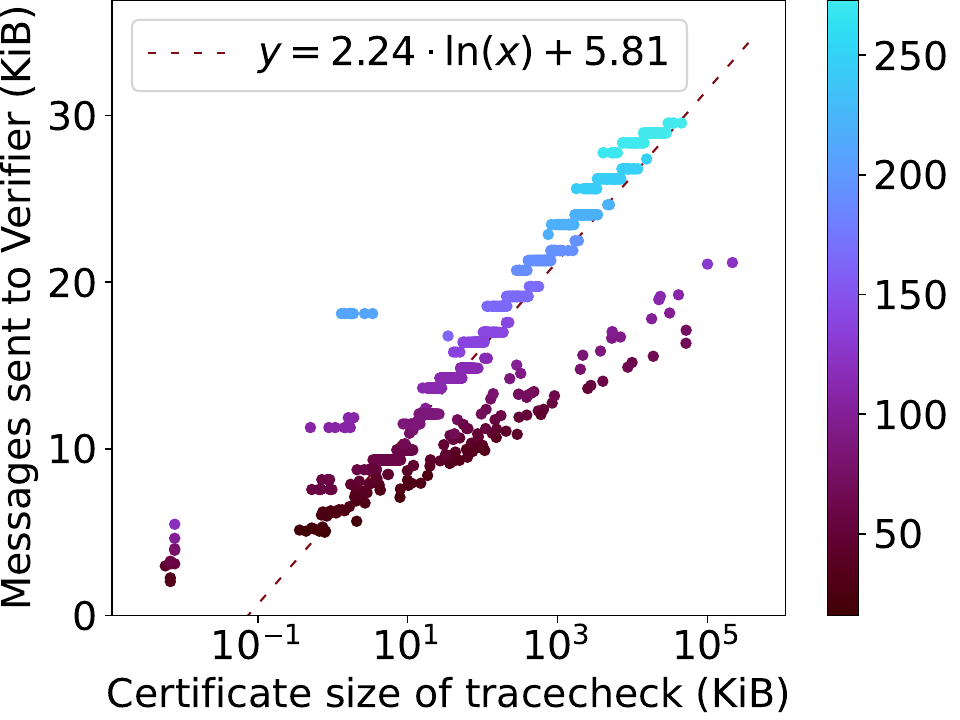}
		\subcaption{Size of messages from the prover compared to size of resolution proof.
            Colors as in \autoref{fig:verifier-tracecheck}.
        }
		\label{fig:proof_sizes}
	\end{subfigure}
	\hfill
	\begin{subfigure}[t]{0.475\textwidth}
		\centering
		\includegraphics[width=\textwidth]{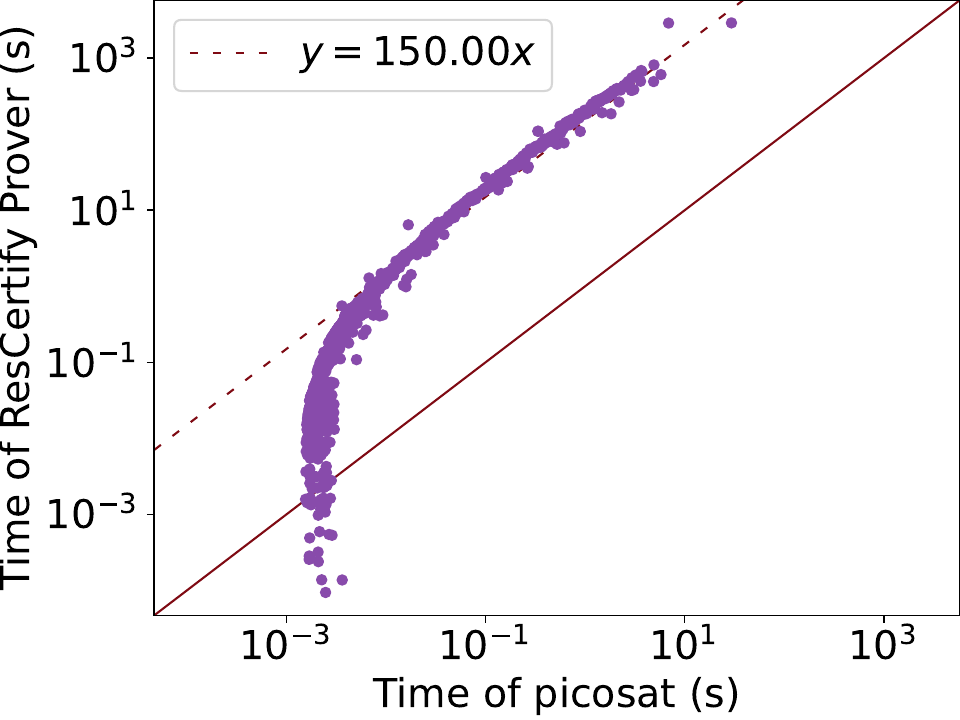}
        \subcaption{Time of \texttt{picosat} to solve instances compared to our prover for answering the verifier's queries.}
		\label{fig:prover_picosat}
	\end{subfigure}
    \caption{Experimental Evaluation of $\ResCertify$.}
\end{figure}

We have implemented both the verifier and the prover of
$\ResCertify$, but not the multilinearity test, i.e., the verifier trusts that the interpretation $\omega$ is multilinear but checks $\Eval{\phi}_{\omega} = 0$. This corresponds to the following scenario: the verifier trusts that the SAT-solver computes an object of type resolution proof, but not that the object itself is a correct refutation of the input formula (see \autoref{rem:oracle-trust-tradeoff} about the oracle assumption). While in theory checking type correctness can be as hard as checking any other property of a program, in practice it is not so:  for simple types, as in our case, type correctness is enforced by many programming languages, or checked at compile time. If, for example, the code of the SAT-solver is publicly available, then the verifier may well be willing to trust type correctness, but not full correctness.

Our implementation takes a \texttt{tracefile} as input,  generated from a formula with the help of \texttt{picosat}
and \texttt{tracecheck}
\cite{picosat}.
From the tracefile we extract a RES-refutation and the induced polynomial refutation.

Note that \texttt{picosat} and \texttt{tracecheck} are not considered state-of-the-art.
Modern solvers use other formats, e.g.~\texttt{LRUP}.
To evaluate the overhead introduced by our (resolution-based) approach,
we chose a format as close to plain resolution as possible.

We conduct our experiments on an AMD EPYC 7702 with 2 TB of system memory under Linux.
As benchmarks we use 2890 unsatisfiable formulas, 2740 from satlib \cite{satlib}, 149 generated with cnfgen \cite{cnfgen} and the last is taken from~\cite{CzernerEK24}.
\autoref{fig:benchmark-stats} shows an overview of the number of clauses and variables of the instances.
A dot represents an instance which $\ResCertify$ could solve under 60 minutes, crosses represent timeouts (which occurred for only 16 formulas). The brighter a dot is, the higher the overall runtime for that instance.
Our instances are small examples compared to the benchmarks of the annual SAT-competition; the limiting factor is the performance of our prover. We use the field $\F_p$ with $p = 2^{61} - 1$.

\subparagraph{Performance of $\ResCertify$'s verifier.}
\autoref{fig:verifier-tracecheck} plots for each benchmark the time \texttt{tracecheck} needs to check a \texttt{tracefile} supplied by \texttt{picosat} against the time needed by $\ResCertify$'s verifier.
In the main branch of the diagram the verifier quickly becomes much faster than \texttt{tracecheck}: for the larger formulas (bright) it is about $10^3$ times faster, and the factor grows rapidly.
A second branch is visible. It is roughly parallel to the main branch and thus the factor is only slightly smaller.
Extrapolating the behavior according to the dashed line --- which is, of course, speculative --- the verifier could check in one second refutations for which \texttt{tracecheck} would need about 40 hours.
This corroborates the theoretical advantage of our verifier, which needs linear time in the \emph{size of the formula}, whereas \texttt{tracecheck} needs at least linear time in the \emph{size of the resolution proof}.

\autoref{fig:proof_sizes} compares the total size of the messages sent from the prover to the verifier against the size of the resolution proof. This corresponds to the communication complexity of checking a refutation in the interactive and conventional setting (where the prover sends the verifier the complete refutation). 
Note that \autoref{fig:proof_sizes} is a log-linear plot.
We observe the same pattern as for the runtime: for the larger formulas (bright), the communication complexity of  \texttt{tracecheck} is about $10^3$ times higher. 
Extrapolating the dashed line (which, again, is speculative), the prover and the verifier would need to exchange about 50 KB of messages to check a 1 TB refutation.
Again, a second branch is visible. It consists of smaller instances where fewer messages are sent, resulting in a smaller slope.

The price to pay for the advantages of an interactive verifier is ``only'' a multiplicative constant factor in the runtime of the prover.  However, this factor is still high.
\autoref{fig:prover_picosat} plots for each benchmark the time needed by \texttt{picosat} to compute the refutation against the time needed by our prover to execute its side of $\ResCertify$.
The results corroborate the theoretical result (after an initial phase for small instances the dots arrange in a line parallel to the diagonal), but give a constant of about $1.5 \cdot 10^2$.
Loosely speaking, the reason is that our prover has to evaluate the polynomial encoding of the resolution proof multiple times, whereas \texttt{picosat} only produces it once. Whether the factor can be reduced is left for future research.
 \section{Conclusion}
We presented $\ResCertify$, the first interactive proof system tailored towards certifying arbitrary resolution proofs with a linear-time verifier in the size of the initial formula. We thus improve on \cite{CzernerEK24}, which only worked for resolution proofs  constructed using the Davis-Putnam procedure, at the price of adding a second prover. Our provers run in quasi-optimal time if both the multilinearity and the zero tests are used, and in optimal time if only the zero test is used. The verifier runs in time $\O(|\varphi| + \log(|\Phi|))$ for a formula $\varphi$ with resolution proof $\Phi$.
By comparison, a conventional certificate checker
runs in total time $\O(|\Phi|)$,
so in our protocol the work done by the trusted component is reduced substantially.
However, the combined cost of the verifier and the prover exceeds that of a conventional certificate checker, since the $\O(|\Phi|)$ runtime of our prover hides a considerably larger constant.

While our work is theoretical, we have also presented a proof-of-concept implementation and compared it to the plain resolution proof checker \texttt{tracecheck}.
Note that modern checkers use more sophisticated proof formats.
Our verifier checked most instances around $10^3$ times faster than \texttt{tracecheck} in the size range we could explore, and the factor should grow rapidly for larger size due to the theoretical bounds.
On the other hand, the zero test introduces a still large constant time factor of about $10^2$ on the side of the prover.

For clarity of presentation, we focused on plain resolution proofs,
which are no longer viewed as state of the art.
Our approach is amenable to other implicational proof formats; we sketch how to adapt it to RUP proofs in \ArxivAppendixRef{C}{sec:rup}.
However, other formats like \texttt{DRAT} remain out of reach.
Constructing interactive certification systems for such proof formats remains open.
 \bibliography{bibliography}
\newpage
 \appendix 

 \section{Proofs for \autoref{sec:gadgets}}
\label{appendix:proof:gadgets}

 \subsection{Proof of Lemma \ref{lem:polynomial-refutation-complete}}
\label{sec:appendix:res-refutation-complete}

A RES-refutation can be transformed into a polynomial refutation as follows:
Let $\varphi=(C_0,\ldots,C_{m-1})$ be a CNF-formula with set of variables $X$
and let $\Phi=(C_0,\ldots,C_{M-1})$ be a RES-refutation of $\varphi$.
Let $Y$ and $Z$ be tuples of new variables, such that $\Abs{Y} = \Abs{Z} > \left\lceil \log M \right\rceil$.
Let $U_1$ and $U_2$ be functions $\{m,\ldots,M-1\} \to \{0,\ldots,M-1\}$ such that,
for all $i = m,\ldots,M-1$,
$(U_1(i), U_2(i))$ is an antecedent list for $C_i$,
i.e., $C_i$ is the resolvent of  $C_{U_1(i)}$ and $C_{U_2(i)}$.
We arithmetize $p$ and the functions $U_1$ and $U_2$ as in \autoref{subsection:multilinear-polynomials},
i.e., $\omega(p) \coloneqq \Arith{\Phi}_Y$, $\omega(u_1) \coloneqq \MultilinEncode{Y,Z}{U_1} = \sum_{U_1(i) = j} \IndexPoly{Y}{i} \IndexPoly{Z}{j}$ and $\omega(u_2) \coloneqq \MultilinEncode{Y,Z}{U_2}$.
This yields a multilinear interpretation $\omega\colon \Wres \to \F[X,Y,Z]_{\le 1}$
which is called the polynomial refutation \emph{induced by} $\Phi$.

\begin{lemma}
    \label{lem:induced-refutation-is-valid}
    Let $\varphi$ denote a CNF-formula. Let $\Phi$ denote a RES-refutation of $\varphi$. Then the polynomial refutation induced by $\Phi$ fulfills \ref{valid:binary}\,--\,\ref{valid:resolution} of \autoref{def:poly-ref}.
\end{lemma}
\begin{proof}
	\ref{valid:binary}\,--\,\ref{valid:regressive} follow immediately from \autoref{def:cnfres} and the choice of the $U_i$.
    Consider $i\in\{m,...,M-1\}$ and write $C_i = \{\ell_0,\ldots,\ell_k\}$.
    Since $(U_1(i), U_2(i))$ is an antecedent list of $C_i$,
    the formula $\psi \coloneqq C_{U_1(i)} \land C_{U_2(i)} \land \lnot \ell_0 \land \ldots \land \lnot \ell_k$
    is unsatisfiable.
    Thus $\Arith{C_{U_1(i)}}\cdot\Arith{C_{U_2(i)}}\cdot(1-\Arith{C_i})$, which is an arithmetization of $\psi$,
    must evaluate to $0$ under all binary assignments, and \ref{valid:resolution} follows.
\end{proof}

Now we can prove:
\polyResComplete*

\begin{proof}
	The “if” direction follows from \autoref{lem:induced-refutation-is-valid} and the well-known fact that a RES-refutation exists for all unsatisfiable CNF-formulas.
For the other direction, let $(p,u_1, u_2)$ denote a polynomial refutation for $\varphi=(C_0,\ldots,C_{m-1})$. Fix $M$ as in \ref{valid-derives-empty}.
    We construct a list of formulas $\Phi=(f_0,\ldots,f_{M-1})$. Conceptually, $\Phi$ is similar to a RES-refutation of $\varphi$, however the $f_i$ are not necessarily clauses.

    Let $i\in\{0,\ldots,M-1\}$. By \ref{valid:binary}, we know that $p_i\coloneqq \Pev{\IndexA{Y}{i}}\,p$ is a binary (and multilinear) polynomial. So we can choose a formula $f_i$
    such that $p_i$ is the multilinear extension of $f_i$
    (viewed as a function $f_i\colon \BooleanDomain^X \to \F$) to $\F^X \to \F$.

	The goal is to prove that $\varphi$ is unsatisfiable. We split this into two steps: we first show that $f_{M-1}$ is unsatisfiable, and then that $\varphi$ implies $f_i$ for each $i$.
	
	The first part follows immediately from \ref{valid-derives-empty}. It remains to show the second. We do this via induction over $i$. For the base case, let $i\in\{0,...,m-1\}$. By \ref{valid:prefix}, we have that $f_i$ is equivalent to $C_i$ and thus implied by $\varphi$.
	
	For the induction step, let $i\in\{m,...,M-1\}$ and assume that $f_0,...,f_{i-1}$ are implied by $\varphi$.
    Towards a contradiction, assume that $f_i$ is not implied by $\varphi$. Then, there is some assignment $\sigma:X\rightarrow\{0,1\}$ which satisfies $\varphi$ but does not satisfy $f_i$. As $f_0,...,f_{i-1}$ are implied by $\varphi$, we get $\Pev{\sigma}p_j=1$ for $j<i$ and $\Pev{\sigma}p_i=0$. Using $U_1(i), U_2(i)<i$ (via \ref{valid:regressive}) we obtain
	\[\Pev{\sigma}\big(\Pev{\IndexA{X}{U_1(i)}}\,p \cdot\Pev{\IndexA{X}{U_2(i)}}\,p\cdot(1-\Pev{\IndexA{X}{i}}\,p)\big)
	=1\cdot1\cdot(1-0)=1\]
	contradicting \ref{valid:resolution}.
\end{proof}

\subsection{Combining multiple EAC gadgets}

\begin{lemma}\label{lem:gadget:combine-eacs}
        Let $\psi_0,\ldots,\psi_{k-1}$ denote EACs over variables $X$ and function variables $W$.
        Let $Y$ denote a disjoint set of variables with $\Abs{Y}\ge\log_2k$.
        Then $\psi \coloneqq \IndexPoly{Y}{0}\psi_0+\ldots+\IndexPoly{Y}{k-1}\psi_{k-1}$ is satisfiable iff there is a multilinear interpretation $\omega$
        that satisfies $\psi_0,\ldots,\psi_{k-1}$ simultaneously,
        i.e., $\Eval{\psi_0}_\omega=\ldots=\Eval{\psi_{k-1}}_\omega=0$.
\end{lemma}
\begin{proof}
    Since $\Eval{\psi}_{\omega}$ is multilinear in $Y$,
    we have that  $\Eval{\psi}_{\omega} = 0$
    iff for all  $i$,
    $0 = \Eval{\Pev{Y \gets i} \psi}_{\omega} = \Eval{\psi_i}_{\omega}$.
\end{proof}

\subsection{Correctness of Gadget \texorpdfstring{$\GadgetBinary$}{binary} for \ref{valid:binary}}

\begin{lemma}
	\label{lem:gadget:binary}
	Let $f\in W$ be a function variable.
	A multilinear interpretation $\omega\colon W \to \F^X$ satisfies $\GadgetBinary(f)\coloneqq \DegreeReduction{X}f(1-f)$
	iff $\omega(f)$ is binary.
\end{lemma}
\begin{proof}
	Assume that $\omega(f)$ is binary and let $\sigma:X\rightarrow\BooleanDomain$ denote a binary assignment.
	Then, $\Pev{\sigma} f \in \BooleanDomain$
	and thus $\Pev{\sigma}\Eval{f(1-f)}_\omega=0$.
	Since multilinear functions are determined by their values on $\BooleanDomain^X$,
	it follows that $\Eval{\DegreeReduction{X} f(1-f)}_\omega=0$.
	
	For the other direction, assume that $\omega(f)$ is not binary,
	and let $\sigma:X\rightarrow\BooleanDomain$ denote an assignment with $k \coloneqq \Pev{\sigma}\omega(f) \notin\BooleanDomain$.
	Then $\Pev{\sigma}\Eval{f(1-f)}_\omega=k(1-k)\ne0$ and thus $\Eval{\DegreeReduction{X}f(1-f)}_\omega\ne0$.
\end{proof}

Note that, by definition of EACs,
function variables are interpreted by functions depending on \emph{all variables} of the EAC. However, it is easy to restrict a function variable $f\in W$ to some subset $X'\subseteq X$:
Choose a partial assignment $\sigma^{X \setminus X'}_{0}\colon X\setminus X'\rightarrow\{0\}$ and consider $\Pev{\sigma}f$ instead of $f$.
Combining $\GadgetBinary$ with the above trick, we can now encode \ref{valid:binary}.
To improve readability, in the following we will not write this down explicitly,
i.e., write $p$ instead of $\Pev{\sigma^{Z}_0} p$ and $u_i$ instead of $\Pev{\sigma^{X}_0} u_i$.

\subsection{Correctness of Gadget \texorpdfstring{$\GadgetInitial$}{initial} for \ref{valid:prefix}}
It is fine to encode \ref{valid:prefix} in the obvious manner,
i.e., by using the EAC $(\Pev{\IndexA{Y}{i}}p)-\Arith{C_i}$ for each $i=0,...,m-1$.
However, we construct an alternative gadget which is more efficient.
For this purpose, we construct an arithmetic circuit $[Y < m]$ that checks whether the binary number $y_k...y_0$, where $\{y_0,...,y_k\}=Y$, is less than $m$,
i.e.,
$\Pev{\IndexA{Y}{i}} [Y < m] = 1$ iff $i < m$
and
$\Pev{\IndexA{Y}{i}} [Y < m] = 0$ otherwise.
Moreover, we want the polynomial described by the circuit to be multilinear,
i.e., \emph{as functions} $[Y < m] = \sum_{i < m} \IndexPoly{Y}{i}$.
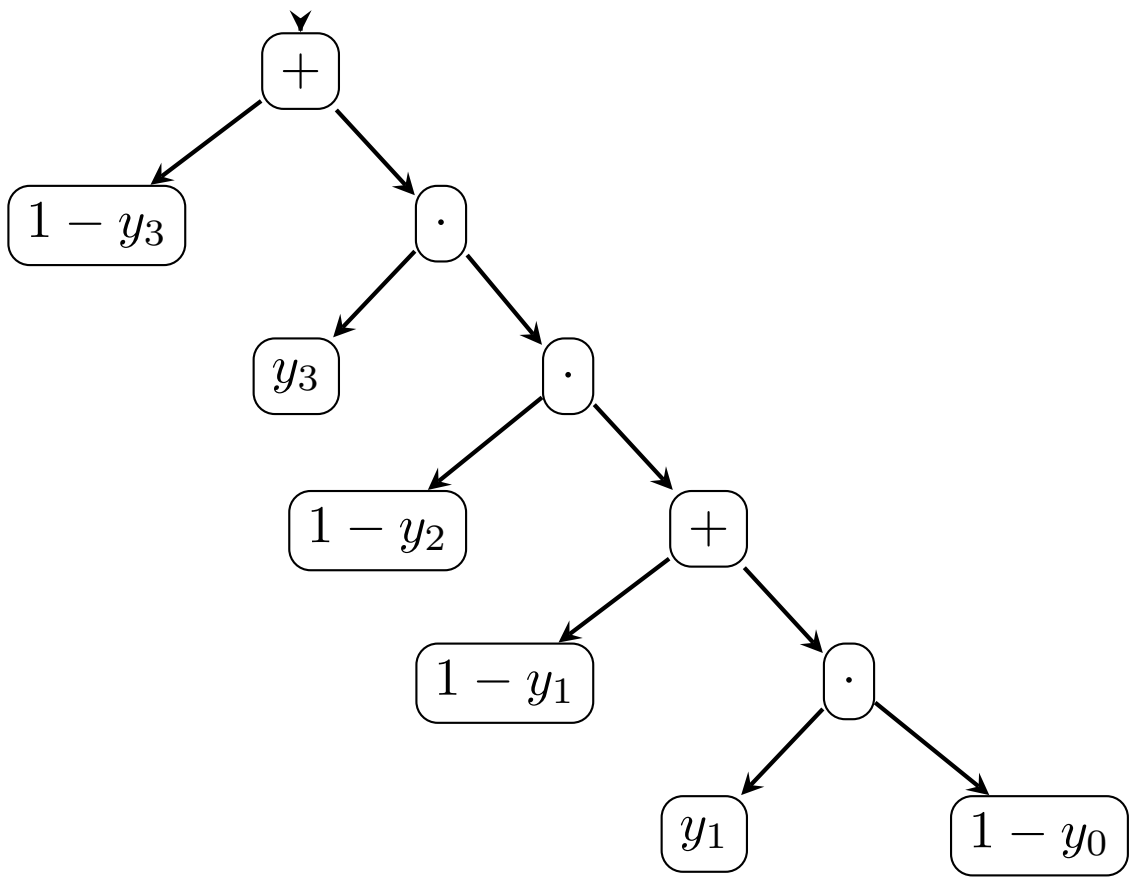
\begin{figure}
	\centering
	\scalebox{1}{
	\begin{tikzpicture}[node distance=0.7cm]
		\node (root) at (0,0) [eacroot] { $+$ };
        \node (top) [above=0.01cm of root] {};
		\node (leaf4neg) [eacleaf, below left=of root] { $1 - y_3$ };
		\node (mult4) [eacnode, below right=of root] { $\cdot$ };
		\node (leaf4pos) [eacleaf, below left=of mult4] { $y_3$ };
		\node (mult3) [eacnode, below right=of mult4] { $\cdot$ };
		\node (leaf3neg) [eacleaf, below left=of mult3] { $1 - y_2$ };
		\node (plus2) [eacnode, below right=of mult3] { $+$ };
		\node (leaf2neg) [eacleaf, below left=of plus2] { $1 - y_1$ };
		\node (mult2) [eacnode, below right=of plus2] { $\cdot$ };
		\node (leaf2pos) [eacleaf, below left=of mult2] { $y_1$ };
		\node (leaf1neg) [eacleaf, below right=of mult2] { $1 - y_0$ };
		\draw [eacedge] (top) -- (root);
		\draw [eacedge] (root) -- (leaf4neg);
		\draw [eacedge] (root) -- (mult4);
		\draw [eacedge] (mult4) -- (leaf4pos);
		\draw [eacedge] (mult4) -- (mult3);
		\draw [eacedge] (mult3) -- (leaf3neg);
		\draw [eacedge] (mult3) -- (plus2);
		\draw [eacedge] (plus2) -- (leaf2neg);
		\draw [eacedge] (plus2) -- (mult2);
		\draw [eacedge] (mult2) -- (leaf2pos);
		\draw [eacedge] (mult2) -- (leaf1neg);
	\end{tikzpicture}
    }
    \caption{The arithmetic circuit $[Y < m]$ for $Y = \{y_3, y_2, y_1, y_0\}$ and $m = (1011)_2$}
	\label{fig:Y-leq-m}
\end{figure}

\begin{lemma}
    For $0 \le m < 2^{|Y|}$,
    an arithmetic circuit $[Y < m]$ as described above can be constructed with size $\O(|Y|)$.
\end{lemma}
\begin{proof}
	Let $Y' \coloneqq Y \setminus \{y_{|Y|-1}\}$
	and consider the $|Y|$-th, i.e., most significant, bit of $m$.
	If this bit is $0$,
	then
    \[
        [Y < m]
	= (1 - y_{|Y| -1}) \sum_{i < m} \IndexPoly{Y'}{i}
    = (1 - y_{|Y| -1}) [Y' < m].
	\]
    	On the other hand, if the most significant bit of $m$ is $1$, we have
	\[
        [Y < m]
	= \sum_{\substack{i < m \\ i < 2^{|Y| - 1}}} \IndexPoly{Y}{i}
	+ \sum_{\substack{i < m \\ i \ge 2^{|Y| - 1}}} \IndexPoly{Y}{i}
= (1 - y_{|Y|-1}) + y_{|Y| -1} \cdot [Y' < m'],
	\]
	where $m' \coloneqq m - 2^{|Y|-1}$.
	We conclude by induction.
    An example of this construction is shown in \autoref{fig:Y-leq-m}.
\end{proof}
\begin{lemma}
	\label{lem:gadget:initial}
	Let $\GadgetInitial(p) \coloneqq \Arith{\varphi}_Y-\DegreeReduction{Y}([Y < m]\cdot p)$.
	Then, $\GadgetInitial(p)$ is satisfied by $\omega$ iff $\omega(p)$ fulfills \ref{valid:prefix}.
\end{lemma}
\begin{proof}
	Note that, for all $i$, we have
	\begin{align*}
		\Pev{\IndexA{Y}{i}}\,\Eval{\GadgetInitial(p)}_\omega
		& =\Pev{\IndexA{Y}{i}}\,\EvalBig{\Arith{\varphi}_Y-\DegreeReduction{Y}([Y < m]\cdot p)}_\omega          \\
		& =\Pev{\IndexA{Y}{i}}\,\Arith{\varphi}_Y-\Pev{\IndexA{Y}{i}}\,\Eval{[Y < m]\cdot p}_\omega \\\qquad
		& = \begin{cases}
			\Arith{C_i}-\Pev{\IndexA{Y}{i}}\,\omega(p) & \text{ if } i < m, \\
			0                                         & \text{ otherwise}.
		\end{cases}
	\end{align*}
	\end{proof}

\subsection{Correctness of Gadget \texorpdfstring{$\GadgetFunction$}{function} for \ref{valid-clause-mapping}}

Similarly to $[Y < m]$ above,
let $[m \le Y < M]$ denote a circuit that checks whether the binary number encoded by $Y$ is at least $m$ and less than $M$.

\begin{lemma}
    \label{lem:gadget:function}
	Consider the constant partial assignment $\sigma^{Z}_{2^{-1}}\colon Z\rightarrow\{2^{-1}\}$,
	and let
    \[\GadgetFunction(u_i)\coloneqq (2^{\Abs{Z}}\Pev{\sigma^{Z}_{2^{-1}}}u_i)-[m\le Y < M]\]
    for $i\in\{1,2\}$.
    Assume that $\omega(u_i)$ is binary.
	Then, $\GadgetFunction(u_i)$ is satisfied by $\omega$ iff $\omega(u_i)$ fulfils \ref{valid-clause-mapping}.
\end{lemma}
\begin{proof}
	Let $\omega$ denote an interpretation
    such that $\omega(u_i) \in \Multilinear{Y,Z}$ is multilinear and binary.
	We write $\omega(u_i)=\sum_{j=0}^{2^{\Abs{Y}}-1}\IndexPoly{Y}{j}q_j$,
	where the $q_j \in \Multilinear{Z}$ are multilinear binary polynomials over $Z$.
	Note that $\omega(u_i)$ encodes a function $f\colon \{m,\ldots,M-1\} \to \{0,\ldots,M-1\}$
	iff $q_j = \IndexPoly{Z}{f(j)}$ for $j \in \{m,\ldots,M-1\}$ and $q_j =0$ for all other $j$.
	Let $k_j$ denote the number of satisfying assignments of $q_j$.
	It suffices to ensure that $k_j = 1$ for all $j \in \{m,\ldots,M-1\}$ and $k_j=0$ otherwise.
	By \autoref{fact:count-satisfying-assignments},
    we have $k_j=2^{\Abs{Z}}\Pev{\sigma^{Z}_{2^{-1}}}q_j$,
	i.e.,
	\begin{align*}
	\Pev{\IndexA{Y}{j}} \EvalBig{\GadgetFunction(u_i)}_{\omega} 
	&= (2^{\Abs{Z}} \Pev{\sigma^{Z}_{2^{-1}}}q_j) -\Pev{\IndexA{Y}{j}} [m \le Y < M]\\
	&=
	\begin{cases}
		k_j - 1 & \text{ if } j \in \{m,\ldots,M-1\}, \\
		k_j     & \text{ otherwise.}
	\end{cases}
	\end{align*}
		
	\end{proof}

\subsection{Correctness of Gadget \texorpdfstring{$\GadgetResolution$}{resolution} for \ref{valid:resolution}}

The gadget for \ref{valid:resolution} is the most involved construction.
The general idea is that we can use $u_i$ to rename the clauses of $p$,
i.e., to store clause $C_{U_i(j)}$ at index $j$.
To do this, we first rename the variables of $p$, replacing $Y$ with $Z$.
We use $\overwriteVars{Y}{Z}$ as a shorthand for $\pev{y_0 \coloneqq z_0} \ldots \pev{y_{|Y|-1} \coloneqq z_{|Y|-1}}$ (recall that $|Y| = |Z|$),\NewNotation{notation:overwriteVars}
    i.e., for $p = \sum_i \IndexPoly{Y}{i} p_i$, $p_i \in \F[X]$,
    we have $\overwriteVars{Y}{Z} p = \sum_i \IndexPoly{Z}{i} p_i$.

Essentially, $u_i$ is a $\F^Y$-$\F^Z$-relation, and $\overwriteVars{Y}{Z}\,p$ is a $\F^Z$-$\F^X$-relation.
Similarly to \autoref{example:join},
we want to compute the join of these two relations.
To eliminate the variables in $Z$, we use the constant assignment $\sigma^{Z}_{2^{-1}}\colon Z\rightarrow\{2^{-1}\}$ from \autoref{lem:gadget:function}.
We obtain the EAC $\psi_i\coloneqq 2^{\Abs{Z}}\Pev{\sigma^{Z}_{2^{-1}}}\DegreeReduction{Z}(u_i\cdot\overwriteVars{Y}{Z}\,p)$ which encodes the CNF-formula $(C_{U_i(m)},...,C_{U_i(M-1)})$.
We need to invert the clauses in $p$, which is simply $1-p$. Then, we take the componentwise product,
$\psi\coloneqq \DegreeReduction{Y}\big(\psi_1\cdot\psi_2\cdot(1-p)\big)$.
Note that $\GadgetResolution(p,u_1,u_2) \coloneqq \DegreeReduction{X}\psi$ is satisfied by $\omega$ iff $(\Eval{\psi}_\omega) \restrict{\BooleanDomain^X} = 0$.
We obtain\ the following:
\begin{lemma}
	\label{lem:gadget:resolution}
    Let $\omega$ be a multilinear interpretation of $\Wres$
	satisfying \ref{valid:binary} and \ref{valid-clause-mapping}.

	Then, $\omega$ satisfies $\GadgetResolution(p,u_1,u_2)$
	iff \ref{valid:resolution} holds for $\omega$.
\end{lemma}
\begin{proof}
    By \ref{valid-clause-mapping},
    $\omega(u_1)$ and $\omega(u_2)$
    encode functions
	$U_1,U_2$,
    i.e., $u_i = \sum_j \IndexPoly{Y}{j} \IndexPoly{Z}{U_i(j)}$.
    Let $p_j \coloneqq \Pev{\IndexA{Y}{j}} \omega(p)$.
    By the same argument as in \autoref{example:join}, we have
	\[
        \Eval{\psi_i}_{\omega} = \sum_{j=m}^{M-1} \IndexPoly{Y}{j} p_{U_i(j)}
	\]
	and thus
	\begin{align*}
		& \Eval{\DegreeReduction{Y} \left( \psi_1 \cdot \psi_2 \cdot (1-p) \right)}_{\omega}                                                                \\
		& = \DegreeReduction{Y} \left( \left(\sum_{i=m}^{M-1} \IndexPoly{Y}{i} p_{U_1(i)}\right)
		\cdot \left(\sum_{i=m}^{M-1} \IndexPoly{Y}{i} p_{U_2(i)}\right)
		\cdot \left( \left( \sum_{i} \IndexPoly{Y}{i} \right) - \left(\sum_{i=m}^{M-1} \IndexPoly{Y}{i} p_i\right) \right)
		\right)                                                                                                                                              \\
		& = \sum_{i=m}^{M-1} \IndexPoly{Y}{i}p_{U_1(i)} \cdot  p_{U_2(i)}\cdot ( 1 - p_i ).
	\end{align*}
    So \ref{valid:resolution} is equivalent to $\Eval{\psi}_{\omega} \restrict{\BooleanDomain^X \times \F^{Y \cup Z}} = 0$ for all $i\in\{m,...,M-1\}$,
    i.e., equivalent to $\DegreeReduction{X}\psi =0$.
\end{proof}
 \section{Prover Strategies}
\label{sec:appendix:prover-strategies}

In this section, we describe how a prover for $\ResCertify$ can answer the verifier's queries efficiently
in the remaining cases not covered in \autoref{sec:efficient-prover}.

 \subsection{Degree Reductions in Y}

\subsubsection{Proof of Lemma \ref{lem:y-index:product:efficient}}
\label{sec:proof:lem:y-index:product:efficient}

In the following, for integers $i$, $j$, we use $i \preceq j$\NewNotation{notation:bit_prefix} to denote
that the binary representation of $i$ (most significant bit first) is a prefix of the binary representation of $j$,
e.g.~$2 = (10)_2 \preceq (10111)_2 = 23$.
Moreover, we denote their concatenation by $i \concat j$.\NewNotation{notation:concat}

\def\cP{\mathcal{P}}
\def\cQ{\mathcal{Q}}
\def\cT{\mathcal{T}}

\degYLinear*

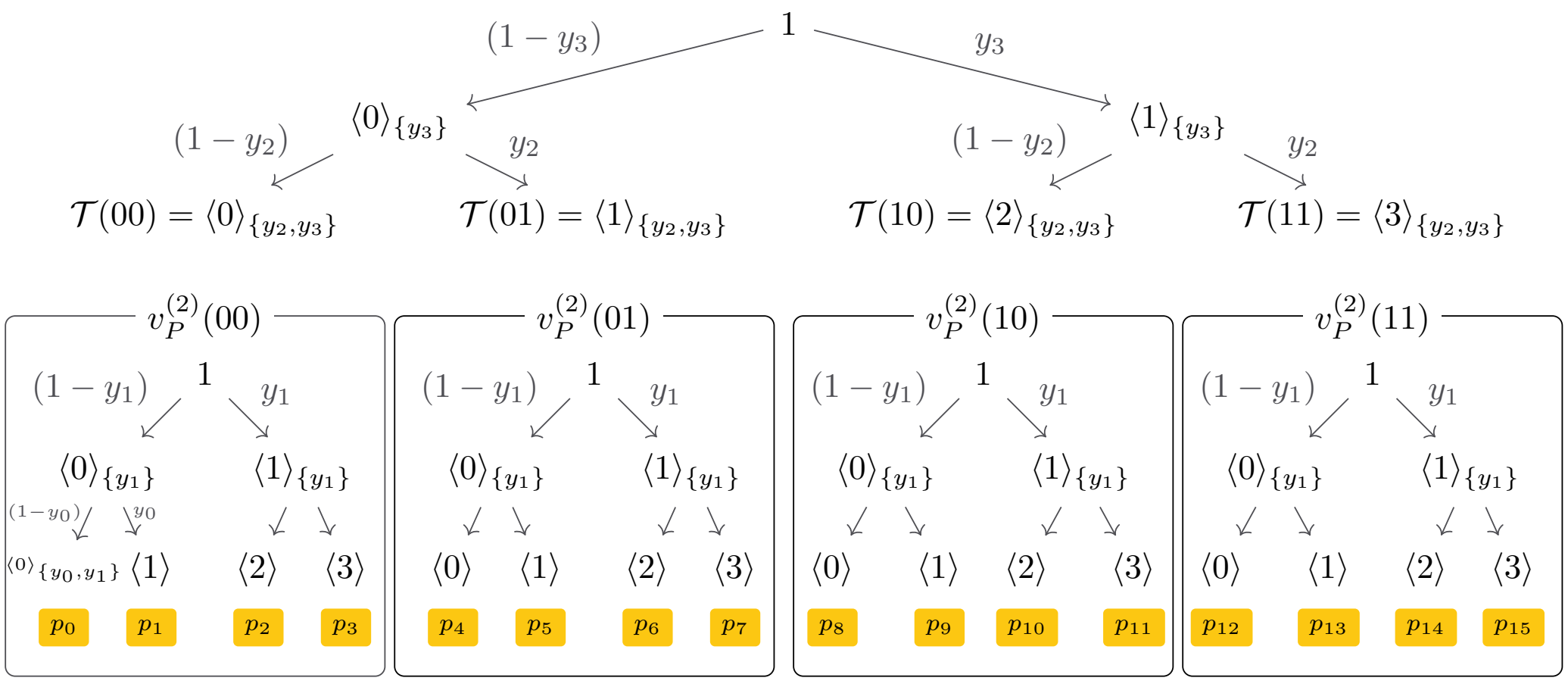
\begin{figure}[t]
	\centering
	\scalebox{0.85}{
        \begin{tikzpicture}
            \def\rootHeight{0}
            \def\layerAHeight{-1}
            \def\layerBHeight{-2}
            \def\layerBbHeight{-3}
            \def\layerBcHeight{-3.6}
            \def\layerCHeight{-4.6}
            \def\layerDHeight{-5.6}
            \def\leafHeight{\layerDHeight-0.6}
            \node[yindexnode] (r) at (0,\rootHeight)        {$1$};
            \node[yindexnode] (r0) at (-4,\layerAHeight)    {$\IndexPoly{\{y_3\}}{0}$};
            \node[yindexnode] (r1) at (4,\layerAHeight)     {$\IndexPoly{\{y_3\}}{1}$};

            \node[yindexnode] (r00) at (-6,\layerBHeight)    {$\cT(00) = \IndexPoly{\{y_2,y_3\}}{0}$};
			\node[yindexnode] (r01) at (-2,\layerBHeight)    {$\cT(01) = \IndexPoly{\{y_2,y_3\}}{1}$};
			\node[yindexnode] (r10) at (2,\layerBHeight)     {$\cT(10) = \IndexPoly{\{y_2,y_3\}}{2}$};
			\node[yindexnode] (r11) at (6,\layerBHeight)     {$\cT(11) = \IndexPoly{\{y_2,y_3\}}{3}$};

            \node[yindexnode] (r00b) at (-6,\layerBcHeight) {$1$}; \node[yindexnode] (r01b) at (-2,\layerBcHeight) {$1$};  \node[yindexnode] (r10b) at (2,\layerBcHeight)  {$1$};  \node[yindexnode] (r11b) at (6,\layerBcHeight)  {$1$};  

            \node[yindexnode] (r000) at (-7,\layerCHeight)   {$\IndexPoly{\{y_1\}}{0}$};
            \node[yindexnode] (r001) at (-5,\layerCHeight)   {$\IndexPoly{\{y_1\}}{1}$};
            \node[yindexnode] (r010) at (-3,\layerCHeight)   {$\IndexPoly{\{y_1\}}{0}$};
            \node[yindexnode] (r011) at (-1,\layerCHeight)   {$\IndexPoly{\{y_1\}}{1}$};
            \node[yindexnode] (r100) at (1,\layerCHeight)    {$\IndexPoly{\{y_1\}}{0}$};
            \node[yindexnode] (r101) at (3,\layerCHeight)    {$\IndexPoly{\{y_1\}}{1}$};
            \node[yindexnode] (r110) at (5,\layerCHeight)    {$\IndexPoly{\{y_1\}}{0}$};
            \node[yindexnode] (r111) at (7,\layerCHeight)    {$\IndexPoly{\{y_1\}}{1}$};

            \node[yindexnode] (r0000) at (-7.45,\layerDHeight) {$\scriptscriptstyle\IndexPoly{ \{y_0,y_1\} }{0}$};
            \node[yindexnode] (r0001) at (-6.55,\layerDHeight) {$\IndexPoly{}{1}$};
            \node[yindexnode] (r0010) at (-5.45,\layerDHeight) {$\IndexPoly{}{2}$};
            \node[yindexnode] (r0011) at (-4.55,\layerDHeight) {$\IndexPoly{}{3}$};
            \node[yindexnode] (r0100) at (-3.45,\layerDHeight) {$\IndexPoly{}{0}$};
            \node[yindexnode] (r0101) at (-2.55,\layerDHeight) {$\IndexPoly{}{1}$};
            \node[yindexnode] (r0110) at (-1.45,\layerDHeight) {$\IndexPoly{}{2}$};
            \node[yindexnode] (r0111) at (-0.55,\layerDHeight) {$\IndexPoly{}{3}$};
            \node[yindexnode] (r1000) at (0.45,\layerDHeight)  {$\IndexPoly{}{0}$};
            \node[yindexnode] (r1001) at (1.55,\layerDHeight)  {$\IndexPoly{}{1}$};
            \node[yindexnode] (r1010) at (2.45,\layerDHeight)  {$\IndexPoly{}{2}$};
            \node[yindexnode] (r1011) at (3.55,\layerDHeight)  {$\IndexPoly{}{3}$};
            \node[yindexnode] (r1100) at (4.45,\layerDHeight)  {$\IndexPoly{}{0}$};
            \node[yindexnode] (r1101) at (5.55,\layerDHeight)  {$\IndexPoly{}{1}$};
            \node[yindexnode] (r1110) at (6.55,\layerDHeight)  {$\IndexPoly{}{2}$};
            \node[yindexnode] (r1111) at (7.45,\layerDHeight)  {$\IndexPoly{}{3}$};

            \node[yindexleaf] (r0000l) at (-7.45,\leafHeight)    {$\scriptstyle p_{0}$};
            \node[yindexleaf] (r0001l) at (-6.55,\leafHeight)    {$\scriptstyle p_{1}$};
            \node[yindexleaf] (r0010l) at (-5.45,\leafHeight)    {$\scriptstyle p_{2}$};
            \node[yindexleaf] (r0011l) at (-4.55,\leafHeight)    {$\scriptstyle p_{3}$};
            \node[yindexleaf] (r0100l) at (-3.45,\leafHeight)    {$\scriptstyle p_{4}$};
            \node[yindexleaf] (r0101l) at (-2.55,\leafHeight)    {$\scriptstyle p_{5}$};
            \node[yindexleaf] (r0110l) at (-1.45,\leafHeight)    {$\scriptstyle p_{6}$};
            \node[yindexleaf] (r0111l) at (-0.55,\leafHeight)    {$\scriptstyle p_{7}$};
            \node[yindexleaf] (r1000l) at (0.45,\leafHeight)     {$\scriptstyle p_{8}$};
            \node[yindexleaf] (r1001l) at (1.55,\leafHeight)     {$\scriptstyle p_{9}$};
            \node[yindexleaf] (r1010l) at (2.45,\leafHeight)     {$\scriptstyle p_{10}$};
            \node[yindexleaf] (r1011l) at (3.55,\leafHeight)     {$\scriptstyle p_{11}$};
            \node[yindexleaf] (r1100l) at (4.45,\leafHeight)     {$\scriptstyle p_{12}$};
            \node[yindexleaf] (r1101l) at (5.55,\leafHeight)     {$\scriptstyle p_{13}$};
            \node[yindexleaf] (r1110l) at (6.55,\leafHeight)     {$\scriptstyle p_{14}$};
            \node[yindexleaf] (r1111l) at (7.45,\leafHeight)     {$\scriptstyle p_{15}$};

			\draw[yindexedge] (r) edge node [above left] {$(1-y_3)$} (r0);
			\draw[yindexedge] (r) edge node [above right] {$y_3$} (r1);

			\draw[yindexedge] (r0) edge node [above left] {$(1-y_2)$} (r00);
			\draw[yindexedge] (r0) edge node [above right] {$y_2$} (r01);
			\draw[yindexedge] (r1) edge node [above left] {$(1-y_2)$} (r10);
			\draw[yindexedge] (r1) edge node [above right] {$y_2$} (r11);

			\draw[yindexedge] (r00b) edge node [above left] {$(1-y_1)$} (r000);
			\draw[yindexedge] (r00b) edge node [above right] {$y_1$} (r001);
			\draw[yindexedge] (r01b) edge node [above left] {$(1-y_1)$} (r010);
			\draw[yindexedge] (r01b) edge node [above right] {$y_1$} (r011);
			\draw[yindexedge] (r10b) edge node [above left] {$(1-y_1)$} (r100);
			\draw[yindexedge] (r10b) edge node [above right] {$y_1$} (r101);
			\draw[yindexedge] (r11b) edge node [above left] {$(1-y_1)$} (r110);
			\draw[yindexedge] (r11b) edge node [above right] {$y_1$} (r111);

			\draw[yindexedge] (r000) edge node [above left=-1mm and -1mm]     {$\scriptscriptstyle(1-y_0)$} (r0000);
			\draw[yindexedge] (r000) edge node [above right=-1mm and -1mm]    {$\scriptscriptstyle y_0$}     (r0001);
            \draw[yindexedge] (r001) edge (r0010);
			\draw[yindexedge] (r001) edge (r0011);
        	\draw[yindexedge] (r010) edge (r0100);
			\draw[yindexedge] (r010) edge (r0101);
            \draw[yindexedge] (r011) edge (r0110);
			\draw[yindexedge] (r011) edge (r0111);
    		\draw[yindexedge] (r100) edge (r1000);
			\draw[yindexedge] (r100) edge (r1001);
            \draw[yindexedge] (r101) edge (r1010);
			\draw[yindexedge] (r101) edge (r1011);
        	\draw[yindexedge] (r110) edge (r1100);
			\draw[yindexedge] (r110) edge (r1101);
            \draw[yindexedge] (r111) edge (r1110);
			\draw[yindexedge] (r111) edge (r1111);

            \draw[draw=lipicsGray, rounded corners=1mm] (-8.05,\layerBbHeight) rectangle ++(3.9, \leafHeight-\layerBbHeight-0.5);
            \draw[draw=black, rounded corners=1mm] (-4.05,\layerBbHeight) rectangle ++(3.9, \leafHeight-\layerBbHeight-0.5);
            \draw[draw=black, rounded corners=1mm] (0.05,\layerBbHeight) rectangle ++(3.9, \leafHeight-\layerBbHeight-0.5);
            \draw[draw=black, rounded corners=1mm] (4.05,\layerBbHeight) rectangle ++(3.9, \leafHeight-\layerBbHeight-0.5);

            \node[yindexnode, rectangle, fill=white] (r00b) at (-6,\layerBbHeight)    {$v_P^{(2)}(00)$};
			\node[yindexnode, rectangle, fill=white] (r01b) at (-2,\layerBbHeight)    {$v_P^{(2)}(01)$};
			\node[yindexnode, rectangle, fill=white] (r10b) at (2,\layerBbHeight)     {$v_P^{(2)}(10)$};
			\node[yindexnode, rectangle, fill=white] (r11b) at (6,\layerBbHeight)     {$v_P^{(2)}(11)$};
	\end{tikzpicture}}
    \caption{
        The data structures in the proof of \autoref{lem:y-index:product:efficient} for $Y=\{y_0,y_1,y_2,y_3\}$, a $Y$-index $P = \sum_{i=0}^{15} \IndexPoly{Y}{i}p_i$ and $k = 2$.
        The $\cT(c)$, $c \in \BooleanDomain^2$, store the evaluations of the factors of the key polynomials corresponding to the $k$ most significant bits.
        The $v_P^{(2)}(c)$ store the evaluations of the subtrees.
    }
	\label{fig:lem:y-index:product:efficient}
\end{figure}

\begin{proof}
 	In the $k$-th step, let $v_P^{(k)} \colon \BooleanDomain^{|Y|-k} \to \F$ be an array such that
    \[
    v_P^{(k)}(c) = \sum_{c \preceq j} \Pev{\sigma_{k}} \IndexPoly{\{y_0, \ldots, y_{k-1}\}}{j \bmod{2^k}} p_j.
    \]
	For $k = 0$, we have $v_P^{(0)}(c) = \Pev{\sigma_0} p_c$ for all $c \in \BooleanDomain^{{|Y|}}$.
	For $k > 0$, we compute $v_P^{(k)}$ from $v_{P}^{(k-1)}$:
    \begin{displaymathifenoughspace}
        v_P^{(k)}(c) = \sigma_{k}(y_{k-1}) \cdot v_P^{(k-1)}(c \concat 1) + (1 - \sigma_{k}(y_{k-1})) \cdot v_P^{(k-1)}(c \concat 0).
    \end{displaymathifenoughspace}
We define $v_Q$ analogously for $Q$.
	Let  $\cT\colon \BooleanDomain^{\le {|Y|}} \to \F$ be a tree
	such that $\cT(\emptyset) = 1$
	and $\cT(c) = \Pev{\sigma_0} \IndexPoly{\{y_{|Y|-|c|}, \ldots, y_{|Y|-1}\}}{c}$.
	Then
    \begin{displaymathifenoughspace}
		\Pev{\sigma_k} \DegreeReduction{\{y_k, \ldots, y_{|Y|-1}\}} (P \cdot Q)
		= \sum_{c \in \BooleanDomain^{|Y|-k}} \cT(c) \cdot v_P^{(k)}(c) \cdot v_Q^{(k)}(c).
    \end{displaymathifenoughspace}
	\autoref{fig:lem:y-index:product:efficient} illustrates this for a small example.
Computing all $v_P^{(0)}$ and $v_Q^{(0)}$
    takes time $\O(|P| + |Q|)$ and computing $\cT$ takes time $\O(2^{|Y|})$.
	For each $k > 0$,
    it takes time $\O(2^{|Y| - k})$ to compute $v_{P}^{(k)}$ from $v_{P}^{(k-1)}$
    and likewise for $Q$.
    Thus we obtain a total runtime
    of  $\O(|P| + |Q| + 2^{|Y|} + \sum_k 2^{|Y|-k}) = \O(|P| + |Q| + 2^{|Y|})$.
\end{proof}

\subsection{Degree Reductions in X}\label{sec:degX}

\subsubsection{Efficient repeated application of \autoref{lem:prover:evaluate-binary-p}}
\label{sec:prover:evaluate-binary-p:optimized}
\begin{lemma} \label{lem:prover:evaluate-binary-p:optimized}
	Let $P$ be as in \autoref{lem:prover:evaluate-binary-p}.
    Let $\sigma_0, \ldots, \sigma_n$ be assignments such that
    $\sigma_k$ and  $\sigma_{k+1}$ differ only in $x_{k}$.
	Then,
    the sequence of evaluations
    \[
        \Pev{\sigma_0} \DegreeReduction{\{x_0,\ldots,x_{n-1}\}} P,
        \Pev{\sigma_1} \DegreeReduction{\{x_1,\ldots,x_{n-1}\}} P,
        \ldots,
        \Pev{\sigma_n} P
    \]
    can be computed online in total time $\O(|P|)$.
\end{lemma}
\begin{proof}
    Using the notation of \autoref{lem:prover:evaluate-binary-p},
    the goal is to compute
    \[\Pev{\sigma_k}\DegreeReduction{\{x_{k},\ldots,x_{n-1}\}} P = \sum_i \Pev{\sigma_k} \IndexPoly{Y}{i} q_i^{\{x_{k},\ldots,x_{n-1}\} }\]
    for all $k$.
    To this end, we maintain an array
    $\overline{v}$ such that in the \emph{$k$-th step}
    \[
        v_i = \Pev{\sigma_k} \IndexPoly{Y}{i} q_i^{\{x_{k},\ldots,x_{n-1}\} } \in \F.
    \]
    Note that in the beginning all entries are $0$.

    During the initialization,
    we build a set
    $U_x \coloneqq \{i : \text{$x$ occurs in  $p_i$}\}$
    for every variable $x \in X$.
    Note that if $i \not\in U_{x_{k}}$,
    we have
    $\Pev{\sigma_{k}} \IndexPoly{Y}{i} q_i^{\{x_{k+1},\ldots,x_{n-1}\} } = \Pev{\sigma_{k-1}} \IndexPoly{Y}{i} q_i^{\{x_{k},\ldots,x_{n-1}\} }$,
    i.e., $U_{x_k}$ contains exactly the indices where
    $\overline{v}$ needs to be updated when proceeding to
    the $k$-th step.
    For each $i \in U_{x_k}$, we precompute the value
    $w_k^i \coloneqq \Pev{\sigma_0} \prod_{\ell \in L_i \setminus \ArithLit{\{x_{k+1},\ldots,x_{n-1}\} }} \ell$.
Moreover, we compute the values
    $\Pev{\sigma_0}\IndexPoly{Y}{i}$ for all $i$.
    In addition to $\overline{v}$,
    we maintain an array $\overline{u}$
    such that in the $k$-th step
    \[u_i = \Pev{\sigma_k} \prod_{\ell \in L_i \cap \ArithLit{\{x_0,\ldots,x_k\}}} \ell.\]When proceeding to the $k$-th step,
    for every $i \in U_{x_k}$,
    we update $u_i \gets u_i \cdot \sigma_{k}(x_{k})$
    (if $x_{k} \in L_i$)
    or $u_i \gets u_i \cdot (1-\sigma_{k}(x_{k}))$
    (if $(1-x_{k}) \in L_i$).
    Having $\Pev{\sigma_{k}} \IndexPoly{Y}{i} = \Pev{\sigma_0} \IndexPoly{Y}{i}$ cached from the initialization,
    the new value for $v_i$ can be computed in constant time:
    \[v_i \gets \left(\Pev{\sigma_{k}}\IndexPoly{Y}{i}\right)\cdot  (1 - u_i) \cdot u_i \cdot w_k^i.\]
    It is clear that the total size of all the $U_x$ is in $\O(|P|)$.
    Thus the time needed for all updates
    to $\overline{v}$ is linear in $|P|$.

    Computing the sum $\sum_i v_i$ naively for every step
    would only yield a quadratic runtime.
    Instead, we compute $s \gets \sum_i v_i$ once after the initialization of $\overline{v}$
    and whenever updating an entry of $\overline{v}$
    we update $s$ accordingly.\end{proof}

\subsubsection{\texorpdfstring{$\GadgetResolution(p, u_1, u_2)$}{resolution(p, u1, u2)}}
In the main part, we did not consider the gadget for \ref{valid:resolution},
\begin{align*}
	&\GadgetResolution(p, u_1, u_2) \\
	&= \DegreeReduction{X} \DegreeReduction{Y} \left(
        \underbrace{\left(2^{\Abs{Z}}\Pev{\sigma^{Z}_{2^{-1}}}\DegreeReduction{Z}(u_1\cdot\overwriteVars{Y}{Z}p)\right)
	}_{\psi_1}
    \cdot \underbrace{\left(2^{\Abs{Z}}\Pev{\sigma^{Z}_{2^{-1}}}\DegreeReduction{Z}(u_2\cdot\overwriteVars{Y}{Z}p)\right)
	}_{\psi_2} \cdot  (1-p) \right).
\end{align*}
The prover needs to evaluate EACs of the following forms:

\FixEnumerateItemRefParentheses{\begin{enumerate}[(i)]
    \item\label{proper:gadget-resolution:z}
    $\DegreeReduction{Z'}(u_i\cdot\overwriteVars{Y}{Z}p)$, where $i \in \{1,2\}$ and $Z' \subseteq Z$,
	\item\label{proper:gadget-resolution:y}
	$\DegreeReduction{Y'} \left( \psi_1 \cdot  \psi_2 \cdot  (1-p) \right)$, where $Y' \subseteq Y$, and
	\item\label{proper:gadget-resolution:x}
	$\DegreeReduction{X'} \DegreeReduction{Y} \left( \psi_1 \cdot  \psi_2 \cdot  (1-p) \right)$, where $X' \subseteq X$.
\end{enumerate}
}

Note that for \ref{proper:gadget-resolution:z}, we store $u_i$ and  $\overwriteVars{Y}{Z} p$ as $Z$-indices and apply \autoref{lem:y-index:product:efficient}.
\ref{proper:gadget-resolution:y} is analogous to \autoref{lem:y-index:product:efficient},
the only difference being that we now consider a product of three $Y$-indices,
which, however, poses no obstacle.
The degree reduction \ref{proper:gadget-resolution:x} remains.
Note that
\begin{IEEEeqnarray*}{rCl}
	\DegreeReduction{Y} \left( \psi_1 \cdot  \psi_2 \cdot  (1-p) \right)
	&=& \sum_{i = m+1}^M \IndexPoly{Y}{i}
	\cdot \underbrace{\Arith{ C_{U_1(i)} } \cdot \Arith{ C_{U_2}(i) } \cdot (1 - \Arith{ C_i })}_{r_i} \\
	&=& \sum_{i = m+1}^M \IndexPoly{Y}{i}
	\cdot \Bigl( 1 - \prod_{\ell \in C_{U_1(i)}} \Arith{ \ell }  \Bigr)
	\cdot \Bigl( 1 - \prod_{\ell \in C_{U_2(i)}} \Arith{ \ell }  \Bigr)
	\cdot \prod_{\ell \in C_{i}} \Arith{ \ell }.
\end{IEEEeqnarray*}
Since $C_i = \Res_{x^\ast}(C_{U_1(i)}, C_{U_2(i)})$,
every variable $x \in X\setminus \{x^\ast\}$ that occurs in $r_i$ also occurs in $C_i$.
In particular, for all $x \in X \setminus \{x^\ast\}$ occurring in $r_i$
at least one of
$\pev{x \coloneqq 0} r_i = 0$
and
$\pev{x \coloneqq 1} r_i = 0$
holds.
Thus, we can apply the following lemma,
which is a generalization of \autoref{lem:prover:evaluate-binary-p}:
\begin{lemma}
	\label{lem:y-index:degree-reduction-x}
    Let $R \in \F[X][Y]_{\le 1}$ be a $Y$-index and $X' \subseteq X$ a subset of variables.
	Suppose that, for every index $i$, there is $x^\ast_i \in X$ such that,
	for all $x \in X' \setminus \{x^\ast_i\}$,
	$\pev{x \coloneqq 0} {r_i} = 0$ or $\pev{x \coloneqq 1} {r_i} = 0$.
	Then an EAC $Q = \DegreeReduction{X'} {P}$
	can be computed in time $\O(\Abs{P})$.
\end{lemma}
\begin{proof}
	The argument is similar to that in \autoref{lem:prover:evaluate-binary-p}:
	Consider a variable $x \in X'$.
    Define a $Y$-index $R' = \sum_i \IndexPoly{Y}{i} r'_i$ by setting
	\[
	r'_i \coloneqq \begin{cases}
		x \cdot  \pev{x\coloneqq 1} r'_i &\text{ if } \pev{x \coloneqq 0}{r_i} = 0,\\
		(1-x) \cdot \pev{x\coloneqq 0} r'_i &\text{ if } \pev{x \coloneqq 1}{r_i} = 0,\\
		x \cdot  \pev{x\coloneqq 1} r'_i +  (1-x) \cdot \pev{x\coloneqq 0} r'_i  &\text{ if } x = x^\ast_i,\\
		r'_i &\text{ if $x$ does not occur in $r'_i$.}
	\end{cases}
	\]
	Then ${R'} = \DegreeReduction{x} {R}$.
	Note that $R'$ still satisfies the assumptions,
	so we can iterate this to obtain $Q$ as desired.

	The runtime of $\O(\Abs{P})$ is achieved by computing the $q_i$ in one single step:
	For $b \in \BooleanDomain$,
	let $X_b \subseteq X' \setminus \{x^\ast_i\}$
	be the set of variables occurring in $r_i$ such that $\pev{x \coloneqq 1-b}{R} = 0$.
	If $X_0 \cap X_1 \neq \emptyset$, we have $q_i = 0$.
	Otherwise,
	let $\sigma\colon X_0 \cup X_1 \to \BooleanDomain$ be the assignment
	sending all $x \in X_b$ to $b$.
	Then, if $x^\ast_i \not\in X'$,
	\[
	q_i =
	\prod_{x \in X_0} (1-x)
	\cdot \prod_{x \in X_1} x
	\cdot \Pev{\sigma} r_i
	\]
	and otherwise, i.e., if $x^\ast_i \in X'$,
	\[
	q_i = x^{\ast} \cdot \pev{x^\ast_i \coloneqq 1}
	\prod_{x \in X_0} (1-x)
	\cdot \prod_{x \in X_1} x
	\cdot \Pev{\sigma} r_i
	+ (1 - x^{\ast}) \pev{x^\ast_i \coloneqq 0}
	\prod_{x \in X_0} (1-x)
	\cdot \prod_{x \in X_1} x
	\cdot \Pev{\sigma} r_i
	.
	\]
\end{proof}

Using the same techniques as in \autoref{lem:prover:evaluate-binary-p:optimized},
we can again improve this to computing an entire sequence of evaluations:
\begin{lemma}
	\label{lem:y-index:degree-reduction-x:optimized}
	Let $R$ be as in \autoref{lem:y-index:degree-reduction-x}.
    Let $\sigma_0, \ldots, \sigma_n$ be assignments such that
    $\sigma_k$ and  $\sigma_{k+1}$ only differ in $x_{k+1}$.
	Then,
    the sequence of evaluations
    \[
        \Pev{\sigma_0} \DegreeReduction{\{x_1,\ldots,x_n\}} {R},
        \Pev{\sigma_1} \DegreeReduction{\{x_2,\ldots,x_n\}} {R},
        \ldots,
        \Pev{\sigma_n} {R}
    \]
    can be computed online in total time $\O(|R|)$.
\end{lemma}
 \subsection{Degree Reductions in Z}\label{sec:degZ}
Recall that the set of variables $Z$ was chosen such that  $|Z| = |Y|$.
We only use the variables $Z$ to arithmetize relations between indices of clauses,
namely the functions $U_1$ and $U_2$ encoding the antecedent lists of the refutation.

Remember that we use $i \preceq j$ to denote that $i$ is a prefix of $j$ in their binary representation starting at their most significant bits.
Moreover, we denote their concatenation by $i \concat j$.
In the entire section, we write $Z = \{z_1,\ldots, z_{|Z|}\}$.

\subsubsection{\texorpdfstring{$\GadgetResolution(p,u_1,u_2)$}{resolution(p, u1, u2)}}
For the two occurrences of $\DegreeReduction{Z}$ in $\GadgetResolution(p, u_1, u_2)$,
the gadget for \ref{valid:resolution},
we can simply store both factors in $Z$-indices and apply \autoref{lem:y-index:product:efficient}.

\subsubsection{\texorpdfstring{$\GadgetBinary(u_i)$}{binary(ui)}}

Consider $\GadgetBinary(u_1)$, the gadget for \ref{valid:binary} for $u_1$;
$\GadgetBinary(u_2)$ is of course analogous.
We cannot apply \autoref{lem:y-index:product:efficient} because building a $(Y \cup Z)$-index is too expensive
and the degree reduction does not act directly on a multiplication of two multilinear polynomials.
Instead, we find a procedure specifically tailored to this case.
First note that
\[
u_1 \cdot (1 - u_1) = \Bigl(\sum_i \IndexPoly{Y}{i} \cdot \IndexPoly{Z}{U_1(i)}\Bigr)
\Bigl(1 - \sum_{j} \IndexPoly{Y}{j} \IndexPoly{Z}{U_1(j)} \Bigr).
\]
Thus
\[
    \DegreeReduction{Y} \left( u_1 \cdot (1-u_1) \right)
    = \sum_i \IndexPoly{Y}{i} \left( \IndexPoly{Z}{U_1(i)} - \IndexPoly{Z}{U_1(i)}^2 \right)
    = \underbrace{\MultilinEncode{Y,Z}{U_1}}_{\alpha} - \underbrace{\sum_i \IndexPoly{Y}{i} \MultilinEncode{Z}{U_1(i)}^2
}_{\beta} \]
We can store $\alpha$ as a $Z$-index
and evaluate it efficiently:
\begin{lemma}
    \label{lem:index:online-evaluation}
    Consider a $Z$-index $P = \sum_{j} \IndexPoly{Z}{j} p_j$,
    where $p_j \in \F[X]$.
    Let $\sigma_0, \sigma_1,\ldots, \sigma_{|Z|}$ be assignments $Z \cup X \to \F$ such that
    $\sigma_i$ and $\sigma_{i+1}$ differ only in $z_{i+1}$.
    Then the list of evaluations
    $\Pev{\sigma_0} P, \ldots, \Pev{\sigma_{|Z|}} P$ can be computed online
    in total time $\O(|P| + 2^{|Z|})$.
\end{lemma}
\begin{proof}
    This is a variant of \autoref{lem:y-index:product:efficient} with only a single factor
    (and using $Z$ instead of $Y$).
    Indeed, compute $\cT$ and  $v_P^{(k)}$ as in the proof of \autoref{lem:y-index:product:efficient}
    (see Appendix \ref{sec:proof:lem:y-index:product:efficient}).
    Then
    \[\Pev{\sigma_k} P = \sum_{c \in \BooleanDomain^{|Z| - k}} \cT(c) \cdot v_P^{(k)}(c). \qedhere\]
\end{proof}

For $\beta$, we need a modification of this, since the key polynomials appear squared,
i.e., we are no longer working with a $Z$-index:
\begin{IEEEeqnarray*}{rCl}
    \sum_i \IndexPoly{Y}{i} \IndexPoly{Z}{U_1(i)}^2
    &=& \sum_j \IndexPoly{Z}{j}^2 \left( \sum_{i \in U_1^{-1}(j)} \IndexPoly{Y}{i} \right).
\end{IEEEeqnarray*}

\begin{lemma}
    \label{lem:squared-index}
    Consider $P = \sum_{j} \IndexPoly{Z}{j}^{2} p_j$,
    where $p_j \in \F[X]$.
    Let $\sigma_0, \sigma_1,\ldots, \sigma_{|Z|}$ be assignments $Z \cup X \to \F$ such that
    $\sigma_i$ and $\sigma_{i+1}$ differ only in $z_{i+1}$.
    Then the list of evaluations
    $\Pev{\sigma_0} \DegreeReduction{Z} P, \Pev{\sigma_1} \DegreeReduction{\{z_1,\ldots, z_{|Z|-1}\}} P,
    \ldots, \Pev{\sigma_{{|Z|}-1}} \DegreeReduction{\{z_{|Z|-1}\}} P, \Pev{\sigma_{|Z|}} P$ can be computed online
    in total time $\O(|P| + 2^{|Z|})$.
\end{lemma}
\begin{proof}
    The idea is again the same as \autoref{lem:index:online-evaluation}.
    However, instead of $v_P^{(k)}$ we compute
    \[\tilde{v}_P^{(k)}(c) = \sum_{c \preceq j} \Pev{\sigma_{k}} \IndexPoly{\{y_0, \ldots, y_{k}\}}{j \bmod{2^k}}^2 p_j.\]
    Then
    \[\Pev{\sigma_k} \DegreeReduction{\{z_{k},\ldots,z_{|Z|-1}\}}{P} = \sum_{c \in \BooleanDomain^k} \cT(c) \cdot \tilde{v}_P^{(k)}(c).\qedhere\]
\end{proof}

\subsubsection{\texorpdfstring{$\GadgetRegressive(u_i)$}{regressive(ui)}}

Only the $\DegreeReduction{Z}$ of $\GadgetRegressive(u_i)$, the gadget for \ref{valid:regressive}, remains.
By \autoref{lem:y-index:product}, we know that
\begin{IEEEeqnarray*}{rCl}
    \DegreeReduction{Y} (u_1 \cdot [Y \le Z])
    &=& \DegreeReduction{Y}\Bigl( \bigl( \sum_i \IndexPoly{Y}{i} \IndexPoly{Z}{U_1(i)} \bigr) \cdot \bigl( \sum_{j \le k} \IndexPoly{Y}{j} \IndexPoly{Z}{k} \bigr) \Bigr) \\
    &=& \sum_i \IndexPoly{Y}{i} \cdot \IndexPoly{Z}{U_1(i)} \cdot \sum_{k \ge i} \IndexPoly{Z}{k}\\
    &=& \sum_j \IndexPoly{Z}{j} \cdot \sum_{U_1(i) = j} \IndexPoly{Y}{i} \cdot \sum_{k \ge i} \IndexPoly{Z}{k}.
\end{IEEEeqnarray*}

As in \autoref{lem:squared-index},
we prove that the sequence of evaluations can be computed in time linear in the size of the $Z$-index.
This is the most elaborate construction of all degree reductions:
\begin{lemma}
	\label{lem:efficient_v5z}
	Consider
    \[
        P = \sum_j \IndexPoly{Z}{j} \cdot \Bigl(\sum_{U_1(i) = j} \IndexPoly{Y}{i} \cdot \bigl( \sum_{k \ge i} \IndexPoly{Z}{k} \bigr)\Bigr),
    \]
    where $U_1$ is regressive.
	Let $\sigma_0, \sigma_1,\ldots, \sigma_{|Z|} \in \F^{Z \cup Y}$ be assignments
    such that
	$\sigma_i$ and $\sigma_{i+1}$ differ only in $z_{i+1}$.
	Then the list of evaluations
	\[
        \Pev{\sigma_0} \DegreeReduction{Z} P, ~
        \Pev{\sigma_1} \DegreeReduction{\{z_1,\ldots, z_{|Z|-1}\}} P,~
        \ldots,~
        \Pev{\sigma_{{|Z|}-1}} \DegreeReduction{\{z_{|Z|-1}\}} P,~
        \Pev{\sigma_{|Z|}} P
    \]
    can be computed online in total time $\O(|P|)$.
\end{lemma}
\begin{proof}
    Remember that $\Bit{t}{j}$ denotes the $t$-th
    bit of $j$ and similarly for other integers.
    In a slight abuse of notation,
    we let $\ArithBit{0}{j}, \ldots, \ArithBit{{|Z|-1}}{j}$
    represent the factors of $\IndexPoly{Z}{j}$
    that correspond to the individual bits,
    i.e., $\IndexPoly{Z}{j} = \prod_t \ArithBit{t}{j}$,
    where $\ArithBit{t}{j} \in \{z_t, 1- z_t\}$.

	We fix an arbitrary $t \in \{0, \ldots, {|Z|-1}\}$.
    In the following, we examine $\DegreeReduction{\{z_{t+1},\ldots, z_{|Z|-1}\}} P$,
    which is a quadratic polynomial in $\{z_0, \ldots,z_t\}$
    and multilinear in $\{z_{t+1}, \ldots, z_{|Z|-1}\} \cup Y$.
Note that we need only consider values $k$ where
    \begin{equation}\label{eq:upper_bits_eq}
    	\ArithBit{{t+1}}{j} = \ArithBit{{t+1}}{k}, \ldots, \ArithBit{{|Z|-1}}{j} = \ArithBit{{|Z|-1}}{k}
    \end{equation}
    hold.
    Otherwise,
    $\IndexPoly{Z}{j} \IndexPoly{Z}{k}$ contains a factor $z_{i} \cdot (1 - z_{i})$ for some $i > t$,
    and thus
    \[\DegreeReduction{\{z_{t+1}, \ldots, z_{|Z|-1}\}} \IndexPoly{Z}{j} \IndexPoly{Z}{k} = 0.\]
    \def\UpperBitsEq#1#2#3{#2 \sim_{#1} #3}
    Let $\UpperBitsEq{t}{j}{k}$ denote that \eqref{eq:upper_bits_eq}
    holds for the pair $(j,k)$ and the bits $t+1, \ldots, \Abs{Z}-1$.
    We obtain
\begin{IEEEeqnarray*}{rCl}
		&& \DegreeReduction{\{z_{t+1},\ldots, z_{|Z|-1}\}}
        \sum_j \ArithBit{{|Z|-1}}{j} \cdot \ldots \cdot \ArithBit{1}{j} \cdot\Bigl(\sum_{U_1(i) = j} \IndexPoly{Y}{i} \cdot \bigl( \sum_{k \ge i} \ArithBit{{|Z|-1}}{k} \cdot \ldots \cdot \ArithBit{1}{k} \bigr)\Bigr)\\
		&=& \DegreeReduction{\{z_{t+1},\ldots, z_{|Z|-1}\}}
    \sum_j \ArithBit{{|Z|-1}}{j}^2 \cdot \ldots \cdot \ArithBit{{t+1}}{j}^2 \cdot \ArithBit{t}{j} \cdot \ldots \cdot \ArithBit{1}{j} \cdot \Bigl(\sum_{U_1(i) = j} \IndexPoly{Y}{i} \cdot \bigl( \sum_{\substack{k \ge i\\\UpperBitsEq{t}{j}{k}}} \ArithBit{t}{k} \cdot \ldots \cdot \ArithBit{1}{k} \bigr)\Bigr)\\
		&=& \DegreeReduction{\{z_{t+1},\ldots, z_{|Z|-1}\}} \sum_j \ArithBit{{|Z|-1}}{j} \cdot \ldots \cdot \ArithBit{1}{j}
        \cdot \Bigl(\sum_{U_1(i) = j} \IndexPoly{Y}{i} \cdot \overbrace{\bigl( \sum_{\substack{k \ge i\\\UpperBitsEq{t}{j}{k}}} \ArithBit{t}{k} \cdot \ldots \cdot \ArithBit{1}{k} \bigr)}^{\xi_{j,i}^{(t)}}\Bigr).
    \end{IEEEeqnarray*}
    Consider one of the $\xi_{j,i}^{(t)}$.
    Since $U_1$ is regressive, we have $j < i$.
    Let ${m(i,j)} \in \{0, \ldots, {|Z|-1}\}$ denote the highest index such that $\Bit{{m(i,j)}}{j}$ and $\Bit{{m(i,j)}}{i}$ differ,
    i.e., $\Bit{{m(i,j)}}{j} = 0$ and $\Bit{{m(i,j)}}{i} = 1$.
    During initialization, we compute ${m(i,j)}$ for each pair $(i,j)$
    with $i \in U_1^{-1}(j)$.
    This can be done in time $\O(M)$.
Consider the following cases:

    \FixEnumerateItemRefParentheses{\begin{enumerate}
        \item $t < m(i,j)$,
            i.e., $\Bit{t+1}{j}, \ldots, \Bit{{|Z|-1}}{j}$ and $\Bit{t+1}{i}, \ldots, \Bit{{|Z|-1}}{i}$ differ in at least one bit:
            In this case we have $\xi_{j,i}^{(t)} = 0$,
            since there does not exist a $k \ge i$ with $\UpperBitsEq{t}{j}{k}$.
        \item\label{lem:efficient_v5z:cases:l=m}
            $t = {m(i,j)}$:
            In this case $\xi_{j,i}^{(t)}$ simplifies to
	        $\sum_{i \bmod 2^{m(i,j)} \le k < 2^{m(i,j)}} \ArithBit{{{m(i,j)}}}{k} \cdot \ldots \cdot \ArithBit{1}{k}$.
            Note that for a fixed ${m(i,j)}$, there are only $2^{m(i,j)}$ possible values for $i$,
            and thus only $2^{m(i,j)}$ possible values for this sum.
            We compute the evaluations for all of those for $\sigma_t$.
	        Since the assignments for variables with an index smaller than ${m(i,j)}$ do not change in future evaluations,
            we can reuse the computed sums of the previous evaluation and only have to account for the updated value of $\Pev{\sigma_t}\ArithBit{{m(i,j)}}{k}$.
	        In total, this requires time $\O(\sum_{{m(i,j)} = 1}^{|Z|-1} 2^{m(i,j)}) = \O(2^{|Z|}) = \O(M)$.
        \item $t > {m(i,j)}$:
            In particular, we have $\UpperBitsEq{t}{i}{j}$.
            Using $\UpperBitsEq{t}{j}{k}$,
            it follows that
            $k = (\Bit{{|Z|-1}}{i} \ldots \Bit{t+1}{i} \Bit{t}{k} \ldots \Bit{1}{k})_2$.
            We proceed with a case analysis on the value of $\Bit{t}{j}$:
            \begin{itemize}
                \item If $\Bit{t}{j} = 1$, then $\Bit{t}{k} = \Bit{t}{i} = 1$
                    because otherwise $k = (\Bit{{|Z|-1}}{i} \ldots \Bit{t+1}{i} 0 \Bit{t-1}{k} \ldots \Bit{1}{k})_2 < i$.
                Thus, $\xi_{j,i}^{(t)}$ simplifies to
	            \begin{IEEEeqnarray*}{rCl}
	            	\xi_{j,i}^{(t)} &=&
                    \sum_{{\substack{k \ge i\\\UpperBitsEq{t}{j}{k}}} } \underbrace{\ArithBit{t}{k} }_{z_t} \cdot \ldots \cdot \ArithBit{1}{k}
                    = z_t \cdot \sum_{{\substack{k \ge i\\\UpperBitsEq{t-1}{j}{k}}} }
                        \ArithBit{{t-1}}{k} \cdot \ldots \cdot \ArithBit{1}{k}
                    = z_t \cdot \xi_{j,i}^{(t-1)},
	            \end{IEEEeqnarray*}
                where $\xi_{j,i}^{(t-1)}$ is the corresponding value from the previously computed degree reduction step,
                $\DegreeReduction{\{z_t,\ldots,z_{|Z|-1}\}} P$.

        \item If $\Bit{t}{j} = 0$,
                    then both $\Bit{t}{k} = \Bit{t}{i} = 0$ and $\Bit{t}{k} = 1$ are possible.
                Thus,
	            \begin{IEEEeqnarray*}{rCl}
	            	\xi_{j,i}^{(t)}
                    &=& (1-z_t) \cdot \sum_{\substack{k\ge i\\\UpperBitsEq{t}{j}{k}\\k_t = 0}} \ArithBit{{t-1}}{k} \cdot \ldots \cdot \ArithBit{1}{k}
                \quad + z_t \cdot \sum_{\substack{k\ge i\\\UpperBitsEq{t}{j}{k}\\k_t = 1}} \ArithBit{{t-1}}{k} \cdot \ldots \cdot \ArithBit{1}{k}\\
                    &=& (1-z_t) \cdot \sum_{{\substack{k\ge i\\\UpperBitsEq{t-1}{j}{k}}}} \ArithBit{{t-1}}{k} \cdot \ldots \cdot \ArithBit{1}{k}
                    \quad + z_t \cdot \underbrace{\sum_{k' < 2^{t-1}} \ArithBit{{t-1}}{k'} \cdot \ldots \cdot \ArithBit{1}{k'}}_{1}\\
                    &=& (1 - z_t) \cdot \xi_{j,i}^{(t-1)} + z_t,
	            \end{IEEEeqnarray*}
                where, again, $\xi_{j,i}^{(t-1)}$ denotes the value from the previous degree reduction.
\end{itemize}
                The key insight is that, for a fixed $j$,
                we can handle all  $\xi_{j,i}^{(t)}$ with $t > {m(i,j)}$
                in a single step.
                Thus it suffices to keep track of the sums
                \[
                    A_{j, t} \coloneqq \sum_{\substack{U_1(i) = j\\{m(i,j)} < t} } \IndexPoly{Y}{i}\xi_{j,i}^{(t)}
                    \quad\text{and}\quad
                    B_{j, t} \coloneqq \sum_{\substack{U_1(i) = j\\{m(i,j)} < t} } \IndexPoly{Y}{i},
                \]
                and we do not need to update each $\xi_{j,i}^{(t)}$ individually.
                Indeed,
                for $\Bit{t+1}{j} = 1$,
                \[
                    A_{j, t+1} = z_{t+1} \cdot A_{j, t} + \sum_{\substack{U_1(i) = j\\t = m(i,j)}} \xi_{j,i}^{(t)},
                \]
                and, for $\Bit{t+1}{j} = 0$,
                \[
                    A_{j, t+1} = (1-z_{t+1}) \cdot A_{j, t} + z_{t+1} \cdot B_{j,t} + \sum_{\substack{U_1(i) = j\\t = m(i,j)}} \xi_{j,i}^{(t)}.
                \]
    \end{enumerate}}
	Putting everything together:
	We start with a preprocessing step to compute ${m(i,j)}$ for each pair $(i,j)$ with $i \in U_1^{-1}(j)$,
    which can be done in $\O(M)$.
As described in case \ref{lem:efficient_v5z:cases:l=m},
    we keep track of a tree
    of all possible sums
    $\sum_{i \bmod 2^{m(i,j)} \le k < 2^{m(i,j)}} \ArithBit{{{m(i,j)}-1}}{k} \cdot \ldots \cdot \ArithBit{1}{k}$,
    which takes time $\O(M)$ in total.

    As in the proof of \autoref{lem:y-index:product:efficient}
    (Appendix \ref{sec:proof:lem:y-index:product:efficient}),
    we use a tree $\cT$
    to store the values $\ArithBit{{|Z|-t}}{j} \cdot \ldots \cdot \ArithBit{|Z|-1}{j}$. Each path from the root to the leaves represents one possible value for $j$.
    For each $j$, we store the sums $A_{j,t}$ and $B_{j,t}$.
	After each degree reduction query from the verifier,
    we advance one level in the tree
    and combine the $A_{j,t}$ and $B_{j,t}$ of several $j$, i.e.,
    using the notation of the proof of \autoref{lem:y-index:product:efficient},
    we store only the sums
    \[v_A^{(t)}(c) \coloneqq \sum_{c \preceq j} \Pev{\sigma_{t}} \IndexPoly{\{z_0, \ldots, z_{t}\}}{j \bmod{2^t}} A_{j,t}\]
    for all prefixes $c \in \BooleanDomain^{|Z| - 1 - t}$ instead of each individual $A_{j,t}$
    and likewise for $B_{j,t}$.
    Note that later case distinctions depend only on the prefix $c$,
    not on the exact value of $j$.
Since we need to visit every node of the tree twice (once during initialization and once for the whole sequence), this takes time $\O(2^{|Z|}) = \O(M)$.
	Hence, the time to answer all degree reduction queries is $\O(M)$.
\end{proof}
  
\section{Adaptation for RUP}\label{sec:rup}

We showed how to construct an EAC based on a Boolean formula which is satisfiable iff there exists a RES-refutation. Analogously, we can construct an EAC which is satisfiable iff there exists a RUP-refutation.
 \subsection{Polynomial RUP-Refutations}

We start by giving a brief definition of RUP-refutations:

\begin{definition}
	Let $\varphi=(C_0,\ldots,C_{m-1})$ denote a CNF-formula with set of variables $X$,
    $|X| = n$.
    We say that a list $(\ell_0,\ldots,\ell_k)$ of literals is a \emph{UP-sequence} if for every $i=0,\ldots,k$ there is a clause $C\in\varphi$ with $C\subseteq\{\neg \ell_0,\ldots,\neg \ell_{i-1},\,\ell_i\}$,
    i.e., we can derive the literals $\ell_0, \ldots, \ell_k$ in this order by unit propagation.
    We say that $\varphi$ is \emph{UP-unsatisfiable (with witness $(\ell_0,\ldots,\ell_k)$)}
    iff $(\ell_0,\ldots,\ell_k)$ is a UP-sequence and there exists a clause $C\in\varphi$ with $C\subseteq\{\neg \ell_0,\ldots,\neg \ell_k\}$.
    To derive a new clause $C_{i+1} = \{\ell_0, \ldots, \ell_k\}$ from a CNF-formula $\varphi = (C_0,\ldots, C_i)$,
    we define the \emph{step formula}
    \[
        \RUPStep_\varphi(C_{i+1})\coloneqq (C_0,\ldots,C_{i},\neg \ell_1,\ldots,\neg \ell_k).
    \]
    A \emph{RUP-refutation of} $\varphi = (C_0,\ldots,C_{m-1})$ is a list of clauses $\Phi=(C_0,\ldots,C_{M-1})$ with $M>m$, s.t.
	\begin{enumerate}
		\item the first $m$ clauses of $\varphi$ and $\Phi$ match,
		\item $C_{M-1}=\emptyset$, and
        \item $\RUPStep_{\Phi^{(i)}}(C_{i})$ is UP-unsatisfiable for $i=m,\ldots,M-1$,
            where $\Phi^{(i)} \coloneqq (C_0,\ldots, C_{i-1})$.
	\end{enumerate}
    Wlog.~we may assume that a UP-sequence contains at most $n+1$ literals,
    and thus to derive $C_{i}$ from $\Phi^{(i)}$ only $n+1$ clauses of  $\Phi^{(i)}$ are actually used.
	We say that $(j_0,\ldots,j_k)$ is an \emph{antecedent list} for $C_{i}$ if $j_0,\ldots,j_k\in\{0,\ldots,i-1\}$
    and $\RUPStep_{\Phi'}(C_{i})$ is UP-unsatisfiable, where $\Phi'\coloneqq (C_{j_0},\ldots,C_{j_k})$.
\end{definition}

Note that a RUP-refutation is --- like a RES-refutation --- a sequence of clauses.
The crucial difference between both refutations is how the clauses are deduced based on the previous ones. Hence, only criterion 3 differs in the definition.

We need to adapt \autoref{def:poly-ref}, the definition for polynomial refutation, to account for clauses derived by UP-sequences instead of clauses derived by resolution.
To this end, we fix the set of function variables $\Wrup = \{p, u_1, \ldots, u_{n+1}\}$.

\begin{definition}
	A multilinear interpretation $\omega\colon \Wrup \to \F[X, Y, Z]_{\le 1}$
	is called a \emph{polynomial RUP-refutation}
	iff it satisfies properties
    \ref{valid:binary}\;--\;\ref{valid:regressive} from \autoref{def:poly-ref} and
    \FixEnumerateItemRefParentheses[V]{\begin{enumerate}[(V1')]
        \setcounter{enumi}{5}
        \item $\Bigl(\Pev{\IndexA{Y}{U_1(i)}}\,p \cdot\ldots\cdot\Pev{\IndexA{Y}{U_{n+1}(i)}}\,p\cdot(1-\Pev{\IndexA{Y}{i}}\,p)\Bigr) \restrict{\BooleanDomain^X} = 0$ for $i=m,...,M-1$.
	\end{enumerate}}
\end{definition}

Note that properties \ref{valid:binary}, \ref{valid-clause-mapping}, and \ref{valid:regressive} also have to hold for all $u_1, \ldots, u_{n+1}$. Hence, the number of EACs increases since we have to check the properties for more function variables. However, the construction of each circuit does not change except for property \ref{valid:resolution}.

 \subsection{EAC-Gadgets}
It is not hard to verify that polynomial RUP-refutations are a sound and complete method for showing unsatisfiability,
i.e., \autoref{lem:induced-refutation-is-valid} and \autoref{lem:polynomial-refutation-complete} hold mutatis mutandis.
As already mentioned, only the EAC-gadget for \ref{valid:resolution} changes to account for larger antecedent lists.

We still use $\psi_i\coloneqq 2^{\Abs{Z}}\Pev{\sigma^{Z}_{2^{-1}}}\DegreeReduction{Z}(u_i\cdot\overwriteVars{Y}{Z}\,p)$ to access the $i$-th antecedent.
However, our antecedent list can now consist of up to $n+1$ indices.
Hence, instead of $\GadgetResolution \coloneqq \DegreeReduction{X}\DegreeReduction{Y}(\psi_1 \cdot \psi_2 \cdot (1-p))$
we use the gadget
\[\GadgetRUP \coloneqq \DegreeReduction{X} \DegreeReduction{Y} \big( \psi_1 \cdot \psi_2 \cdot \ldots  \cdot \psi_{n+1} \cdot (1-p)\big).\]
The proof of correctness of $\GadgetRUP$ is nearly identical to that in the case of $\GadgetResolution$, \autoref{lem:gadget:resolution},
and thus omitted.
The nontrivial changes lie in the implementation of an efficient prover:
Recall from \autoref{lem:y-index:degree-reduction-x} that for $\GadgetResolution$,
the degree reduction in $X$ can be easily computed
since the term $\bigl(1-\Pev{\IndexA{Y}{i}}\,\omega(p)\bigr) = 1 - \Arith{C_i}$ of \ref{valid:resolution} evaluates to $0$ for all except one binary assignment to the variables of $C_i$
  and -- apart from the pivot variable -- all variables of the antecedents of $C_i$
  occur in $C_i$.

On the other hand, in $\GadgetRUP$ the antecedents can contain more variables than $\RUPStep_{\Phi^{(i)}}(C_i)$.
To ensure that we always have a unit clause during degree reduction,
which allows us to consider only one assignment of the next variable,
we need to handle the variables in the same order as they appear in the UP-sequence.
Thus we need to keep track of this order for every $i = m,\ldots, M-1$.
More precisely, for every $i$ let $\tau_i\colon [n] \to [n]$ be a permutation
such that in the UP-sequence deriving $\RUPStep_{\Phi^{(i)}}(C_i)$,
the variables occur in the order $x_{\tau_i(0)}, x_{\tau_i(1)}, \ldots, x_{\tau_i(n-1)}$.

Fix a \emph{sorting network} $E = ((\varepsilon_0, \delta_0), \ldots, (\varepsilon_{e-1}, \delta_{e-1}))$, $\varepsilon_k \neq \delta_k \in [n]$,
i.e., $E$ is such that every permutation $\tau\colon [n] \to [n]$
can be written as a composition of transpositions
$\tau = (\varepsilon_{s_l}, \DegreeReduction{s_l}) \circ \ldots \circ (\varepsilon_{s_0}, \DegreeReduction{s_0})$
for some sequence $0 \le s_0 < s_2 < \ldots < s_l \le e-1$.
An example is the \emph{zig-zag sorting network} of Goodrich,
which has size $e = \mathcal{O}(n \log n)$
\cite{zig-zag-sorting}.

Using this sorting network, all the $\tau_i$ with  $i = m, \ldots, M-1$
can be encoded with $e$ binary multilinear polynomials $v_j \in \Multilinear{Y}$.
Let $v_j = \sum_{i \in I_j} \IndexPoly{Y}{i}$,
where $I_j$ contains the indices $i$ of all permutations $\tau_i$ such
that $j$ occurs in the sequence $s_0, \ldots, s_l$ corresponding to $\tau_i$.
We introduce new function symbols for the $v_j$.
Moreover, we allow for an additional type of node in our EACs,
$\pev{x \leftrightarrow y}$, which swaps the variables $x$ and $y$.
In a round of $\EACertify$ for $\pev{x \leftrightarrow y}$, the verifier simply adapts its assignment and leaves the claims unchanged.
(This does not change the definition of EACs substantially, since $\pev{x \leftrightarrow y}$ can also be written as $\pev{x \leftrightarrow y} = \pev{\hat{x} \coloneqq y}\pev{y \coloneqq x}\pev{x \coloneqq \hat{x}}$, where $\hat{x}$ is a fresh variable.)
For an EAC $\phi$, let $\GadgetSort_0(\phi) \coloneqq \phi$,
\[\GadgetSort_{j+1}(\phi) \coloneqq
    \bigl(\pev{x_{\varepsilon_{j}} \leftrightarrow x_{\DegreeReduction{j}}} \DegreeReduction{Y}(v_{j} \cdot \GadgetSort_{j}(\phi))\bigr)
    +
    \bigl(\DegreeReduction{Y}((1-v_{j}) \cdot \GadgetSort_{j}(\phi))\bigr)
\]
and $\GadgetSort(\phi) \coloneqq \GadgetSort_e(\phi)$.
Note that when viewing the polynomial represented by $\phi$ as a $Y$-index,
$\phi = \sum_{i} \IndexPoly{Y}{i} \cdot p_i$,
then $\GadgetSort(\phi)$ has the effect of permuting the variables of  $p_i$
according to  $\tau_i$.
Replacing the gadget $\GadgetRUP$ by
\[
\GadgetRUP' \coloneqq \DegreeReduction{X} \DegreeReduction{Y} \big( \GadgetSort(\psi_1 \cdot \psi_2 \cdot \ldots \cdot \psi_{n+1} \cdot (1-p))\big)
\]
allows the prover to employ the same techniques as for $\GadgetResolution$.

We only implemented the prover for resolution refutation,
since this already demonstrates the feasibility of the central ideas, which are identical for both approaches.
Implementing the more involved prover for RUP-refutations is a possible direction for future research.
 \section{Overview of Notation}
\label{sec:notation}

\def\Entry#1#2#3{#1&#2&\pageref{#3}\\
}
\begin{tabular}{llr}
\Entry{$\F$}{finite prime field}{notation:F}
    \Entry{$\F[X]$}{polynomial ring in $X$ over $\F$}{notation:polynomial_ring}
    \Entry{$\Multilinear{X}$}{multilinear polynomials with variables from $X$}{notation:Multilinear}
    \Entry{$\BooleanDomain$}{the set $\{0,1\} \subseteq \F$}{notation:B}
    \Entry{$f\restrict{A'}$}{restriction of function $f\colon A \to B$ to $A' \subseteq A$}{notation:restriction}
\Entry{$\Lit(X)$}{set of literals,  $\Lit(X) = X \cup \{\overline{x_i} : x_i \in X\}$}{notation:Lit}  
    \Entry{$m$}{number of clauses in the formula}{notation:m}  
    \Entry{$M$}{number of clauses in the resolution refutation}{notation:M}
    \Entry{$\pev{x_i \coloneqq a} p$}{partial evaluation of $p$ at  $x_i \coloneqq a$}{notation:pev}
    \Entry{$\pev{x_i \coloneqq y} p$}{substitution of variable $y$ for  $x_i$ in $p$}{notation:pev-var}
    \Entry{$\overwriteVars{Y}{Z} p$}{shorthand for $\pev{y_1 \coloneqq z_1}\ldots\pev{y_{|Y|} \coloneqq z_{|Y|}}$}{notation:overwriteVars}
    \Entry{$\DegreeReduction{x_i} p$}{degree reduction in  $x_i$;
        $\DegreeReduction{x_i} p \coloneqq x_i \cdot \pev{x_i \coloneqq 1} p + (1-x_i) \cdot \pev{x_i \coloneqq 0} p$
    }{notation:degx}
    \Entry{$\DegreeReduction{X}$}{degree reduction multiple variables.
        $\DegreeReduction{X} = \DegreeReduction{x_1} \ldots\DegreeReduction{x_n}$.
    }{notation:degX}
    \Entry{$\deg_x(p)$}{degree of  $p \in \F[X]$ in variable  $x$}{notation:deg}
\Entry{$\Bit{k}{i}$}{$k$-th  bit of $i$}{notation:Bit}
    \Entry{$(b_{n-1} \ldots b_0)_2$}{binary number corresponding to bits $b_{n-1}, \ldots, b_0$}{notation:binary_number}
    \Entry{$i \concat j$}{concatenation of  $i$ and $j$}{notation:concat}
    \Entry{$i \preceq j$}{the most significant bits of $i$ are a prefix of the most significant bits of $j$}{notation:bit_prefix}
\Entry{$\IndexPoly{X}{i}$}{$i$-th key polynomial ($i \in \{0,\ldots,2^{|X|}-1\}$)}{notation:IndexPoly}
    \Entry{$\Arith{\ell}$}{arithmetization of a literal $\ell \in \Lit{X}$}{notation:Arith}
    \Entry{$\ArithLit{X}$}{set of arithmetized literals,  $\ArithLit{X} \coloneqq \{\Arith{\ell} : \ell \in \Lit(X)\}$}{notation:ArithLit}
    \Entry{$\Arith{C}$}{arithmetization of a CNF clause}{notation:ArithClause}
    \Entry{$\Arith{\varphi}_Y$}{arithmetization of a CNF formula}{notation:ArithCNF}
    \Entry{$\MultilinEncode{X}{M}$}{arithmetization of a set $M \subseteq \{0,\ldots,2^{|X|}-1\} \cong \BooleanDomain^X$}{notation:MultilinEncodeSet}
    \Entry{$\MultilinEncode{X,Y}{f}$}{arithmetization of a function $M \subseteq \{0,\ldots,2^{|X|}-1\} \to \{0,\ldots,2^{|Y|}-1\}$}{notation:MultilinEncode}
\Entry{$\sigma[a/x_i]$}{substitution of $a$ for $x_i$ in $\sigma$}{notation:substitution}
\Entry{$\sigma[{-}/x_i]$}{assignment agreeing with $\sigma$, but undefined at  $x_i$,  i.e., $\sigma[{-}/x_i] \coloneqq \sigma\restrict{X \setminus \{x_i\}}$}{notation:unbind}
    \Entry{$\sigma^{X}_{2^{-1}}$}{constant assignment $x_i \mapsto 2^{-1}$}{notation:assignment:const} \Entry{$\Pev{\sigma}p$}{(partial) evaluation of  $p \in \F[X]$ at  $\sigma\colon X' \to \F$}{notation:Pev}
    \Entry{$\IndexA{X}{k}$}{assignment $x_i \mapsto k_i$, ($k \in \{0,\ldots,2^{|X|}-1\}$)}{notation:IndexA}
\Entry{$\omega$}{mapping of function variables to functions, i.e.\ $\omega\colon W \to \left( \F^{X} \to \F\right)$}{notation:omega}
    \Entry{$\Eval{\psi}_{\omega}$}{evaluation of EAC $\psi$ with interpretation $\omega$}{notation:Eval}
    \Entry{$\Wres$}{set of function variables used in  $\ResCertify$,  $\Wres = \{p, u_1,u_2\}$}{notation:Wres}
\end{tabular}
 
\end{document}